\UseRawInputEncoding
 \documentclass[a4paper,11pt]{article}
\pdfoutput=1
\usepackage{jheppub}

\usepackage{amsmath}
\usepackage{mathtools}
\usepackage{multirow}
\usepackage{array}
\usepackage{bm}

\usepackage{graphicx}
\usepackage{lscape}
\usepackage{subfig}

\usepackage{bbm}
\usepackage{hyperref}
\usepackage{booktabs}
\usepackage{float}

\DeclareMathOperator{\shuffle}{\sqcup\mathchoice{\mkern-2.9mu}{\mkern-2.9mu}{\mkern-3mu}{\mkern-3.5mu}\sqcup}

\newcommand{\be}{\begin{equation}}
\newcommand{\ee}{\end{equation}}
\newcommand{\bea}{\begin{eqnarray}}
\newcommand{\eea}{\end{eqnarray}}
\newcommand{\bei}{\begin{itemize}}
\newcommand{\eei}{\end{itemize}}
\newcommand{\bean}{\begin{eqnarray*}}
\newcommand{\eean}{\end{eqnarray*}}
\newcommand{\nn}{\nonumber \\}

\def\eps{\epsilon}

\def\top #1{\mathcal{T}_{#1}}

\allowdisplaybreaks[1]

\newcommand\scalemath[2]{\scalebox{#1}{\mbox{\ensuremath{\displaystyle #2}}}}

\newcommand{\gammaAB}[2]{\ensuremath\gamma}

\newcommand{\dA}{\ensuremath d\AA}

\newcommand{\dAk}[1]{\ensuremath \underbrace{\dA \ldots \dA}_{\text{#1 times}}}

\newcommand{\Den}{\ensuremath D}
\newcommand{\dd}{\ensuremath \mathrm{d}}
\DeclareMathOperator{\dlog}{\mathit{d}log}

\newcommand{\minus}{\ensuremath \scalebox{0.5}[1.0]{\( - \)}}
\newcommand{\pminus}{\hphantom{\minus}}
\newcommand{\FF}{\ensuremath \text{F}}

\newcommand{\GG}{\ensuremath \text{I}}
\newcommand{\GGvec}{\ensuremath \mathbf{I}}

\newcommand{\MM}{\ensuremath \mathbb{M}}
\renewcommand{\AA}{\ensuremath \mathbb{A}}

\newcommand{\unipd}{Dipartimento di Fisica ed Astronomia, Universit\`a di Padova, Via Marzolo 8, 35131 Padova, Italy}
\newcommand{\pdinfn}{INFN, Sezione di Padova, Via Marzolo 8, 35131 Padova, Italy}

\newcommand{\argonne}{High Energy Physics Division, Argonne National Laboratory, Argonne, IL 60439, USA}

\allowdisplaybreaks

\author[a,b]{Pierpaolo Mastrolia,}
\author[b]{Massimo Passera,}
\author[a,b]{Amedeo Primo,}
\author[c]{Ulrich Schubert}

\affiliation[a]{\unipd}
\affiliation[b]{\pdinfn}
\affiliation[c]{\argonne}

\emailAdd{pierpaolo.mastrolia@pd.infn.it}
\emailAdd{massimo.passera@pd.infn.it}
\emailAdd{amedeo.primo@pd.infn.it}
\emailAdd{schubertmielnik@anl.gov}

\title{\boldmath Master integrals for the NNLO virtual corrections to $\mu e$ scattering in QED: the planar graphs}


\abstract{ 
We evaluate the master integrals for the two-loop, planar box-diagrams
contributing to the elastic scattering of muons and electrons at next-to-next-to 
leading-order in QED. We adopt the method of differential equations
and the Magnus exponential series to determine a canonical set of
integrals, finally expressed as a Taylor series around four
space-time dimensions, with coefficients written as combination of
generalised polylogarithms. The electron is treated as massless, while
we retain full dependence on the muon mass. The considered integrals
are also relevant for crossing-related processes, such as di-muon
production at $e^+ e^-$-colliders, as well as for the QCD
corrections to $top$-pair production at hadron colliders.
}

\begin{document}

\maketitle
\flushbottom
\section{Introduction}

The elastic scattering of muons and electrons is one of the simplest and cleanest processes in particle physics. In spite of this simplicity, $\mu e$ scattering measurements are scarse. In the 60s, experiments at CERN and Brookhaven measured this scattering cross section using accelerator-produced muons~\cite{Backenstoss:1961zz,Backenstoss:1963,Kirk:1968,Jain:1969wt}. At the same time, $\mu e$ collisions were measured by cosmic-ray experiments~\cite{Deery:1961,McDiarmid:1962,Chaudhuri:1965,Kearney:1965}. The scattering of muons off polarized electrons was then proposed as a polarimeter for high-energy muon beams in the late 80s~\cite{Schuler:1988nc} and measured by the SMC collaboration at CERN a few years later~\cite{Adams:1999af}.

Recently, a new experiment, MUonE, has been proposed at CERN to measure the differential cross section of the elastic scattering of high-energy muons on atomic electrons as a function of the spacelike (negative) squared momentum transfer~\cite{Abbiendi:2016xup}. This measurement will provide the running of the effective electromagnetic coupling in the spacelike region and, as a result, a new and independent determination of the leading hadronic contribution to the muon $g$-2~\cite{Abbiendi:2016xup,Calame:2015fva}. In order for this new determination to be competitive with the present dispersive one, which is obtained via timelike data, the $\mu e$ differential cross section must be measured with statistical and systematic uncertainties of the order of 10ppm. This high experimental precision demands an analogous accuracy in the theoretical prediction.

Until recently, the process $\mu e \to \mu e$ had received little attention also on the theory side. The few existing theoretical studies mainly focused on its QED corrections at next-to-leading order (NLO)~\cite{Nikishov:1961,Eriksson:1961,Eriksson:1963,VanNieuwenhuizen:1971yn,Kukhto:1987uj,Bardin:1997nc,Kaiser:2010zz} and tests of the Standard Model (SM)~\cite{Derman:1979zc,DAmbrosio:1984abj,Montero:1998sv}. The QED corrections at next-to-next-to-leading order (NNLO), crucial to interpret the high-precision data of future experiments like MUonE, are not known, although some of the two-loop corrections which were computed for Bhabha scattering in QED~\cite{Gehrmann:2001ck,Bonciani:2003hc}, 
for the heavy-to-light quark decay \cite{Bonciani:2008wf,Asatrian:2008uk,Beneke:2008ei,Bell:2008ws,Huber:2009se}
and the $t{\bar t}$ production~\cite{Bonciani:2008az,Bonciani:2009nb,Bonciani:2010mn,Bonciani:2013ywa} in QCD can be applied to elastic $\mu e$ scattering as well.

In this work we take a first step towards the calculation of the full NNLO QED corrections to $\mu e$ scattering. In particular, we consider the evaluation of the master integrals (MIs) occurring in the decomposition of the genuine two-loop $2 \to 2$ planar box-diagrams, namely all the two-loop four-point topologies for $\mu e$ scattering except for the crossed double box diagram. 
Given the small value of the electron mass $m_e$ when compared to the muon one $m$, we work in the approximation $m_e=0$. In this case, integration-by-parts identities~\cite{Tkachov:1981wb,Chetyrkin:1981qh,Laporta:2001dd} yield the identification of a set of 65 MIs, which we compute analytically by means of the differential equation method~\cite{Kotikov:1990kg,Remiddi:1997ny,Gehrmann:1999as}. Elaborating on recent ideas to simplify the system-solving strategy~\cite{Henn:2013pwa,Argeri:2014qva}, we choose a set of MIs obeying a system of first-order differential equations~(DEQs) in the kinematical variables $s/m^2$ and $t/m^2$ which is linear in the space-time dimension $d$, and, by means of Magnus exponential matrix \cite{Argeri:2014qva}, we derive an equivalent system of equations in {\it canonical form} \cite{Henn:2013pwa}, where the $d$-dependence of the associated matrices is factorized from the kinematics. Let us emphasize that the use of Magnus exponential matrix to identify a canonical basis of master integrals turned out to be very effective in the context of multi-loop integrals involving several scales~\cite{Argeri:2014qva,DiVita:2014pza,Bonciani:2016ypc,DiVita:2017xlr}. The matrices associated with the canonical systems admit a logarithmic-differential ($\dlog$) form, whose entries are rational functions of the kinematics; therefore, the canonical MIs can be cast in a Taylor series around $d=4$, with coefficients written as combinations of generalised polylogarithms (GPLs)~\cite{Goncharov:polylog,Remiddi:1999ew,Gehrmann:2001pz,Vollinga:2004sn}. The final determination of the MIs is achieved after imposing the boundary conditions, implemented by requiring the regularity of the solutions at special kinematics points, and by using simpler integrals as independent input.

The analytic expressions of the MIs have been numerically evaluated with the help of \texttt{GiNaC}~\cite{Bauer:2000cp} and were successfully tested against the values provided by the computer code \texttt{SecDec}~\cite{Borowka:2015mxa}. The package \texttt{Reduze}~\cite{vonManteuffel:2012np} has been used throughout the calculations.

It is important to observe that the MIs of the QED corrections to $\mu e \to \mu e$ scattering are related by crossing to the MIs of the QCD corrections to the $t{\bar t}$-pair production at hadron colliders. The analytic evaluation of the MIs for the leading-color corrections to $pp \to t {\bar t}$, due to planar diagrams only, 
was already considered in refs.~\cite{Bonciani:2008az,Bonciani:2009nb,Bonciani:2010mn,Bonciani:2013ywa}. They correspond to the MIs appearing in the evaluation of the Feynman graphs associated to the topologies $T_i$ with $i \in \{1,2,3,7,8,9,10\}$ in figure~\ref{fig:Feyndiag}, which we (re)compute here independently. The MIs for the planar topology $T_4$ and $T_5$, instead, would correspond to the MIs of subleading-color contributions to $t{\bar t}$-pair production, and were not considered previously.

For certain classes of MIs, like the ones of the processes $\mu e \to \mu e$ and $pp \to t {\bar t}$
, the choice of the boundary conditions may still constitute a challenging problem. In some cases considered in refs.~\cite{Bonciani:2008az,Bonciani:2009nb,Bonciani:2010mn,Bonciani:2013ywa}, the direct integration of the MIs in special kinematic configurations was addressed by using techniques based on Mellin-Barnes representations~\cite{Smirnov:1999gc,Tausk:1999vh}. Alternatively, here we exploit either the regularity conditions at pseudo-thresholds or the expression of the integrals at well-behaved kinematic points. The latter might be obtained by solving simpler auxiliary systems of differential equations, hence limiting the use of direct integration only to a simple set of input integrals. Our preliminary studies make us believe that the strategy we adopt for the determination of the considered integrals is not only limited to the planar contributions, but it can be applied to the non-planar graphs as well. 
In particular, we show its application for the determination of the MIs for the non-planar vertex graph~\cite{Bonciani:2008wf,Asatrian:2008uk,Beneke:2008ei,Bell:2008ws,Huber:2009se}.
Moreover, due to the similarity of the cases, we are confident that it can be very helpful for completing the analytic evaluation of the MIs needed for the two-loop QCD corrections to $pp \to t {\bar t}$, which are currently known only numerically \cite{Czakon:2008zk,Czakon:2012pz,Czakon:2012zr,Baernreuther:2012ws,Czakon:2013goa}.

The paper is organized as follows. In sec.~2 we describe the kinematics of $\mu e$ scattering and we give a brief review of the LO and NLO QED contributions to the cross section.
In sec.~3 we fix our notation and conventions for the four-point topologies.
In sec.~4 we discuss the general features of the systems of differential equations satisfied by the MIs and their general solution in terms of generalised polylogarithms. 
In sec.~5 we describe the
computation of the one-loop MIs and in sec.~6 we present the results for the planar two-loop MIs. Finally, in sec.~7 we compute the MIs for the non-planar two-loop vertex. In sec.~8 we give our conclusions. The information provided in the text is complemented by two appendices: in appendix A we discuss the computation of the auxiliary integrals which have been used to extract some of the boundary constants and, in appendix B, we give the expressions of the dlog-form of the matrices associated
to canonical systems. 

The analytic expressions of the considered MIs are given in the ancillary files accompanying the arXiv version of this publication.
\section{LO cross section and NLO QED corrections}
\label{sec:crosssec}

Let us consider the elastic scattering
\be
	\mu^{+} (p_1) + e^- (p_2)  \to  e^- (p_3) + \mu^{+} (p_4),
\label{eq:treelevel}
\ee
and define the Mandelstam variables
\begin{align}
	&s \, =  (p_1+p_2)^2, \quad t \, =  (p_2 - p_3)^2, \quad u \, = (p_1-p_3)^2,
\end{align}	
satisfying $s+t+u=2m^2+2m_e^2$, with the physical requirements $s> (m_e + m)^2$, $-\lambda(s,m^2,m_e^2)/s < t < 0$, and $\lambda(x,y,z)=x^2+y^2+z^2-2xy-2xz-2yz$ is the K{\"a}llen function.

The LO QED prediction for the differential cross section of the scattering in~(\ref{eq:treelevel}) is
\be
	\frac{d \sigma_0}{dt}  =  4 \pi \alpha^2 \, \frac{\left(m^2+m_e^2\right)^2 - su + t^2/2}{t^2 \lambda \left(s,m^2,m_e^2 \right)},
\label{eq:sigmaLO}
\ee
where $\alpha$ is the fine-structure constant. The NLO QED corrections to this cross section were computed long time ago~\cite{Nikishov:1961,Eriksson:1961,Eriksson:1963,VanNieuwenhuizen:1971yn,Kukhto:1987uj,Bardin:1997nc} and revisited more recently~\cite{Kaiser:2010zz}. As a first check, we recalculated these corrections and found perfect agreement with ref.~\cite{Kaiser:2010zz}, both for the virtual corrections and the soft photon emissions. We note that some of the pioneering publications, like~\cite{Eriksson:1961,VanNieuwenhuizen:1971yn}, contain typos or errors, so that they cannot be directly employed.

In the rest of this paper we will work in the approximation of
vanishing electron mass, $m_e=0$, i.e.\ with the kinematics specified
by $p_1^2=p_4^2=m^2$ and $p_2^2=p_3^2=0$. The master integrals will be
conveniently evaluated in the non-physical region $s < 0$, $t < 0$.

\section{Four-point topologies}
\begin{figure}
\centering
\includegraphics[scale=0.85]{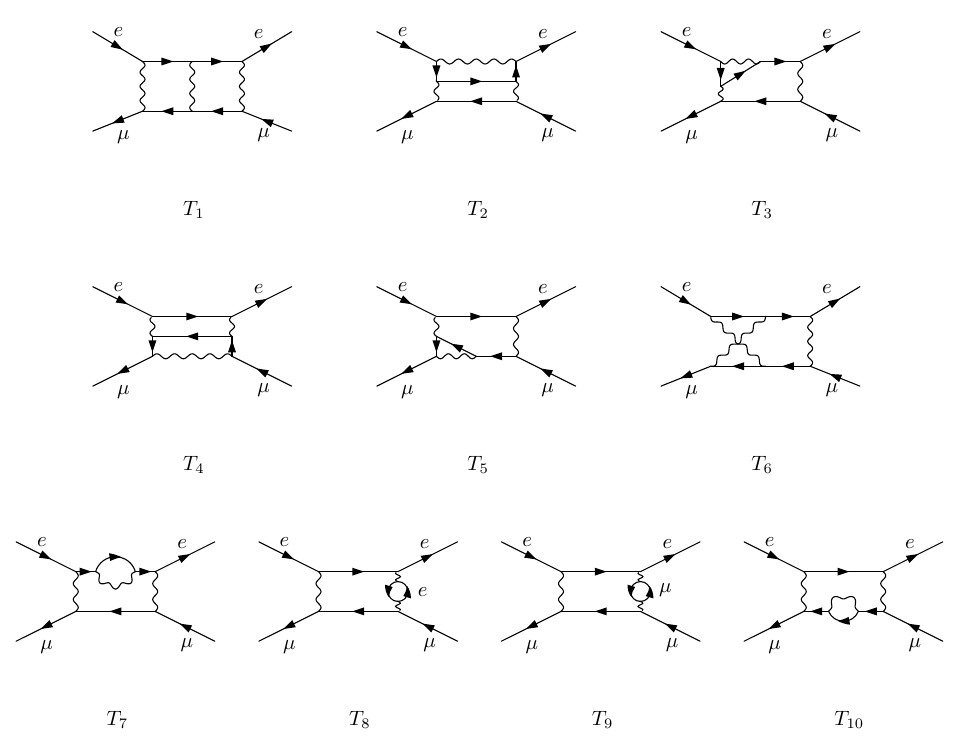}
\caption{Two-loop four-point topologies for $\mu e$ scattering}
\label{fig:Feyndiag}
\end{figure}

The main goal of this work is the evaluation of the master integrals
(MIs) of the planar two-loop four-point functions contributing to 
$\mu e$ scattering, drawn in
figure~\ref{fig:Feyndiag}. For completeness, we will discuss also the
evaluation of the MIs of the one-loop four-point function in figure
\ref{fig:Feyndiag1L}. \\

We consider $\ell$-loop $m$-denominator Feynman integrals in $d=4-2\eps$ dimensions of the type
\begin{gather}
  \int \prod_{i=1}^{\ell}\widetilde{\dd^d k_i}\,
  \frac{1}{\Den_{1}^{n_1} \ldots \Den_{m}^{n_m}}\,, \quad n_i\in \mathbb{Z}\,.
\end{gather}
In our conventions, the integration measure is defined as
\begin{equation}
  \widetilde{\dd^dk_i} = \frac{\dd^d k_i}{(2\pi)^d} \left(\frac{i \, S_\eps}{16 \pi^2} \right)^{-1} \left( \frac{m^2}{\mu^2} \right)^{\eps}, 
\label{eq:intmeasure1}
\end{equation}
with $\mu$ being the 't Hooft scale of dimensional regularization and
\begin{equation}
  S_\eps = (4\pi)^\eps \, \Gamma(1+\eps) \, .
  \label{eq:intmeasure2}
\end{equation}

We choose the following set of propagators for the relevant planar four-point topologies at one- and two-loop:
\begin{itemize}
\item For the one-loop integral family, depicted in figure \ref{fig:Feyndiag1L},
\begin{gather}
\Den_1 = k_1^2-m^2\,,\quad
\Den_2 = (k_1+p_1)^2\,,\nn
\Den_3 = (k_1+p_1+p_2)^2\,,\quad
\Den_4 = (k_1+p_4)^2\, .
\label{eq:1Lfamily}
\end{gather}

\item For the first two-loop integral family, which includes the topologies $T_1$, $T_2$, $T_3$, $T_7$ and $T_8$ of figure \ref{fig:Feyndiag},
\begin{gather}
\Den_1 = k_1^2-m^2,\quad
\Den_2 = k_2^2-m^2,\quad
\Den_3 = (k_1+p_1)^2,\quad
\Den_4 = (k_2+p_1)^2, \nn
\Den_5 = (k_1+p_1+p_2)^2,\quad
\Den_6 = (k_2+p_1+p_2)^2,\quad
\Den_7 = (k_1-k_2)^2, \nn
\Den_8 = (k_1+p_4)^2,\quad
\Den_9 = (k_2+p_4)^2\quad
\, .
\label{eq:2Lfamily1}
\end{gather}
\item For the second two-loop family, which contains topologies $T_4$, $T_5$, $T_9$ and $T_{10}$ shown in figure \ref{fig:Feyndiag},
\begin{gather}
\Den_1 = k_1^2-m^2,\quad
\Den_2 = k_2^2,\quad
\Den_3 = (k_2+p_2)^2,\quad
\Den_4 = (k_1+p_2)^2, \nn
\Den_5 = (k_2+p_2-p_3)^2,\quad
\Den_6 = (k_1+p_2-p_3)^2-m^2,\quad
\Den_7 = (k_1-p_1)^2, \nn
\Den_8 = (k_2-p_1)^2-m^2,\quad
\Den_9 = (k_1-k_2)^2-m^2\quad
\, .
\label{eq:2Lfamily2}
\end{gather}
\end{itemize}
For all families, $k_1$ and $k_2$ denote the loop momenta. 
In the following sections, MIs will be represented by
diagrams where thick lines stand for massive particles (muon), whereas
thin lines stand for massless ones (electron, photon).

\section{System of differential equations}
In order to determine all MIs appearing in the three integral families
defined above, we initially derive their DEQs
in the dimensionless variables $-s/m^2$ and $-t/m^2$. 
Upon the change of variable,
\begin{equation}
-\frac{s}{m^2} = x, \quad  -\frac{t}{m^2} = \frac{(1-y)^2}{y} \, , 
\end{equation}
the coefficients of the DEQs are rational functions of $x$ and $y$.  
  According to our system solving strategy, by means of
  integration-by-parts identities (IBPs), we choose an initial set
  of MIs {\bf F} that fulfills a system of DEQs 
\bea
{\partial {\bf F} \over \partial x} = {\mathbb A}_x(\eps,x,y) {\bf F} \
, \qquad 
{\partial {\bf F} \over \partial y} = {\mathbb A}_y(\eps,x,y) {\bf F}
\ ,
\eea
where the matrices ${\mathbb A}_x(\eps,x,y)$ and ${\mathbb
  A}_y(\eps,x,y)$ are {\it linear} in the
 dimensional regularization parameter $\eps = (4 - d)/2$, being $d$
 the number of space-time dimensions. 
According to the algorithm described in \cite{Argeri:2014qva,DiVita:2014pza,Bonciani:2016ypc,DiVita:2017xlr}, by
means of Magnus exponential matrix, we identify a set of MIs $\GGvec$
obeying canonical systems of DEQs \cite{Henn:2013pwa}, where the
dependence on $\epsilon$ is factorized from the kinematics,
\bea
{\partial \GGvec \over \partial x} = \eps {\hat {\mathbb A}}_x(x,y) \GGvec \
, \qquad 
{\partial \GGvec \over \partial y} = \eps {\hat{\mathbb A}_y(x,y)}
\GGvec\ .
\eea
After combining both systems of
DEQs into a single total differential, we arrive at the following
canonical form
\begin{equation}
d \GGvec = \eps \dA \GGvec \, , \qquad 
\dA \equiv {\hat {\mathbb A}}_x dx + {\hat {\mathbb A}}_y dy \ ,
\label{eq:canonicalDEQ}
\end{equation}
where the generic form of the total differential matrix for the
considered MIs reads as, 
\begin{equation}
\dA = \sum_{i=1}^9 \MM_i  \dlog(\eta_i) \, ,
\end{equation}
with $\MM_i$ being constant matrices.
 The arguments $\eta_i$ of this $\dlog$-form, which contain all the
 dependence of the DEQ on the kinematics, are referred to as the {\it alphabet} and they consist in the following 9 letters:
 \begin{align}
 \begin{alignedat}{2}
\eta_1 & =x\,,&\quad
\eta_2 & =1+x\,, \nn 
\eta_3 & =1-x\,, &\quad
\eta_4 & =y\,, \nn
\eta_5 & =1+y\,,&\quad
\eta_6 & =1-y,,\nn
\eta_7 & =x+y  \,,&\quad
\eta_8 & =1+x \, y, \nn
\eta_9 & = 1-y\,(1-x-y)\,. &  
\end{alignedat} \stepcounter{equation}\tag{\theequation}
\label{alphabet}
\end{align}

Let us observe that, currently, there is neither a proof of existence, nor any systematic algorithm to build a basis of
integrals whose system of DEQs is linear in $\eps$. Nevertheless, by
trial and error, we have been always able to find it within the physical
contexts we have so far studied
\cite{Argeri:2014qva,DiVita:2014pza,Bonciani:2016ypc,DiVita:2017xlr},
as well as for the $\mu e$ scattering. 
We believe it is a very important property which could be considered a
prerequisite for the existence a canonical basis: in fact, a system of
DEQs whose matrix is linear in $\eps$ can be brought into canonical form
by a rotation matrix built either by means of Magnus exponential, or
equivalently by means of the Wronskian matrix (formed by
the solutions of the associated homogenous equations, and their
derivatives), as shown for the case of systems of DEQs involving
elliptic solutions \cite{Remiddi:2016gno,Primo:2016ebd,Primo:2017ipr}.

The MIs presented in this paper are computed in the kinematic region where all letters are real and positive, 
\begin{equation}
  \label{eq:positivityx}
  x>0\,, \quad 0<y<1   \,,
\end{equation}
which corresponds to the Euclidean region $s<0$, $t<0$.
All MIs are chosen to be
finite in the $\eps\to 0$ limit, in such a way that $\GGvec(x,y)$
admits a Taylor expansion in $\epsilon$,
\begin{align}
  \GGvec(\eps,x,y) =  \GGvec^{(0)}(x,y) + \epsilon\, \GGvec^{(1)}(x,y) + \epsilon^2 \GGvec^{(2)}(x,y) + \ldots\,,
\end{align}
with the $n$-th order coefficient given by
\begin{align}
\GGvec^{(n)}(x,y) = \sum_{i=0}^n \Delta^{(n-i)} (x,y; x_0,y_0)  \GGvec^{(i)}(x_0,y_0),
\end{align}
where $ \GGvec^{(i)}(x_0,y_0)$ is a vector of boundary constants and $\Delta^{(k)}$ the weight-$k$ operator
\begin{align}
\Delta^{(k)} (x,y; x_0,y_0) =\int_\gamma \dAk{k},\qquad \Delta^{(0)} (x,y; x_0,y_0) = 1\,,
\label{eq:delta}
\end{align}
which iterates $k$ ordered integrations of the matrix-valued 1-form
$\dA$ along a piecewise-smooth path $\gamma$ in the $xy$-plane.
Since the alphabet given in eq.~\eqref{alphabet} is rational and has only algebraic roots, the iterated integrals~\eqref{eq:delta} can be directly expressed
in terms of GPLs, which are defined as 
\begin{align}
  G({\vec w}_{n} ; x) &\equiv G(w_1, {\vec w}_{n-1} ; x) \equiv
  \int_0^x dt \frac{1}{t-w_1}
  G({\vec w}_{n-1};t) , \\
  G(\vec{0}_n;x)& \equiv \frac{1}{n!}\text{log}^n(x) ,
\end{align}
with ${\vec w}_n$ being a vector of $n$ arguments. The number $n$ is referred to as
the {\it weight} of $G({\vec w}_{n} ; x)$ and amounts to the number of iterated integrations
needed to define it.
Equivalently one has
\begin{equation}
  {\partial \over \partial x} G(\vec{w}_{n}; x) = {\partial \over \partial x} G(w_1,\vec{w}_{n-1}; x) = \frac{1}{x-w_1} G(\vec{w}_{n-1};x).
\end{equation}
GPLs fulfill shuffle algebra relations of the form 
\begin{equation}
  G(\vec{m};x)\,G(\vec{n};x) = G(\vec{m};x) \shuffle G(\vec{n};x)
  = \sum_{\vec{p}=\vec{m} \shuffle \vec{n}} \: G(\vec{p};x),
\end{equation}
where the shuffle product $\vec{m} \shuffle \vec{n}$ denotes all possible merges
of $\vec{m}$ and $\vec{n}$ while preserving their respective orderings.

The analytic continuation of the MIs to the physical region defined in
sec.~\ref{sec:crosssec} can be obtained through by-now standard techniques.

\subsection{Constant GPLs}
Many of the boundary values $\GGvec^{(i)}(x_0,y_0)$ of the MIs have been determined by taking special kinematics limits on the general solution of the DEQs written in terms of GPLs. Through this procedure, the boundary constants are expressed as combinations of constant GPLs of argument $1$, with weights drawn from six different sets:
\begin{itemize}
\item $\{-1,0,1,3,-(-1)^{\frac{1}{3}},(-1)^{\frac{2}{3}} \}$,
\item $\{-\frac{1}{2},-\frac{2}{7},0,\frac{1}{7},\frac{1}{2},\frac{4}{7},1, -\frac{1}{2}(-1)^{\frac{1}{3}},\frac{1}{2}(-1)^{\frac{2}{3}} \}$\,,
\item $\{ -1, -\frac{1}{2}, 0, \frac{1}{2}, 1, 2, 3, 4, \frac{1}{2}(-1)^{\frac{1}{3}}, -\frac{1}{2}(-1)^{\frac{2}{3}}  \}$\,,
\item $\{ -1, -\frac{1}{2}, 0, \frac{1}{4}, \frac{1}{2}, 1, \frac{7}{4}, \frac{1}{2}(-1)^{\frac{1}{3}}, -\frac{1}{2}(-1)^{\frac{2}{3}}  \}$\,,
\item $\{-2,-\frac{1}{2},0,\frac{1}{2},1,4,7, -\frac{1}{2}(-1)^{\frac{1}{3}},\frac{1}{2}(-1)^{\frac{2}{3}}  \}$\,,
\item $\{-2,-1,-\frac{1}{2},0,\frac{1}{4},1,\frac{7}{4},2, -2(-1)^{\frac{1}{3}},2(-1)^{\frac{2}{3}}  \}$\,.
\end{itemize}
Each set arises from a different kinematic limit imposed on the
alphabet given in eq. (\ref{alphabet}). 
We used \texttt{GiNaC} to
numerically verify that at each order in $\eps^n$ 
(up to the order $n=4$), the corresponding
combination of constant GPLs  is proportional to Riemann $\zeta_n$. 
In particular, $\zeta_n$ functions are known to be {\it primitive} \cite{Brown:2011ik,Duhr:2012fh,Duhr:2014woa}, $i.e.$
they have irreducible coproducts, $\Delta(\zeta_n) = 1 \otimes \zeta_n + \zeta_n \otimes 1$.
Therefore, they must have vanishing coproducts components 
$\Delta_{p} (\zeta_n) = 0$
with $p \in \Pi(n)$, where 
\bea
\Pi(n) &=& \{
(n-1,1), 
(n-2,2), 
 (n-2,1,1),
\ldots, \nn
&& \qquad \ldots,
(1,1,1,\ldots,1,1,1),
\ldots,
(1,1,n-2),
(2,n-2),
(1,n-1)
\}  
\eea
is the set of the integer partitions of $n$ with dimensions larger
than one.
We explicitly verified that the combinations of GPLs of argument 1,
appearing in the boundary conditions, are also primitive, although
the considered coproducts components do not necessarily vanish when acting
separately on each GPL involved in those combinations.
Therefore, we made a simple {\it ansatz} that these combinations could be proportional
to $\zeta_n$, and we checked it by means of high-precision
arithmetic.
Below we show some examples for these identities,
\begin{align}
  \zeta_2 = {} & -\frac{1}{2}G(-1;1)^2 +G(0,-2;1) + G(0,-\frac{1}{2} ;1 )\,, \\
  -59 \zeta_4 = {} & \pi^2 \left( G(-1;1)^2-2 \, G(0,-(-1)^{\frac{1}{3}};1)-2 \, G(0,(-1)^{\frac{2}{3}};1)  \right)  -21 \, \zeta_3  \, G(-1;1) \nn  -&G(-1;1)^4  -18 \, G(0,0,0,-(-1)^{\frac{1}{3}};1) 
  -18 \,G(0,0,0,(-1)^{\frac{2}{3}};1) \nn +&12 \, G(0,0,-(-1)^{\frac{1}{3}},-1;1)
  +12 \, G(0,0,(-1)^{\frac{2}{3}},-1;1) \nn +&12 \, G(0,-(-1)^{\frac{1}{3}},-1,-1;1) 
   +12 \, G(0,(-1)^{\frac{2}{3}},-1,-1;1) 
   +24 \, G(0,0,0,2;1)  \,  .
\end{align}
 For related studies see also 
\cite{Broadhurst:1998rz,2007arXiv0707.1459Z,Henn:2015sem}.

\section{One-loop master integrals}
\label{sec:1Loop}
\begin{figure}[h]
\centering
\includegraphics[scale=0.8]{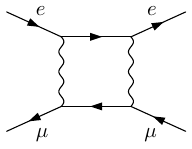}
\caption{One-loop four-point topology for $\mu e$ scattering}
\label{fig:Feyndiag1L}
\end{figure}

\begin{figure}[h]
  \centering
  \captionsetup[subfigure]{labelformat=empty}
  \subfloat[$\mathcal{T}_1$]{%
    \includegraphics[width=0.12\textwidth]{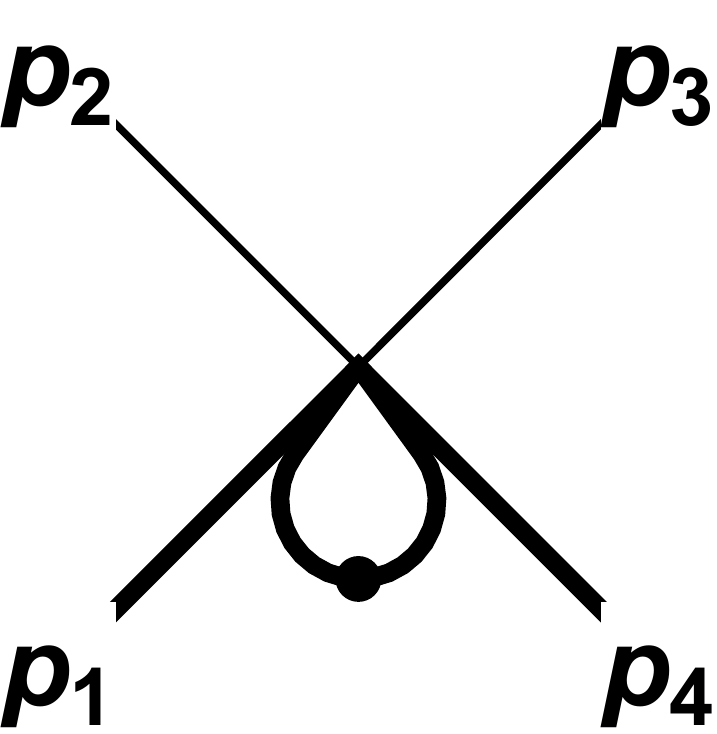}
  }
  \subfloat[$\mathcal{T}_2$]{%
    \includegraphics[width=0.12\textwidth]{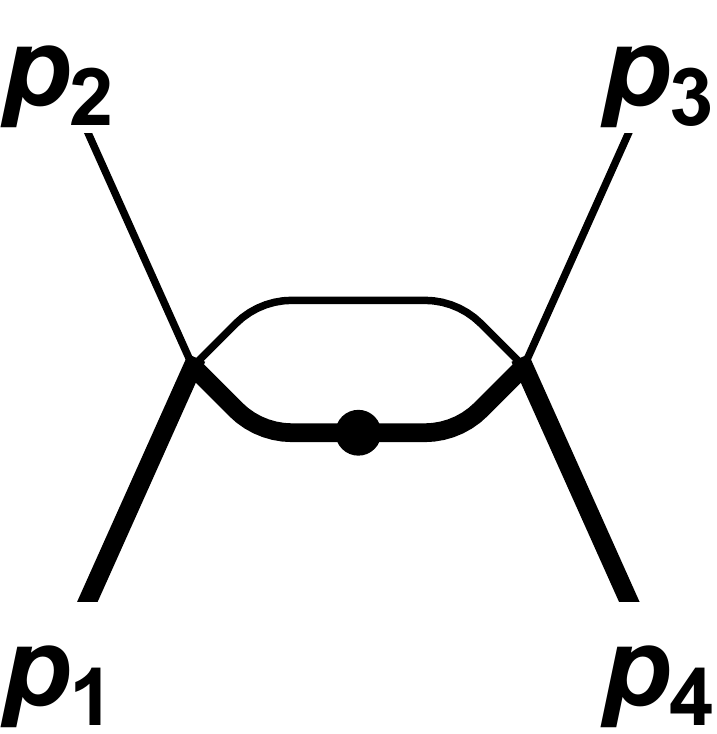}
  }
  \subfloat[$\mathcal{T}_3$]{%
    \includegraphics[width=0.12\textwidth]{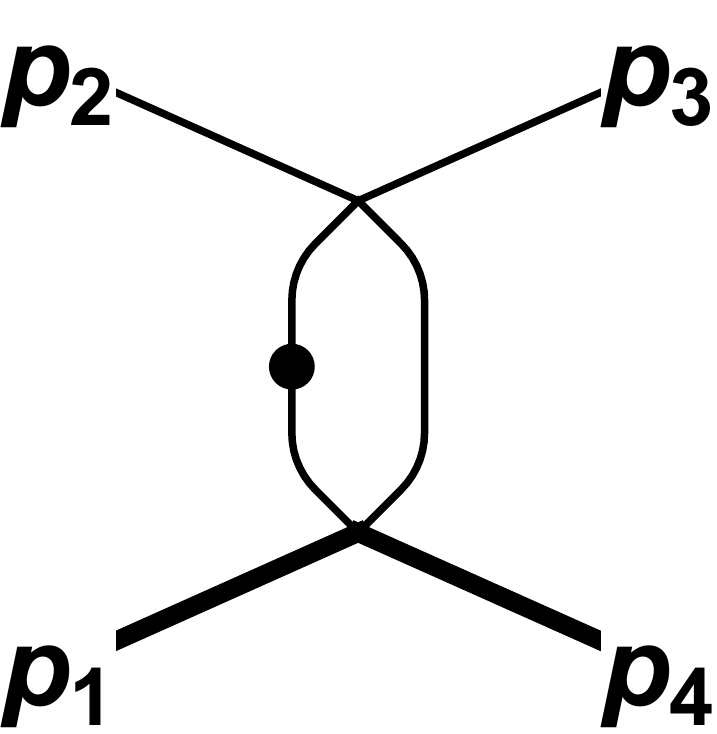}
  }
  \subfloat[$\mathcal{T}_4$]{%
    \includegraphics[width=0.12\textwidth]{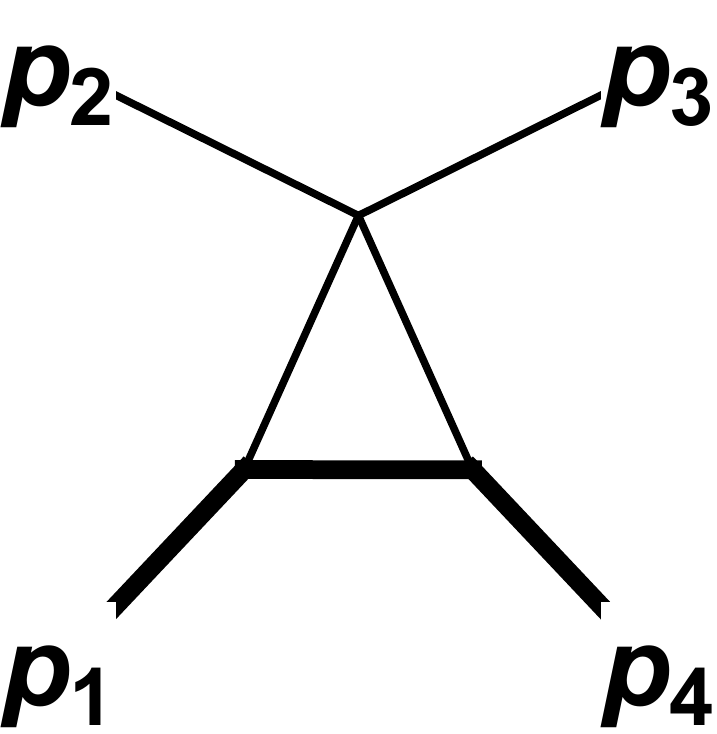}
  }
  \subfloat[$\mathcal{T}_5$]{%
    \includegraphics[width=0.12\textwidth]{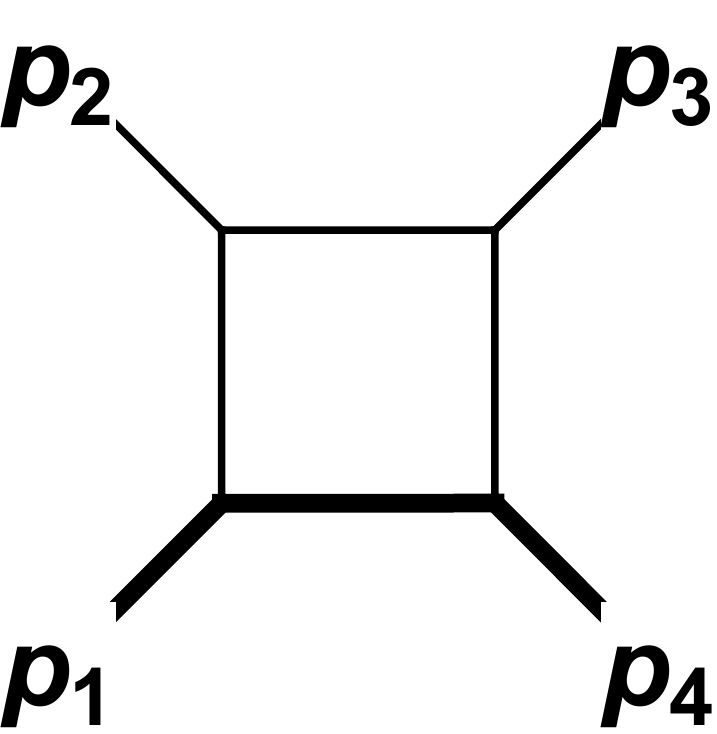}
  }
\caption{One-loop MIs $\mathcal{T}_{1,\ldots,5}$.}
 \label{fig:MIs1L}
\end{figure}

In this section we briefly discuss the computation of the master
integrals of the one-loop four-point graph shown in figure
\ref{fig:Feyndiag1L}, corresponding to the integral family defined in
eq.~\eqref{eq:1Lfamily}. 
We choose the following set of MIs, which satisfy an $\eps$-linear DEQ, 
\begin{align*}
\FF_{1}&=\eps \, \top{1}\,,&\qquad
\FF_{2}&=\eps \, \top{2}\,, &\quad
\FF_{3}&=\eps \, \top{3}\,,  &\qquad
\FF_{4}&=\eps^2 \, \top{4}\,,&\qquad
\FF_{5}&=\eps^2 \, \top{5}\,,  
\stepcounter{equation}\tag{\theequation}
\label{def:LBasis1L}
\end{align*}
where the $\mathcal{T}_i$ are depicted in
figure~\ref{fig:MIs1L}. 
With the help the Magnus algorithm we can identify the corresponding canonical basis
\begin{align*}
  \begin{alignedat}{2}
  \GG_{1}&=   \FF_1\,, & \qquad
  \GG_{2}&= -s  \,  \FF_2\,,\nn
   \GG_{3}&= -t   \FF_3\,, & \qquad
  \GG_{4}&= \lambda_t\, \FF_4\,,   \nn
  \GG_{5}&=(s-m^2)(-t) \FF_5 \,.
\label{def:CanonicalBasis1L}
   \end{alignedat}\stepcounter{equation}\tag{\theequation}
 \end{align*}
 with $\lambda_t=\sqrt{-t} \sqrt{4m^2-t}$. 

This set of MIs satisfies a canonical DEQ of the form given in
eq.~(\ref{eq:canonicalDEQ}), whose coefficient matrices read (in this
case, $\mathbb M_3$ and $\mathbb M_{9}$ vanish),
 \begin{align}
\MM_{1}=&\scalemath{0.6}{\left(
\begin{array}{ccccc}
 0 &\pminus 0 &\pminus 0 &\pminus 0 &\pminus 0 \\
 0 &\pminus 1 &\pminus 0 &\pminus 0 &\pminus 0 \\
 0 &\pminus 0 &\pminus 0 &\pminus 0 &\pminus 0 \\
 0 &\pminus 0 &\pminus 0 &\pminus 0 &\pminus 0 \\
 0 &\pminus 0 &\pminus 0 &\pminus 0 &\pminus 0 \\
\end{array}
\right)}\,,\quad 
\MM_{2}=\scalemath{0.6}{\left(
\begin{array}{ccccc}
 0 &\pminus 0 &\pminus 0 &\pminus 0 &\pminus 0 \\
 -1 & -2 &\pminus 0 &\pminus 0 &\pminus 0 \\
 0 &\pminus 0 &\pminus 0 &\pminus 0 &\pminus 0 \\
 0 &\pminus 0 &\pminus 0 &\pminus 0 &\pminus 0 \\
 2 &\pminus 4 &\pminus 0 & 0 & -2 \\
\end{array}
\right)}\,,\quad
\MM_{4}=\scalemath{0.6}{\left(
\begin{array}{ccccc}
 0 &\pminus 0 &\pminus 0 &\pminus 0 &\pminus 0 \\
 0 &\pminus 0 &\pminus 0 &\pminus 0 &\pminus 0 \\
 0 &\pminus 0 &\pminus 1 &\pminus 0 &\pminus 0 \\
 1 &\pminus 0 & -1 &\pminus 0 &\pminus 0 \\
 1 &\pminus 2 &\pminus 0 &\pminus 0 &\pminus 0 \\
\end{array}
\right)}\,,\quad
\MM_{5}=\scalemath{0.6}{\left(
\begin{array}{ccccc}
 0 &\pminus 0 &\pminus 0 &\pminus 0 &\pminus 0 \\
 0 &\pminus 0 &\pminus 0 &\pminus 0 &\pminus 0 \\
 0 &\pminus 0 &\pminus 0 &\pminus 0 &\pminus 0 \\
 0 &\pminus 0 &\pminus 0 &\pminus 2 &\pminus 0 \\
 0 &\pminus 0 &\pminus 0 &\pminus 0 &\pminus 0 \\
\end{array}
\right)}\,,\nn
\MM_{6}=&\scalemath{0.6}{\left(
\begin{array}{ccccc}
 0 &\pminus 0 &\pminus 0 &\pminus 0 &\pminus 0 \\
 0 &\pminus 0 &\pminus 0 &\pminus 0 &\pminus 0 \\
 0 &\pminus 0 & -2 &\pminus 0 &\pminus 0 \\
 0 &\pminus 0 &\pminus 0 & -2 &\pminus 0 \\
 0 &\pminus 0 &\pminus 2 &\pminus 0 & -2 \\
\end{array}
\right)}\,,\quad
\MM_{7}=\scalemath{0.6}{\left(
\begin{array}{ccccc}
 0 &\pminus 0 &\pminus 0 &\pminus 0 &\pminus 0 \\
 0 &\pminus 0 &\pminus 0 &\pminus 0 &\pminus 0 \\
 0 &\pminus 0 &\pminus 0 &\pminus 0 &\pminus 0 \\
 0 &\pminus 0 &\pminus 0 &\pminus 0 &\pminus 0 \\
 -1 & -2 & -1 & -1 &\pminus 1 \\
\end{array}
\right)
}\,,\quad
\MM_{8}=\scalemath{0.6}{\left(
\begin{array}{ccccc}
 0 &\pminus 0 &\pminus 0 &\pminus 0 &\pminus 0 \\
 0 &\pminus 0 &\pminus 0 &\pminus 0 &\pminus 0 \\
 0 &\pminus 0 &\pminus 0 &\pminus 0 &\pminus 0 \\
 0 &\pminus 0 &\pminus 0 &\pminus 0 &\pminus 0 \\
 -1 & -2 & -1 &\pminus 1 &\pminus 1 \\
\end{array}
\right)} \ . 
 \end{align}
%
The integration of the DEQ in terms of GPLs as well as the fixing of
boundary constants is straightforward. $\GG_{1,3}$ are obtained by
direct integration and, by using the normalization of eq.(\ref{eq:intmeasure1}), are given by
\begin{align}
\GG_{1}(\eps)=&1\,,\qquad \GG_{3}(\eps,y)=\left(\frac{(1-y)^2}{y}\right)^{-\eps} \left(1-\zeta_2\eps^2-2 \zeta_3\eps^3
   +\mathcal{O}\left(\eps^4\right)\right)
   \,.
\end{align}
The boundary constants for $\GG_{2}$, $\GG_{4}$ and  $\GG_{5}$ can be
fixed by respectively demanding regularity at pseudothresholds $s\to
0$, at $t\to4m^2$, and at $s=-t\to m^2/2$.
The final expression of the other MIs are,
\begin{align}
\GG_{i}(\eps,x,y)=\sum_{k=0}^{2}\GG_{i}^{(k)}(x,y)\eps^k+\mathcal{O}(\eps^3)\,,
\end{align}
with
\begin{align}
\GG_{2}^{(0)}(x)=&0\,,\nn
\GG_{2}^{(1)}(x)=&-G(-1;x)\,,\nn
\GG_{2}^{(2)}(x)=&2G(-1,-1;x)-G(0,-1;x)\,, 
\end{align}
\begin{align}
\GG_{4}^{(0)}(y)=&0\,,\nn
\GG_{4}^{(1)}(y)=&0\,,\nn
\GG_{4}^{(2)}(y)=&-4\zeta_2-G(0,0;y)+2G(0,1;y)\,,
\end{align}
\begin{align}
\GG_{5}^{(0)}(x,y)=&\quad2\,,\nn
\GG_{5}^{(1)}(x,y)=&-2G(-1;x)+G(0;y)-2G(1;y)\,,\nn
\GG_{5}^{(2)}(x,y)=&-5\zeta_2+2G(-1;x)\left(2G(1;y)-G(0;y)\right)\,. 
 \end{align}

\begin{figure}
  \centering
  \captionsetup[subfigure]{labelformat=empty}
  \subfloat[$\mathcal{T}_1$]{%
    \includegraphics[width=0.14\textwidth]{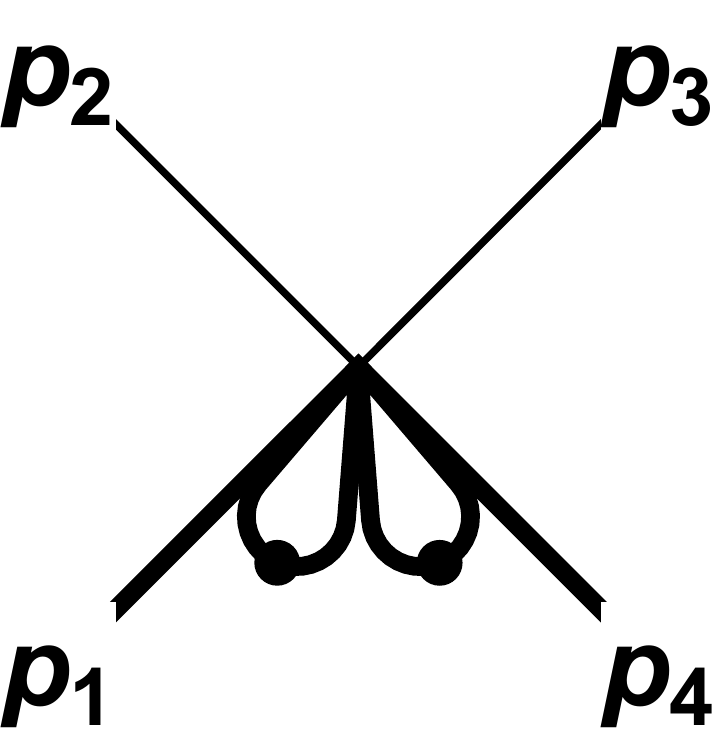}
  }
  \subfloat[$\mathcal{T}_2$]{%
    \includegraphics[width=0.14\textwidth]{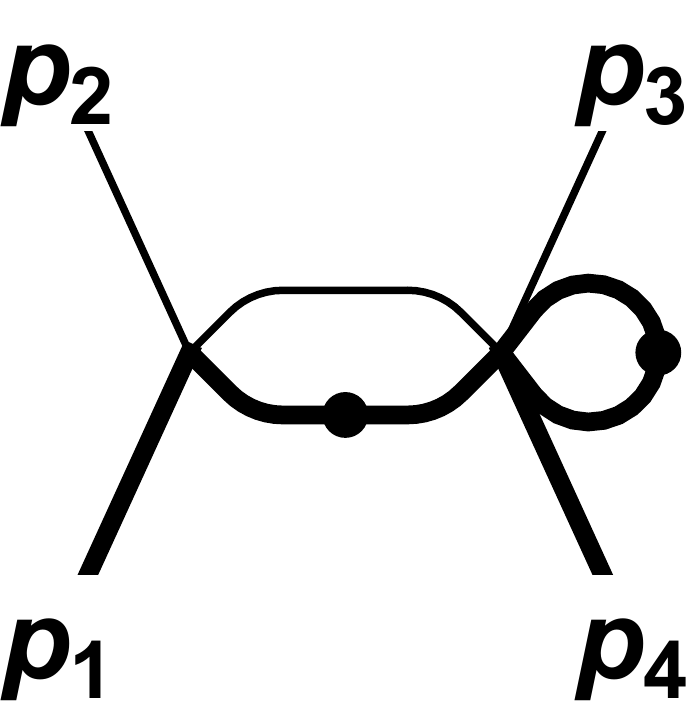}
  }
  \subfloat[$\mathcal{T}_3$]{%
    \includegraphics[width=0.14\textwidth]{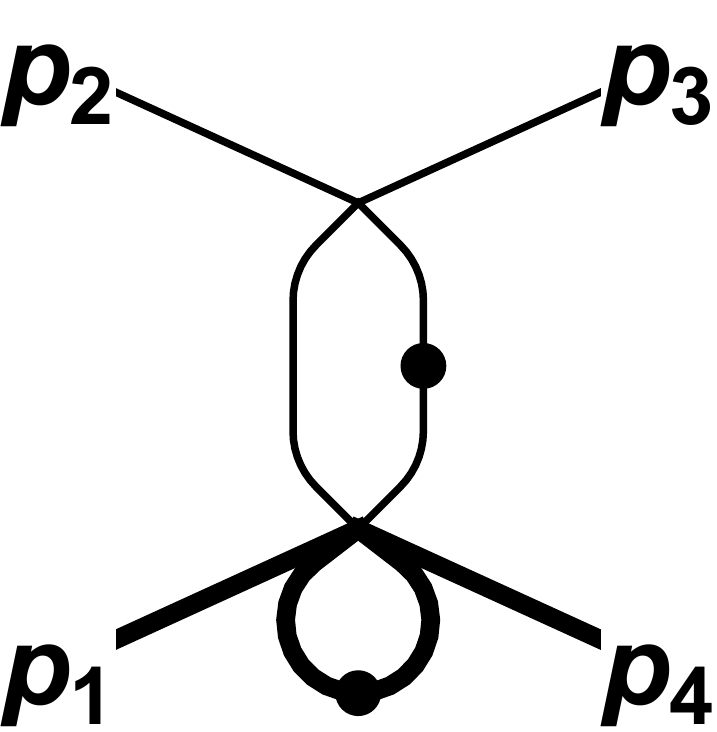}
  }
  \subfloat[$\mathcal{T}_4$]{%
    \includegraphics[width=0.14\textwidth]{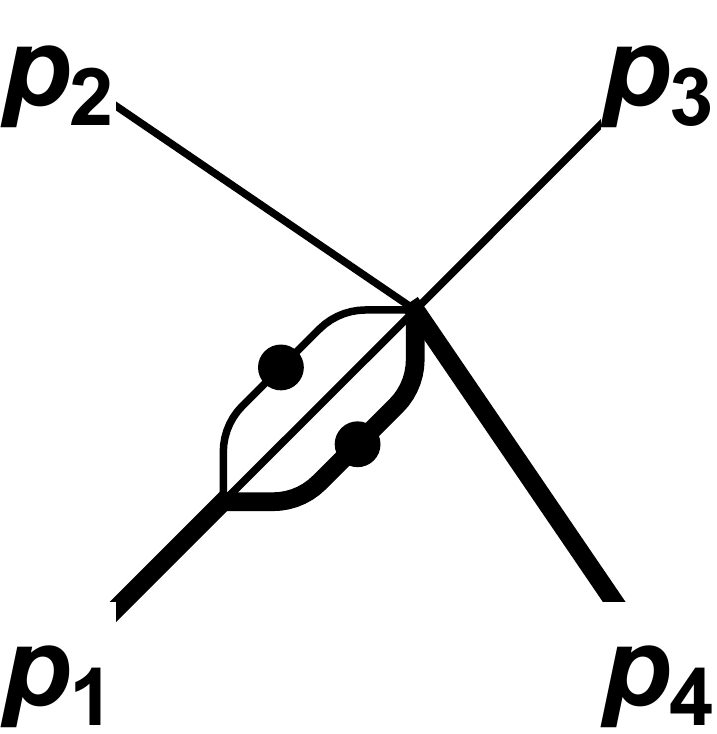}
  }
  \subfloat[$\mathcal{T}_5$]{%
    \includegraphics[width=0.14\textwidth]{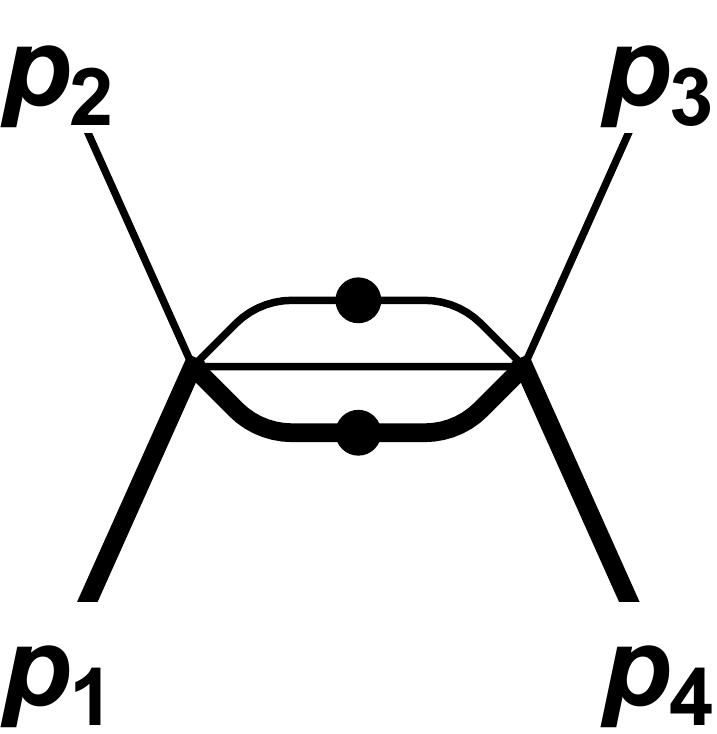}
  }
  \subfloat[$\mathcal{T}_6$]{%
    \includegraphics[width=0.14\textwidth]{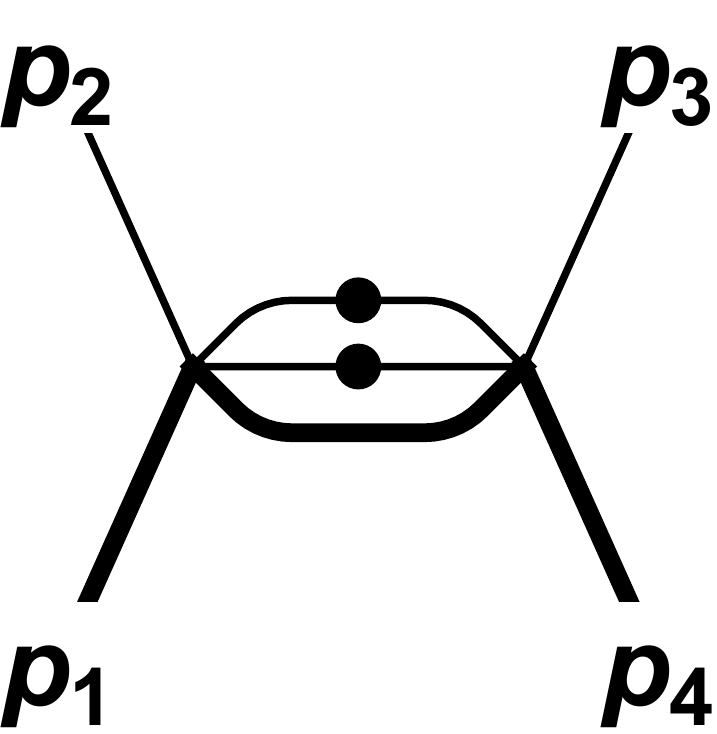}
  }
  \\
  \subfloat[$\mathcal{T}_7$]{%
    \includegraphics[width=0.14\textwidth]{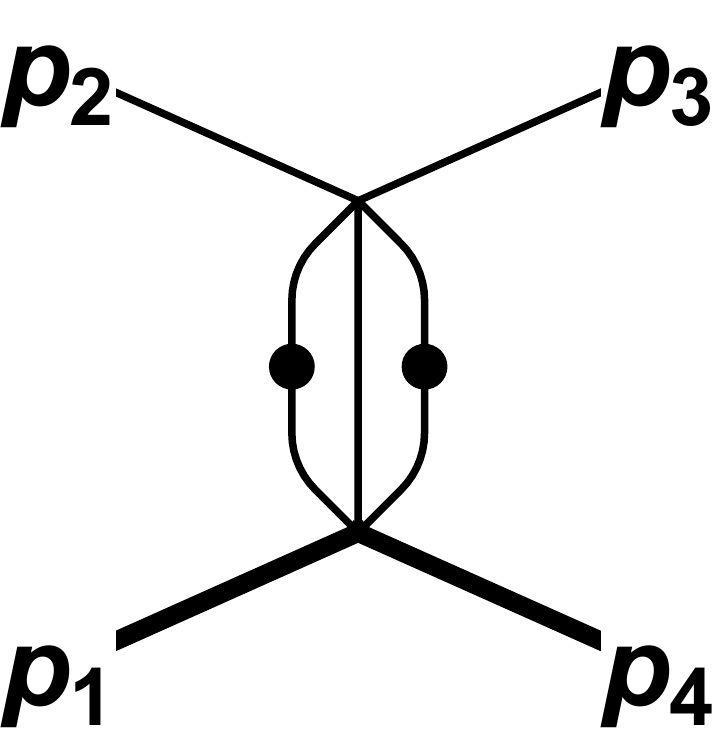}
  }
  \subfloat[$\mathcal{T}_8$]{%
    \includegraphics[width=0.14\textwidth]{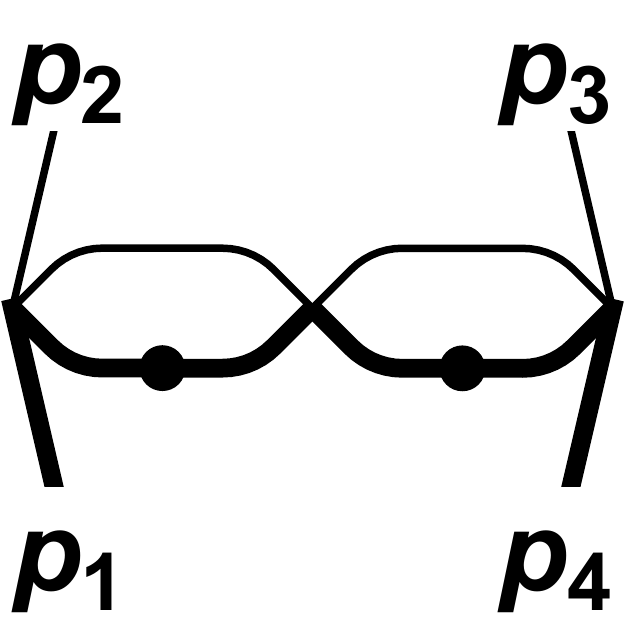}
  }
  \subfloat[$\mathcal{T}_9$]{%
    \includegraphics[width=0.14\textwidth]{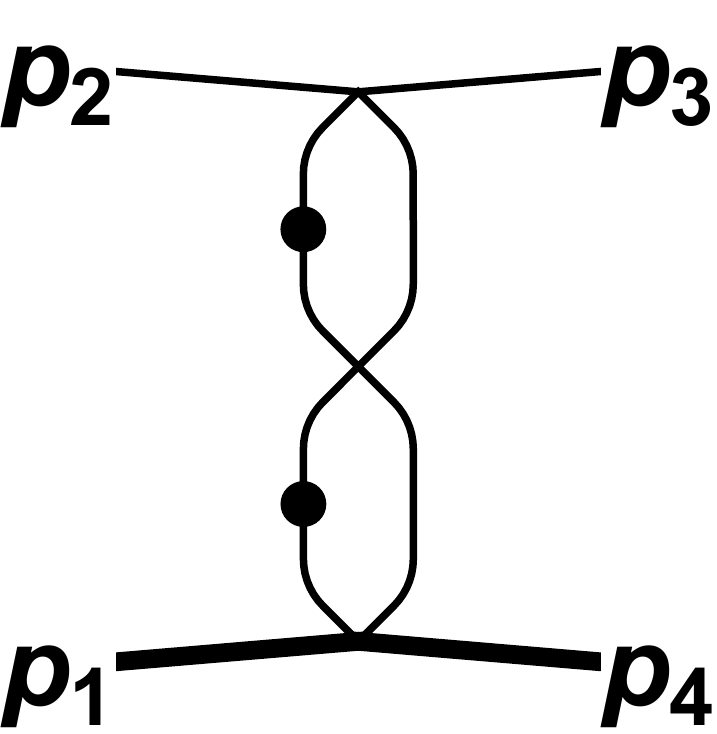}
  }
  \subfloat[$\mathcal{T}_{10}$]{%
    \includegraphics[width=0.14\textwidth]{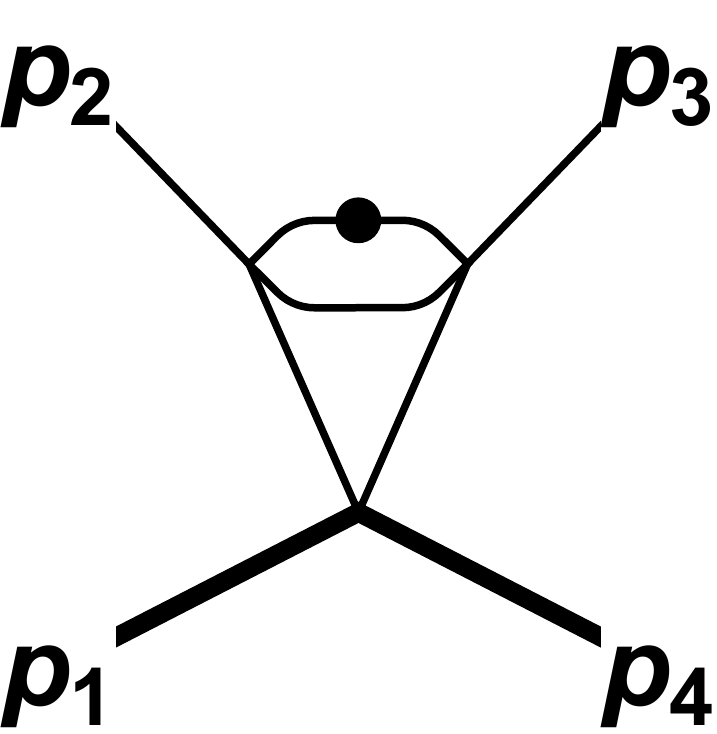}
  }
  \subfloat[$\mathcal{T}_{11}$]{%
    \includegraphics[width=0.14\textwidth]{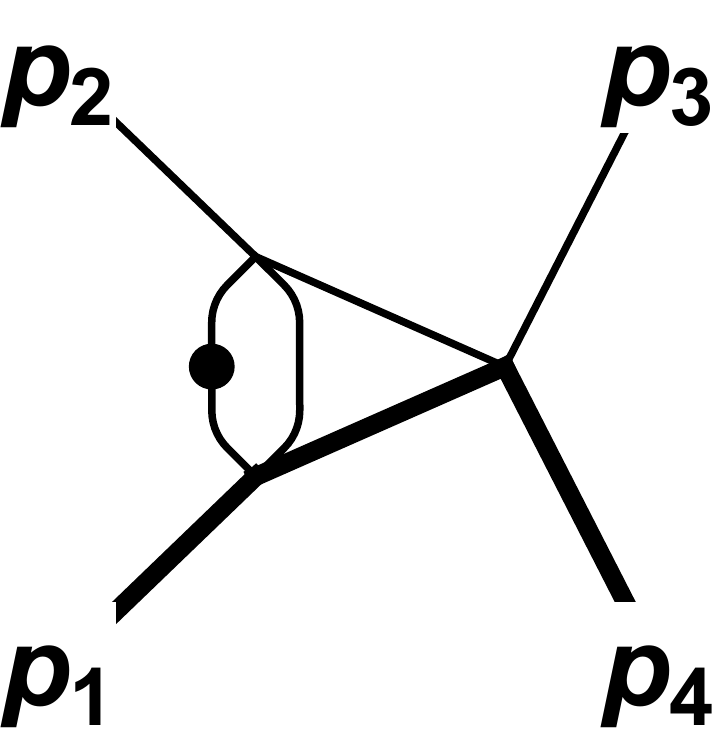}
  }
  \subfloat[$\mathcal{T}_{12}$]{%
    \includegraphics[width=0.14\textwidth]{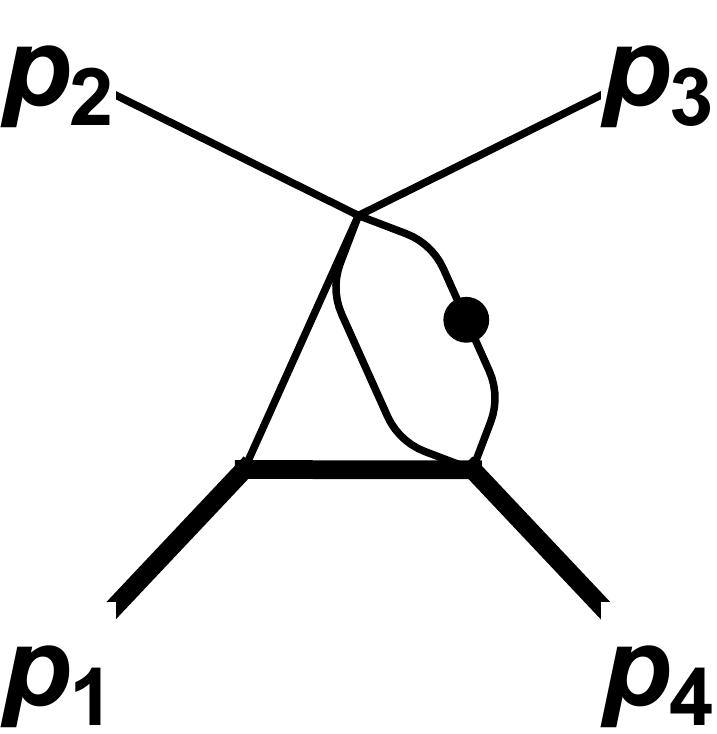}
  }
  \\
  \subfloat[$\mathcal{T}_{13}$]{%
    \includegraphics[width=0.14\textwidth]{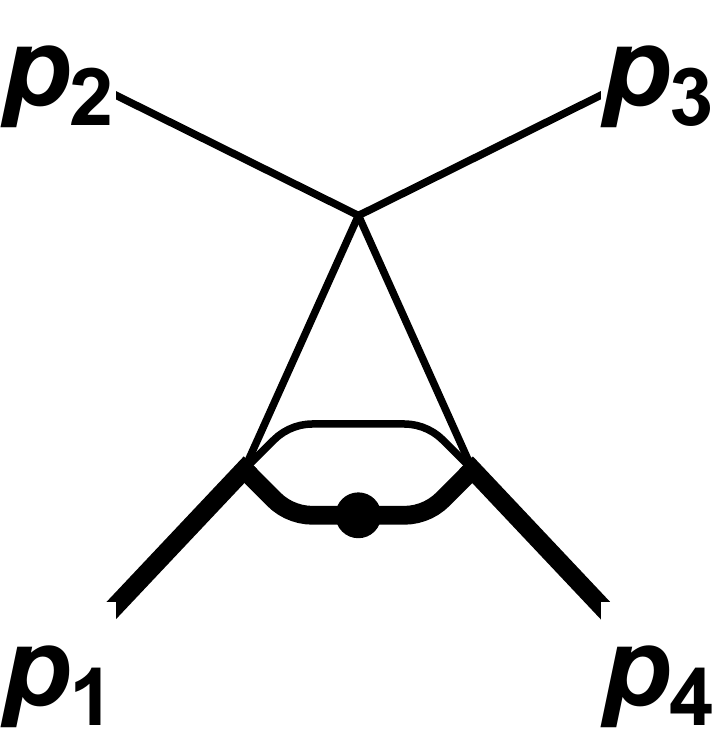}
  }
  \subfloat[$\mathcal{T}_{14}$]{%
    \includegraphics[width=0.14\textwidth]{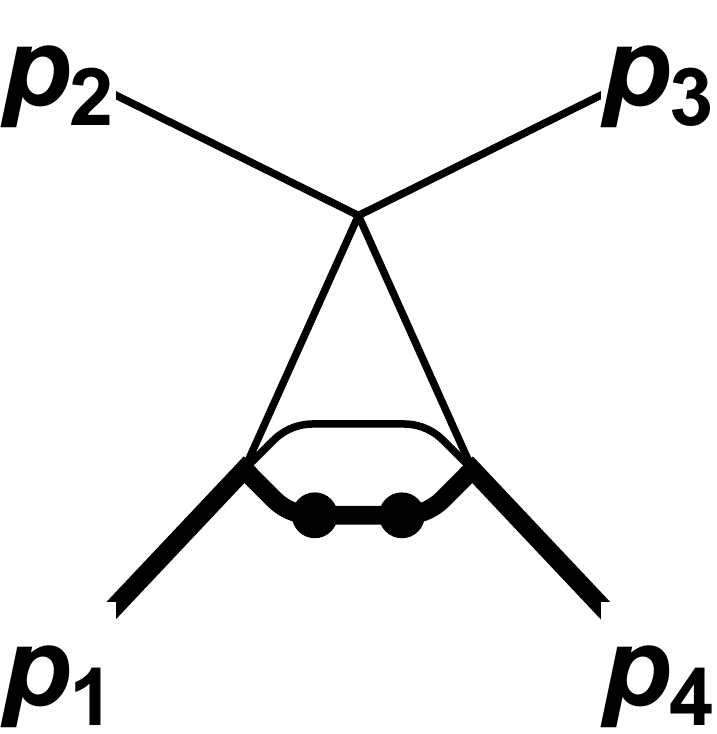}
  }
  \subfloat[$\mathcal{T}_{15}$]{%
    \includegraphics[width=0.14\textwidth]{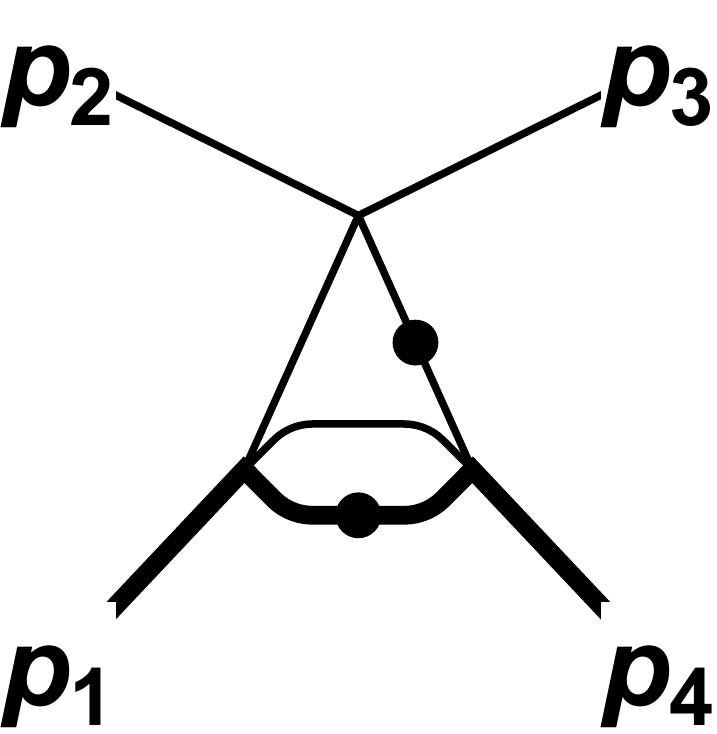}
  }
  \subfloat[$\mathcal{T}_{16}$]{%
    \includegraphics[width=0.14\textwidth]{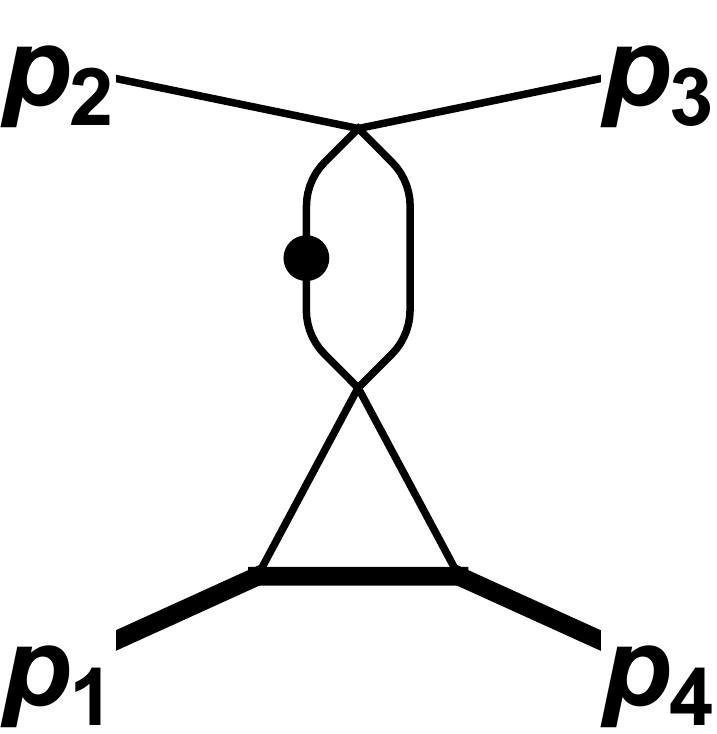}
  }
  \subfloat[$\mathcal{T}_{17}$]{%
    \includegraphics[width=0.14\textwidth]{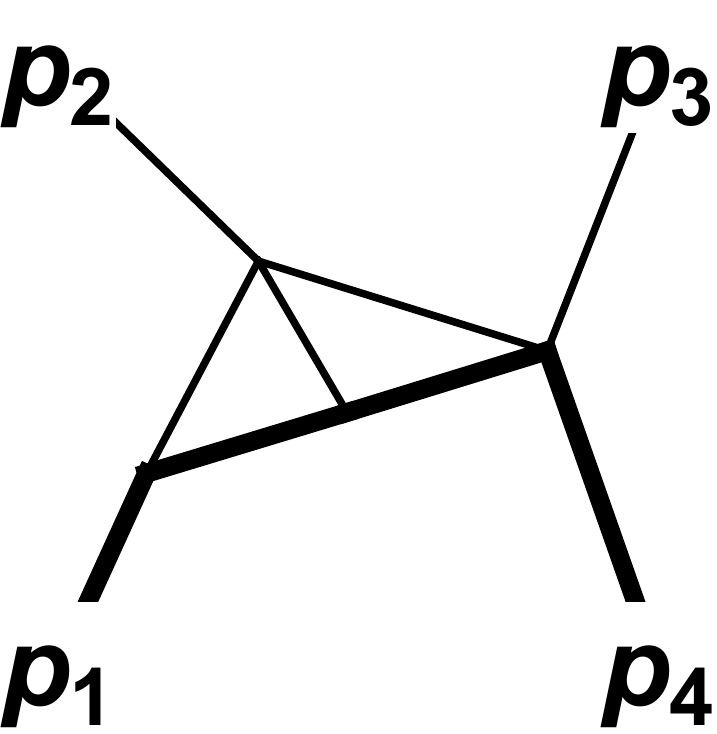}
  }
  \subfloat[$\mathcal{T}_{18}$]{%
    \includegraphics[width=0.14\textwidth]{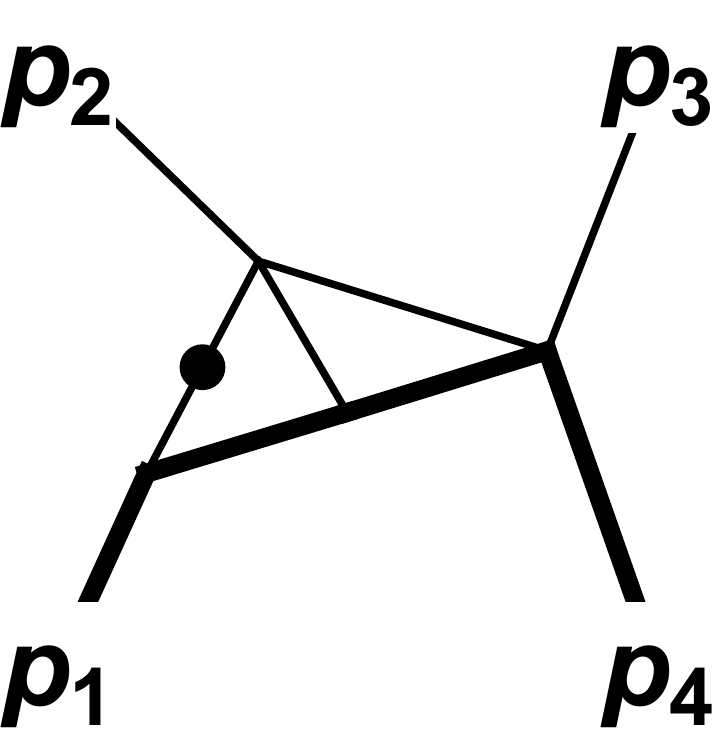}
  }
  \\
  \subfloat[$\mathcal{T}_{19}$]{%
    \includegraphics[width=0.14\textwidth]{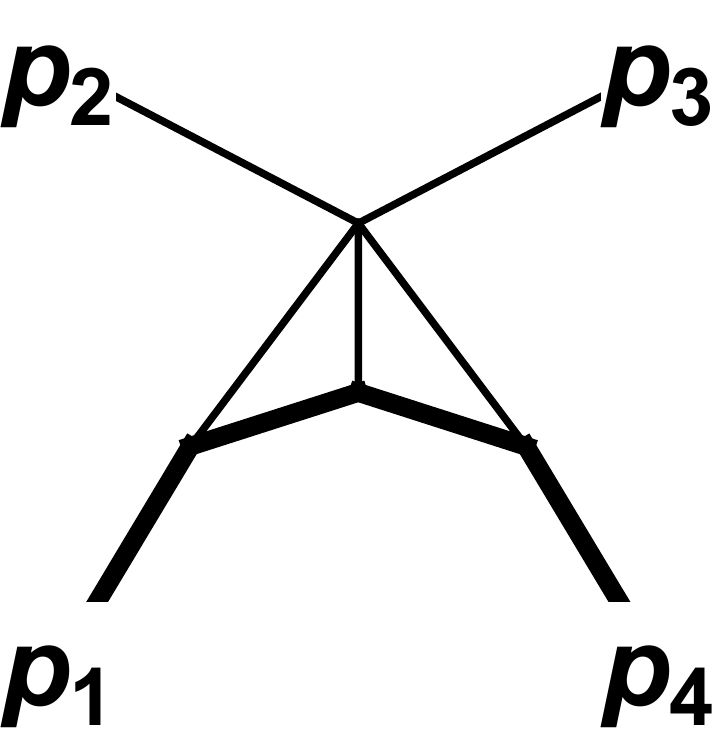}
  }
  \subfloat[$\mathcal{T}_{20}$]{%
    \includegraphics[width=0.14\textwidth]{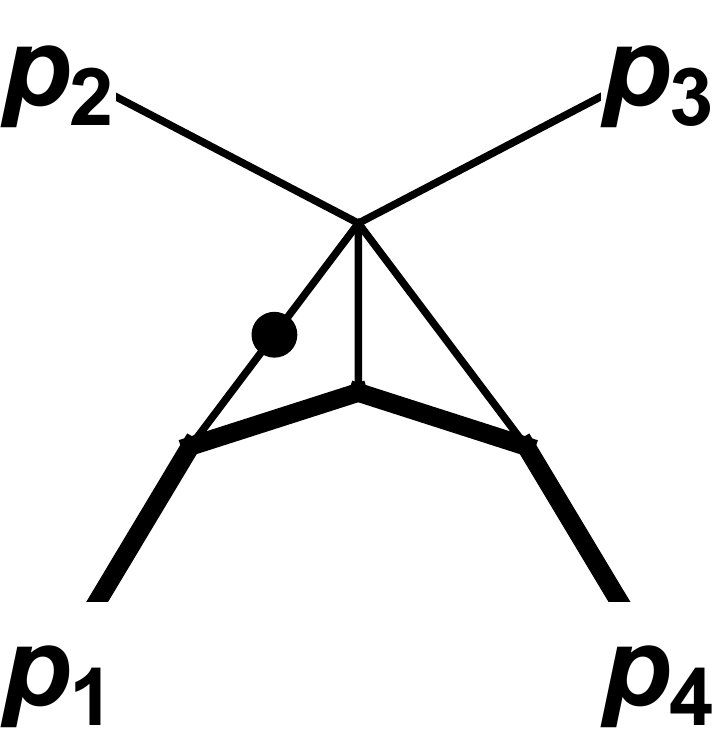}
  }
  \subfloat[$\mathcal{T}_{21}$]{%
    \includegraphics[width=0.14\textwidth]{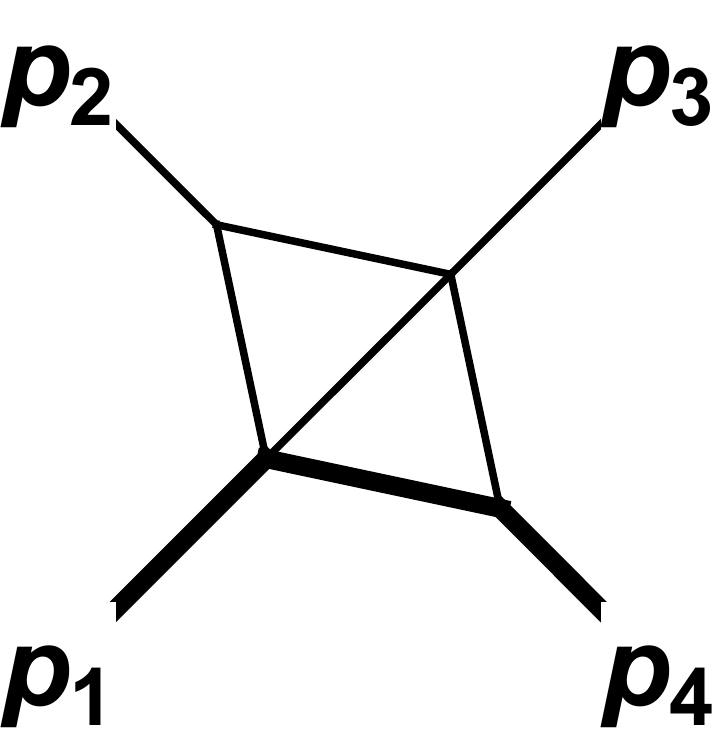}
  }
  \subfloat[$\mathcal{T}_{22}$]{%
    \includegraphics[width=0.14\textwidth]{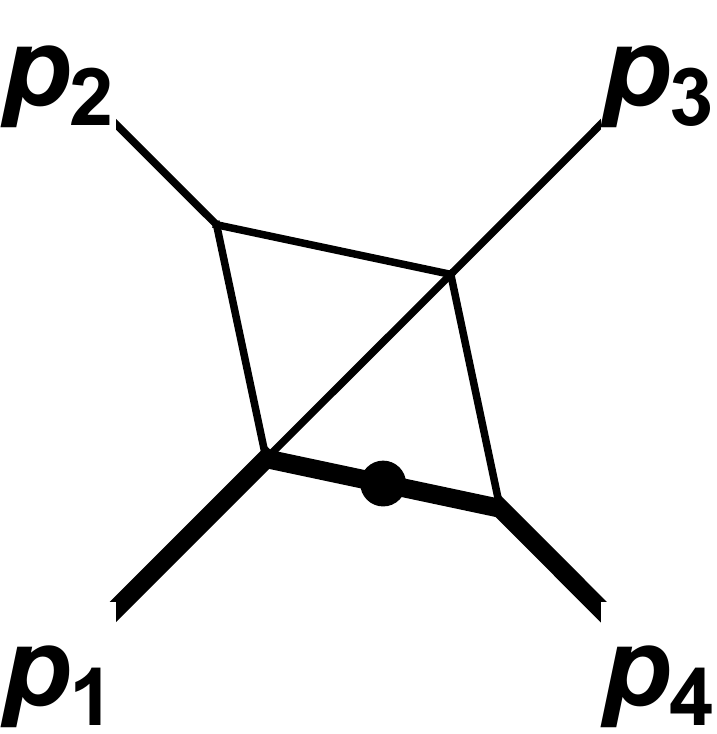}
  }
  \subfloat[$\mathcal{T}_{23}$]{%
    \includegraphics[width=0.14\textwidth]{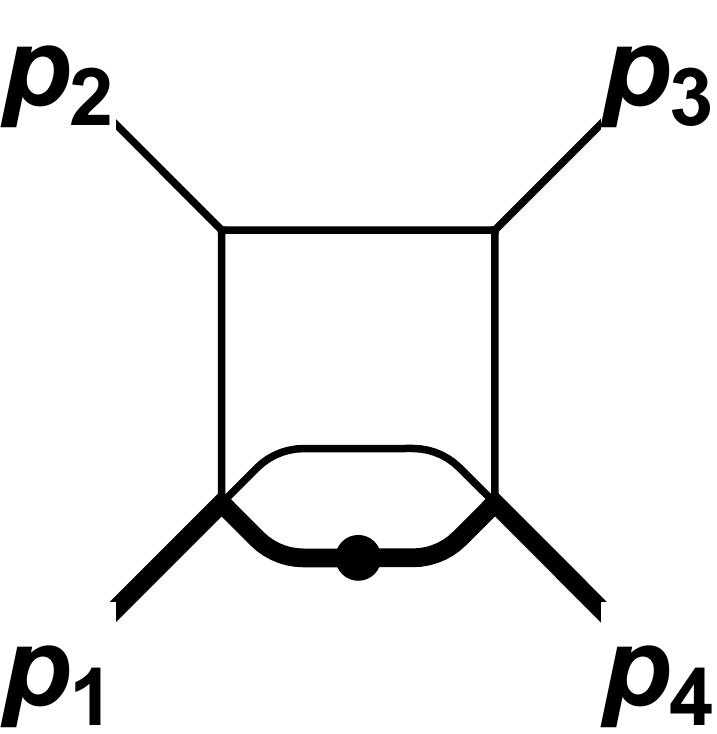}
  }
  \subfloat[$\mathcal{T}_{24}$]{%
    \includegraphics[width=0.14\textwidth]{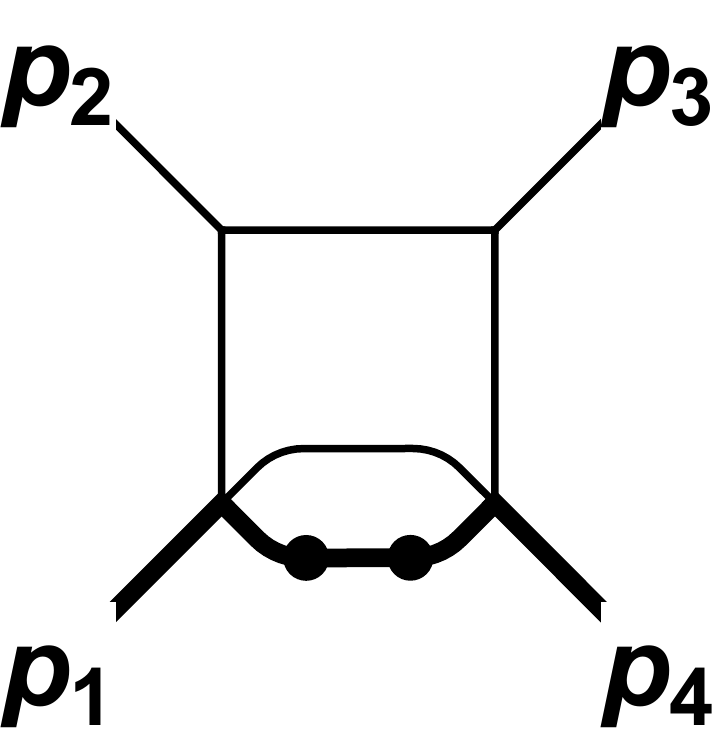}
  }
  \\
    \subfloat[$\mathcal{T}_{25}$]{%
    \includegraphics[width=0.14\textwidth]{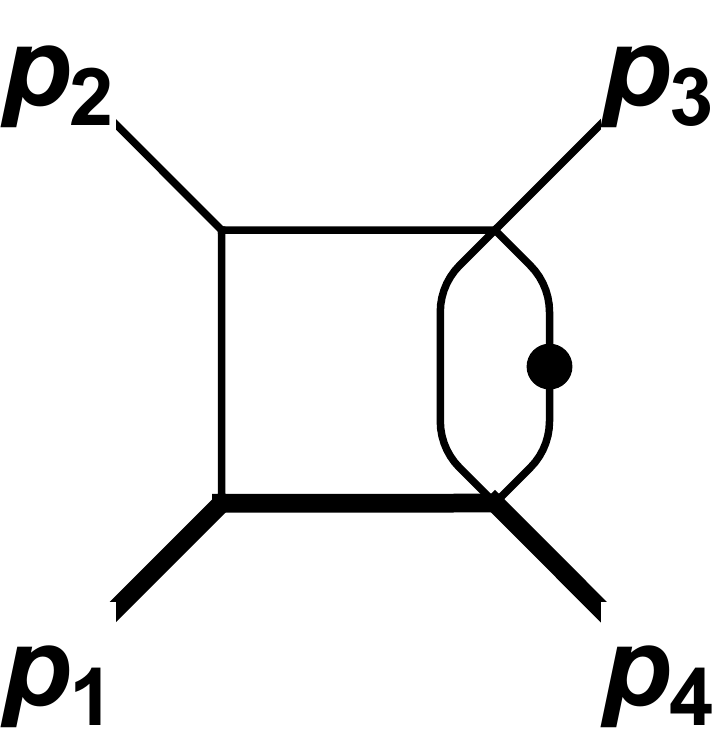}
  }
  \subfloat[$\mathcal{T}_{26}$]{%
    \includegraphics[width=0.14\textwidth]{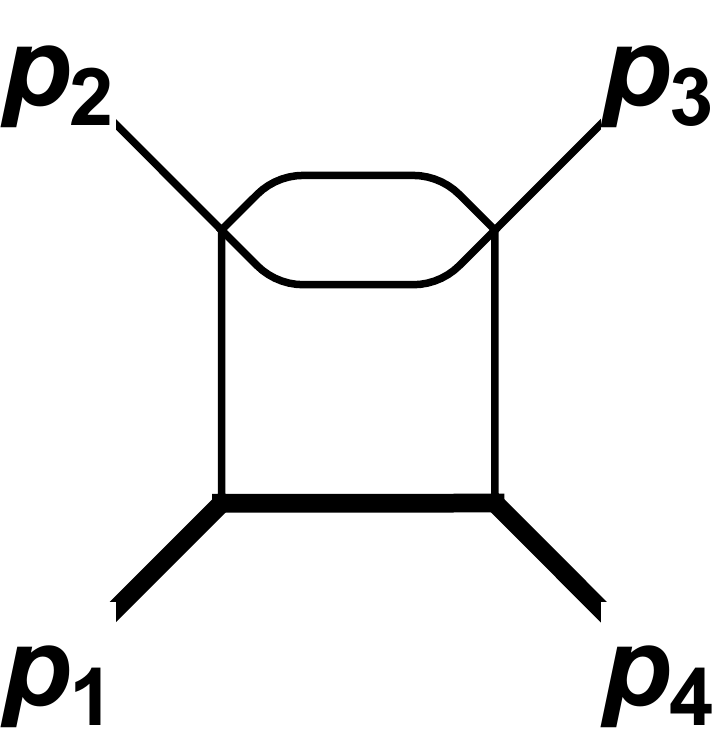}
  }
  \subfloat[$\mathcal{T}_{27}$]{%
    \includegraphics[width=0.14\textwidth]{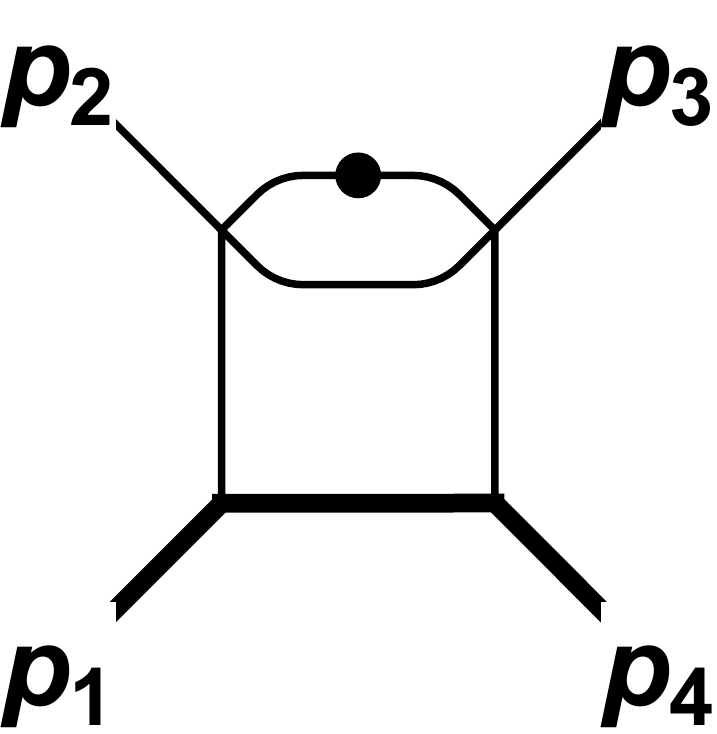}
  }
  \subfloat[$\mathcal{T}_{28}$]{%
    \includegraphics[width=0.14\textwidth]{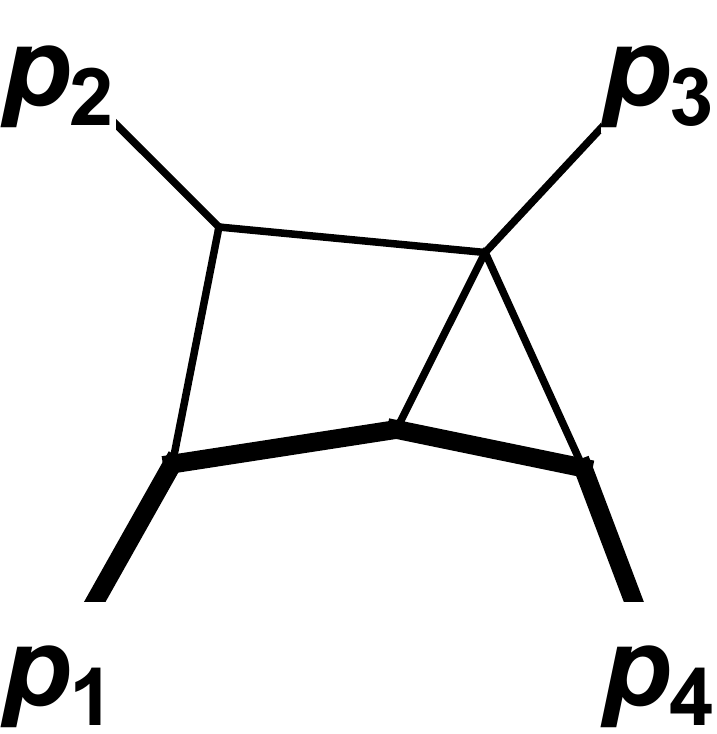}
  }
  \subfloat[$\mathcal{T}_{29}$]{%
    \includegraphics[width=0.14\textwidth]{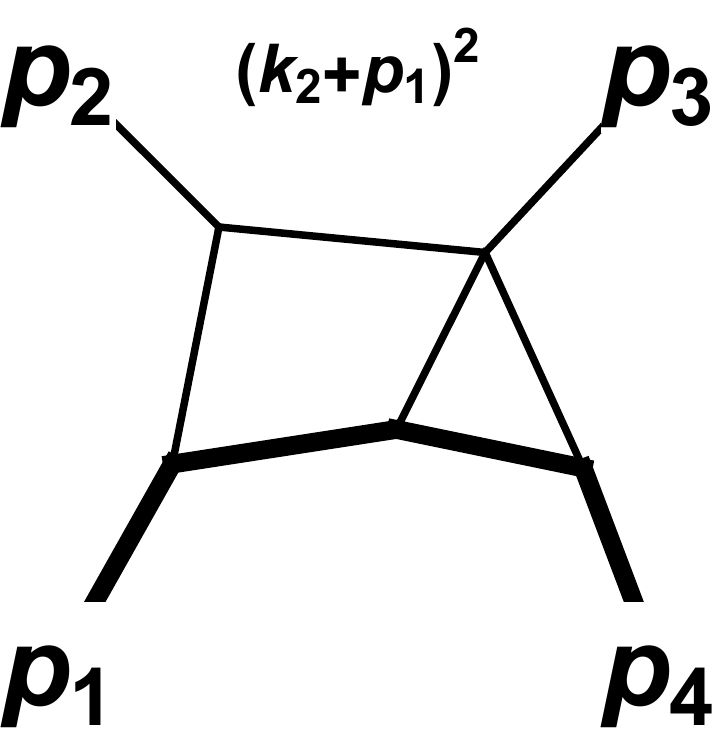}
  }
  \subfloat[$\mathcal{T}_{30}$]{%
    \includegraphics[width=0.14\textwidth]{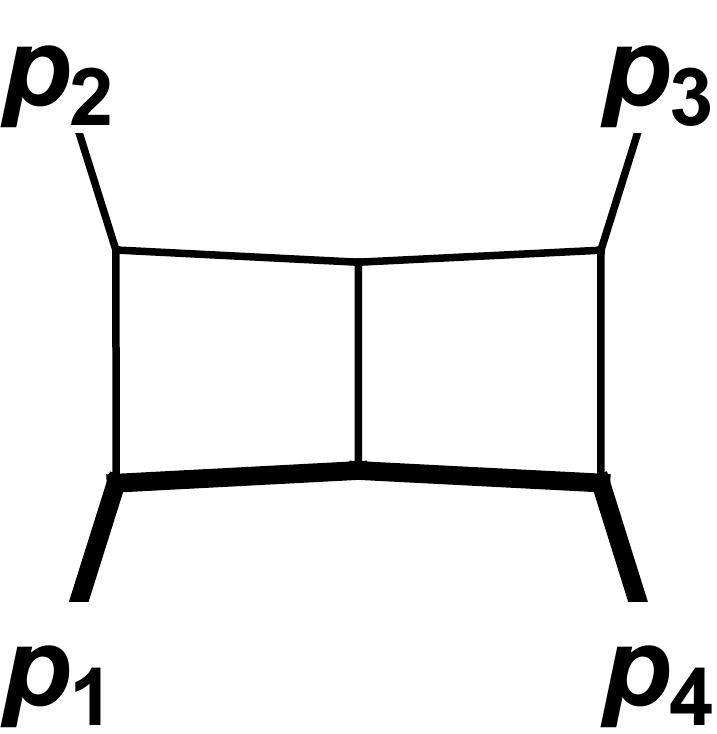}
  }
  \\
    \subfloat[$\mathcal{T}_{31}$]{%
    \includegraphics[width=0.14\textwidth]{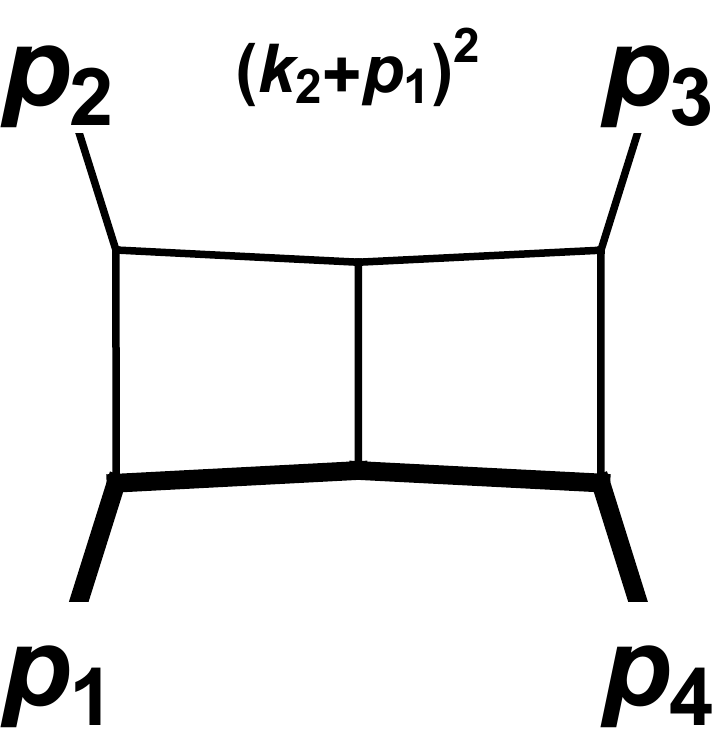}
  }
  \subfloat[$\mathcal{T}_{32}$]{%
    \includegraphics[width=0.14\textwidth]{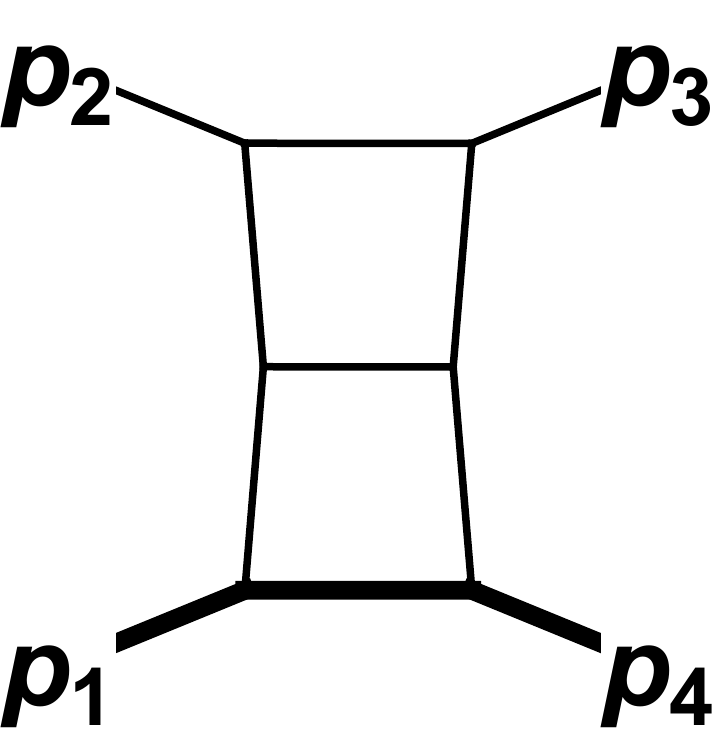}
  }
  \subfloat[$\mathcal{T}_{33}$]{%
    \includegraphics[width=0.14\textwidth]{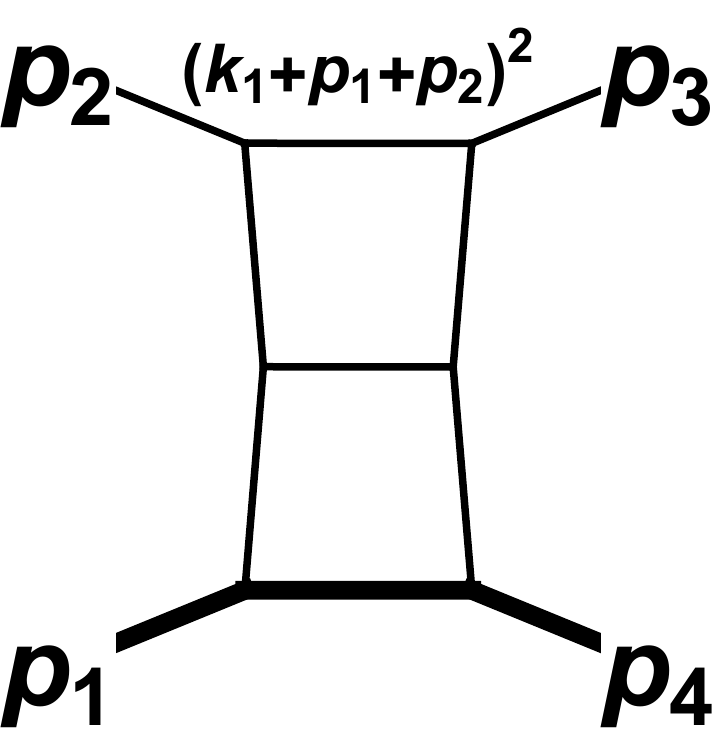}
  }
  \subfloat[$\mathcal{T}_{34}$]{%
    \includegraphics[width=0.14\textwidth]{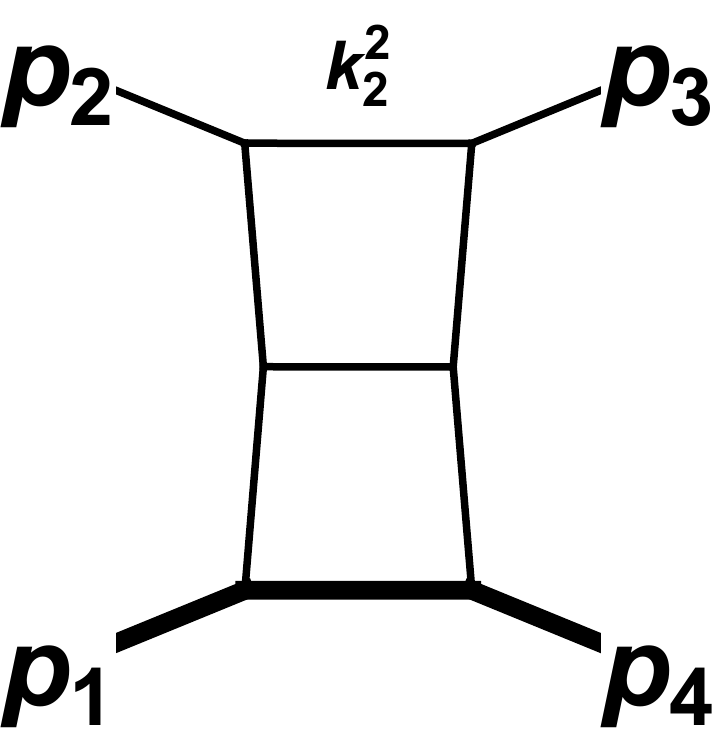}
  }
\caption{Two-loop MIs $\mathcal{T}_{1,\ldots,34}$ for the first integral family.}
 \label{fig:MIsT1}
\end{figure}

\section{Two-loop master integrals}
In this section we present the results for the planar two-loop MIs
contributing to the NNLO virtual QED corrections to $\mu e$
scattering, which are the main results of this work.
We first discuss the computation of the MIs belonging to the integral
family defined in eq.~\eqref{eq:2Lfamily1}, which is associated to the
topologies $T_1$, $T_2$, $T_3$, $T_7$ and $T_8$ of figure \ref{fig:Feyndiag}, and then the MIs belonging to the integral family defined by eq.~\eqref{eq:2Lfamily2}, which groups the topologies $T_4$, $T_5$, $T_9$ and $T_{10}$.
\subsection{The first integral family}
\label{sec:Topo1}

For the two-loop family defined in eq.~\eqref{eq:2Lfamily1}, the
following set of  34 MIs fulfill an $\eps$-linear system of DEQs,
\begin{align*}
\FF_{1}&=\eps^2 \, \top{1}\,,  &
\FF_{2}&=\eps^2 \, \top{2}\,,  &
\FF_{3}&=\eps^2 \, \top{3}\,,  \\
\FF_{4}&=\eps^2 \, \top{4}\,,  &
\FF_{5}&=\eps^2 \, \top{5}\,,  &
\FF_{6}&=\eps^2 \, \top{6}\,,  \\
\FF_{7}&=\eps^2 \, \top{7}\,,  &
\FF_{8}&=\eps^2 \, \top{8}\,,  &
\FF_{9}&=\eps^2 \, \top{9}\,,  \\
\FF_{10}&=\eps^3 \, \top{10}\,,  &
\FF_{11}&=\eps^3 \, \top{11}\,,  &
\FF_{12}&=\eps^3 \, \top{12}\,,  \\
\FF_{13}&=\eps^3 \, \top{13}\,,  &
\FF_{14}&=\eps^2 \, \top{14}\,,  &
\FF_{15}&=\eps^2 \, \top{15}\,,  \\
\FF_{16}&=\eps^3 \, \top{16}\,,  &
\FF_{17}&=\eps^4 \, \top{17}\,,  &
\FF_{18}&=\eps^3\, \top{18}\,,  \\
\FF_{19}&=\eps^4\, \top{19}\,,  &
\FF_{20}&=\eps^2(1+2\eps) \, \top{20}\,, &
\FF_{21}&=\eps^4 \, \top{21}\, , \\
\FF_{22}&=\eps^3\, \top{22}\,,  &
\FF_{23}&=\eps^3 \, \top{23}\,, &
\FF_{24}&=\eps^2 \, \top{24}\, , \\
\FF_{25}&=\eps^3\, \top{25}\,,  &
\FF_{26}&=\eps^3(1-2\eps) \, \top{26}\,, &
\FF_{27}&=\eps^3 \, \top{27}\, , \\
\FF_{28}&=\eps^4\, \top{28}\,,  &
\FF_{29}&=\eps^3(1-2\eps) \, \top{29}\,, &
\FF_{30}&=\eps^4 \, \top{30}\, , \\
\FF_{31}&=\eps^4\, \top{31}\,,  &
\FF_{32}&=\eps^4 \, \top{32}\,, &
\FF_{33}&=\eps^4 \, \top{33}\, , \\
\FF_{34}&=\eps^4\, \top{34}\,,  &
\stepcounter{equation}\tag{\theequation}
\label{def:LBasisT1}
\end{align*}
where the $\mathcal{T}_i$ are depicted in
figure~\ref{fig:MIsT1}. 
	
Through the Magnus exponential, we rotate this set of integrals to the canonical basis
\begingroup
\allowdisplaybreaks[1]
\begin{align*}
  \GG_{1}&=   \FF_1\,, & \qquad
  \GG_{2}&= -s  \,  \FF_2\,,\nn
   \GG_{3}&= -t   \FF_3\,, & \qquad
  \GG_{4}&= m^2\, \FF_4\,,   \nn
  \GG_{5}&=- s \FF_5 \,, &\qquad
  \GG_{6}&= 2m^2\,\FF_5+(m^2-s) \,  \FF_6\,, \nn
  \GG_{7}&= -t  \FF_7\,,  &\qquad
  \GG_{8}&=s^2\,\FF_8\,,  \nn
   \GG_{9}&= t^2  \FF_9\,, & \qquad
  \GG_{10}&=-t\,\FF_{10}\,, \nn
  \GG_{11}&= (m^2-s)\,\FF_{11}\,,& \qquad
  \GG_{12}&= \lambda_t\, \FF_{12}\,,   \nn  
  \GG_{13}&= \lambda_t\, \FF_{13}\,,  & \qquad  
  \GG_{14}&= \lambda_t \, m^2 \, \FF_{14}\,,\nn
  \GG_{15}&= (t - \lambda_t) \left(\frac{3}{2}  \FF_{13} +m^2 \FF_{14} \right)-  m^2 \, t \, \FF_{15}\,, & \qquad 
  \GG_{16}&= -t \, \lambda_t \, \FF_{16}\,, \nn
  \GG_{17}&= (m^2-s)\, \FF_{17}\,, &\qquad
  \GG_{18}&= m^2(m^2-s) \, \FF_{18}\,,  \nn  
  \GG_{19}&= \lambda_t \, \FF_{19}\,,& \qquad 
   \GG_{20}&=\frac{\lambda_t-t}{2}\left( \FF_{12}-4\, \FF_{19}\right) -m^2t\,  \FF_{20}\,,  \nn  
  \GG_{21}&=(m^2-s-t) \, \FF_{21}\,,  & \qquad  
  \GG_{22}&= -m^2\, t \, \FF_{22}\,, \nn
   \GG_{23}&=s\, t \, \FF_{23}\,,  & \qquad  
  \GG_{24}&= -m^2 \, t \, \FF_{23} +  (s-m^2) \, m^2 \, t\, \FF_{24}\,, \nn 
    \GG_{25}&=-(m^2-s)\, t \, \FF_{25}\,, & \qquad 
     \GG_{26}&= \lambda_t \, \FF_{26}\,,  \nn  
   \GG_{27}&=- (m^2-s)\,t \, \FF_{27} \,,  & \qquad  
  \GG_{28}&= (m^2-s)\, \lambda_t \, \FF_{28}\,,  \nn  
   \GG_{29}&=-2t\, \FF_{21}-(m^2-s)(2(\lambda_t -t)\FF_{28}-  \FF_{29})\,, &  \qquad  
  \GG_{30}&= - (m^2-s)^2t \, \FF_{30}\,,  \nn  
   \GG_{31}&=  (m^2-s)^2 \, \FF_{31}\, ,& \qquad
     \GG_{32}&=  (m^2-s)  \, t^2 \, \FF_{32}\,,  \nn  
   \GG_{33}&= -\lambda_t \, t  \, \FF_{33}\,, & \qquad
  \GG_{34}&= -m^2 \, t^2 \, \FF_{32} +  t^2  \, \FF_{34}\,.
\label{def:CanonicalBasisT1}
\stepcounter{equation}\tag{\theequation}
 \end{align*}%
 \endgroup
This set of MIs $\GGvec$ satisfies a system of DEQ of the form given
in eq.(\ref{eq:canonicalDEQ}), which can easily be integrated in terms
of GPLs. 
The $34 \times 34$ coefficient matrices are collected in appendix~\ref{dlog1stIF}. \\
To determine the solution of the DEQ, we need to choose proper
boundary values for each master integral. The boundary fixing can be
achieved either by knowing the integral at some special kinematic
point or by demanding the absence of unphysical thresholds that appear
in the alphabet of the generic solution, defined in
eq. (\ref{alphabet}).  

Below we describe in detail how the boundary constants for each integral were chosen:
\begin{itemize}
\item The boundary values of $\GG_{1,3,4,7,9}$ were obtained by direct
  integration,
  \begin{align}
    \GG_1(\eps
    )&=1, \\
    \GG_3(\eps
    )&=\left(\frac{(1-y)^2}{y}\right)^{-\eps} \left( 1-\zeta_2\eps^2  -2\zeta_3\eps^3 -\frac{9}{4}\zeta_4\eps^4  + \mathcal{O}(\eps^5) \right) , \\
   \GG_{4}(\eps)&=  -\frac{1}{4}-\zeta_2\eps^2-2 \zeta_3\eps^3-16\zeta_4\eps^4+\mathcal{O}\left(\eps^5\right)
   \, , \\  
     \GG_{7}(\eps,y)&=
   \left(\frac{(1-y)^2}{y}\right)^{-2 \eps}  \left(- 1+2\zeta_2\eps^2+10 \zeta_3 \eps^3+11\zeta_4\eps^4+\mathcal{O}\left(\eps^5\right)\right) \, , \\
        \GG_9(\eps
    )&= \left(\frac{(1-y)^2}{y}\right)^{-2 \eps} \left( 1-2 \, \zeta_2\, \eps^2-4 \, \zeta_3 \, \eps^3- 2 \, \zeta_4 \,\eps^4 + \mathcal{O}(\eps^5) \right)\,, \\
        \GG_{10}(\eps
    )&=\left(\frac{(1-y)^2}{y}\right)^{-2 \eps} \left(  \frac{1}{4}-2 \, \zeta_3 \, \eps^3- 3 \, \zeta_4 \,\eps^4 + \mathcal{O}(\eps^5)\right) \,.
   \end{align}
   \item The boundary constants of $\GG_{2,8,11,23}$ are fixed by demanding finiteness in the limit $s \rightarrow 0$.
   
    \item In the regular limit $s \to 0$, $\GG_5$ and $\GG_6$ become, respectively,
     \begin{align}
    \GG_5(\eps, 0)&=0\,,\nn
     \GG_6(\eps, 0)&=\eps^2m^2\left(2 \FF_5(\eps, 0)+\FF_6(\eps, 0)\right)\,.	
     \end{align}
     $ \FF_5(\eps, 0)$ and $\FF_6(\eps, 0)$ correspond to two-loop vacuum diagrams which can be reduced via IBPs to a single integral which can be analytically computed
 \begin{align}
    \FF_{5}(\eps,0)&=\frac{2 \eps (2 \eps-1)}{m^4} \raisebox{-19pt}{\includegraphics[scale=0.11]{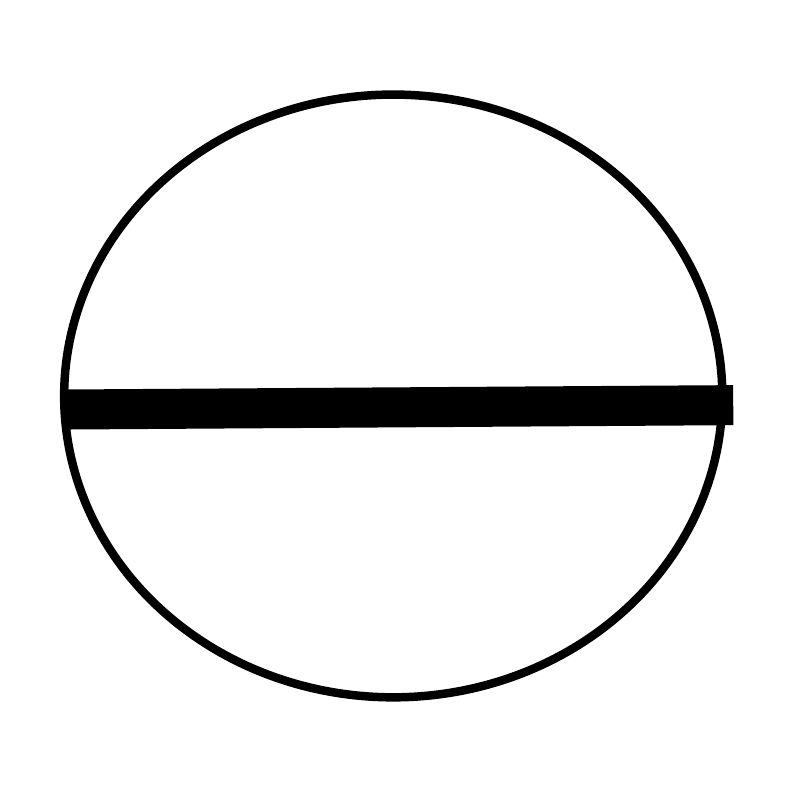}}\,,\nn
    \FF_{6}(\eps,0)&=-\frac{2 (\eps+1) (2\eps-1)}{m^4} \raisebox{-19pt}{\includegraphics[scale=0.11]{figures/Fig_Vac.pdf}}.
    \label{fig:FigVac.pdf}
  \end{align}
  In this way, we obtain the boundary values
  \begin{align}
  \GG_5(\eps, 0)&=0\,,\quad\GG_6(\eps, 0)=-1-2\zeta_2\eps^2+2
                  \zeta_3\eps^3 -9\zeta_4
                  \eps^4+\mathcal{O}\left(\eps^5\right)\ . 
  \end{align} 	
   \item The integration constants of $\GG_{12\dots 16,19,20,26,27,32,33}$ are fixed by demanding finiteness in the $t \rightarrow 4m^2$ limit and by demanding that the resulting boundary constants are real.
  \item The integrals $\GG_{17}$ and $\GG_{18}$ are regular in the $s \to 0$ limit. By imposing the regularity on their DEQ we can only fix the constant of one of them, say $\GG_{18}$.
    The boundary constants of $\GG_{17}$ must be then computed in an
    independent way. We observe that the value of  $\GG_{17}(\eps, 0)$ can
    be obtained in the limit $p_1^2\to m^2$ of a similar vertex integral with off-shell momentum $p_1^2$ and $s\equiv (p_1+p_2)^2=p_2^2=0$,
 \begin{align}
 \GG_{17}(\eps, 0)=\eps^4m^2\lim_{p_1^2\to m^2} \raisebox{-40pt}{\includegraphics[scale=0.20]{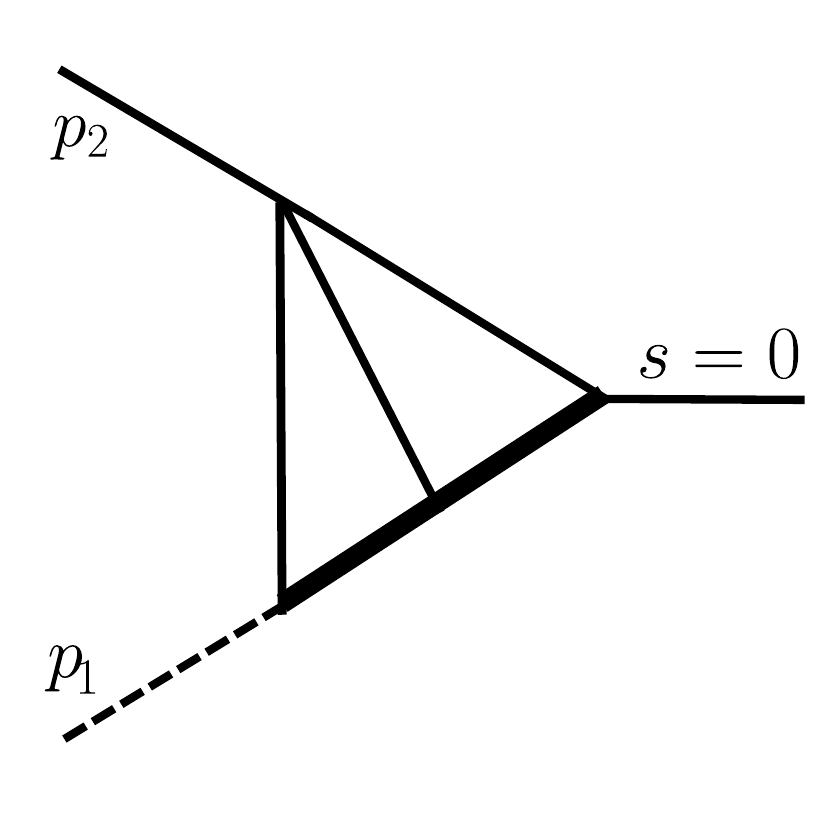}}\,.
 \label{eq:limit_i10}
 \end{align}
We discuss the computation of the auxiliary vertex integral in
appendix~\ref{sec:AuxVert}, where we show that the limit appearing in
the r.h.s of eq.~\eqref{eq:limit_i10} is indeed smooth 
and gives, 
  \begin{align}
  \GG_{17}(\eps,
    0)=-\frac{27}{4}\zeta_4\eps^4+\mathcal{O}\left(\eps^5\right)\, .
  \label{eq:bc_i10}
  \end{align}
  \item The regularity of the four-point integrals $\GG_{21,22,25,28\,\dots\,,31}$ in either $s \to 0$ or $t \to 4m^2$ provides two boundary conditions, which can be complemented with additional relations obtained by imposing the regularity of the integrals at $s=-t=m^2/2$.
\item  The boundary constants of integral $\GG_{24}$ are determined by demanding regularity in the limit $s \rightarrow -m^2$ and $t \rightarrow 4m^2 $.
\item The boundary constants of $\GG_{34}$ are found by demanding finiteness in the limit $u \rightarrow  \infty $.
\end{itemize}
All results have been numerically checked with the help of the computer codes
\texttt{GiNaC} and \texttt{SecDec}, and the analytic expressions
of the MIs are given in electronic form in the ancillary files
attached to the \texttt{arXiv} version of this manuscript.

\subsection{The second integral family}
\label{sec:Topo4}
\begin{figure}[H]
  \centering
  \captionsetup[subfigure]{labelformat=empty}
  \subfloat[$\mathcal{T}_1$]{%
    \includegraphics[width=0.14\textwidth]{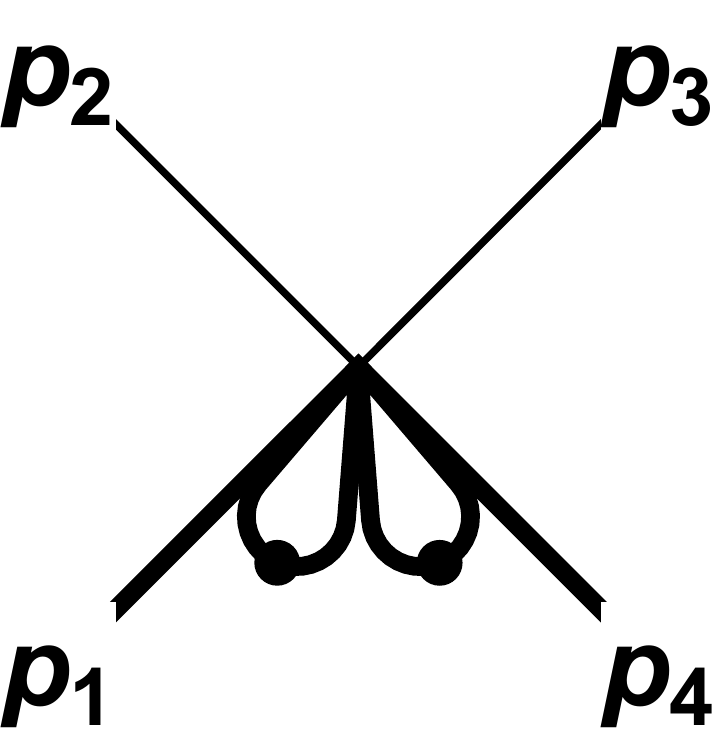}
  }
  \subfloat[$\mathcal{T}_2$]{%
    \includegraphics[width=0.14\textwidth]{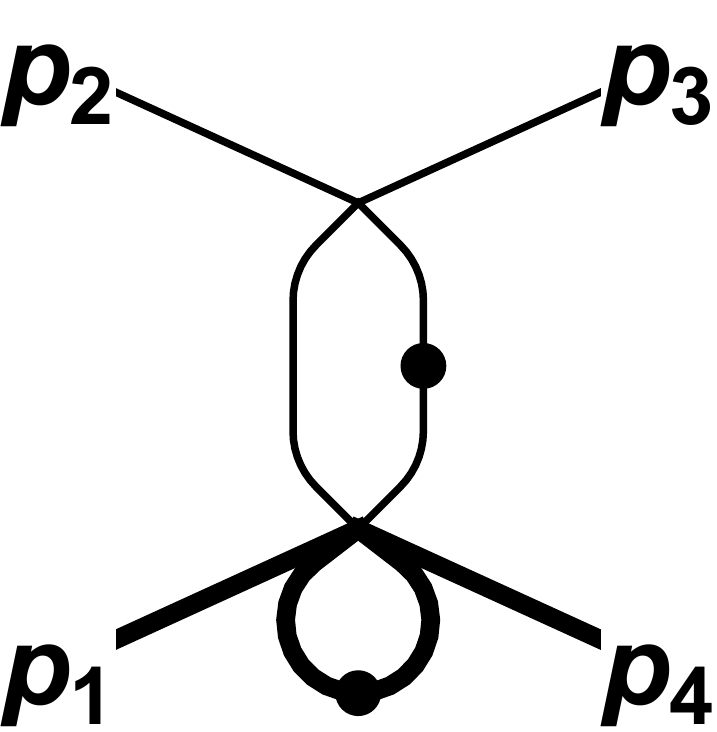}
  }
  \subfloat[$\mathcal{T}_3$]{%
    \includegraphics[width=0.14\textwidth]{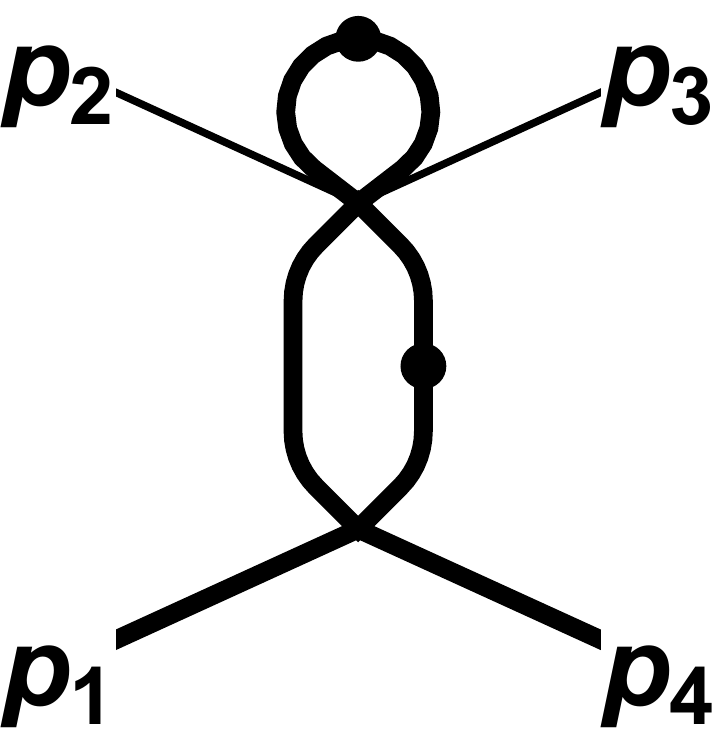}
  }
  \subfloat[$\mathcal{T}_4$]{%
    \includegraphics[width=0.14\textwidth]{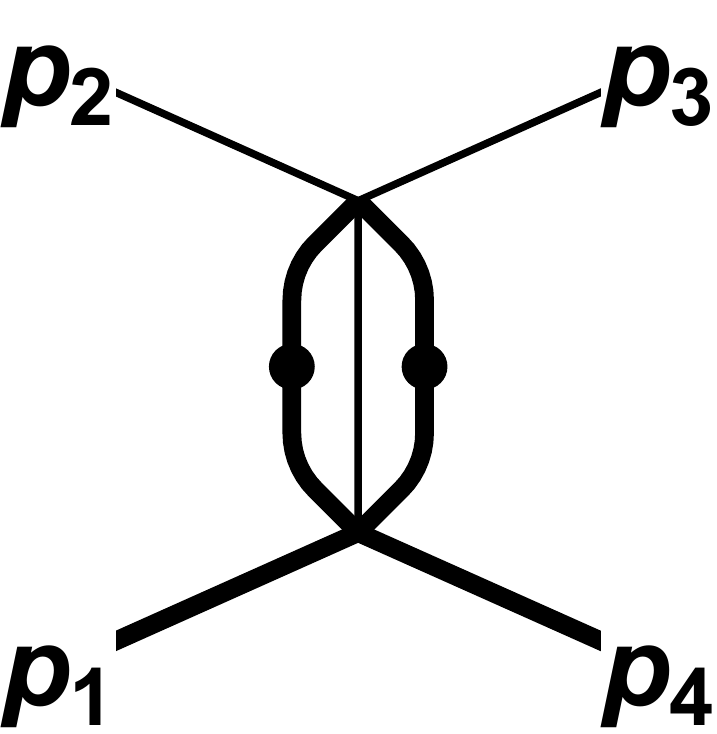}
  }
  \subfloat[$\mathcal{T}_5$]{%
    \includegraphics[width=0.14\textwidth]{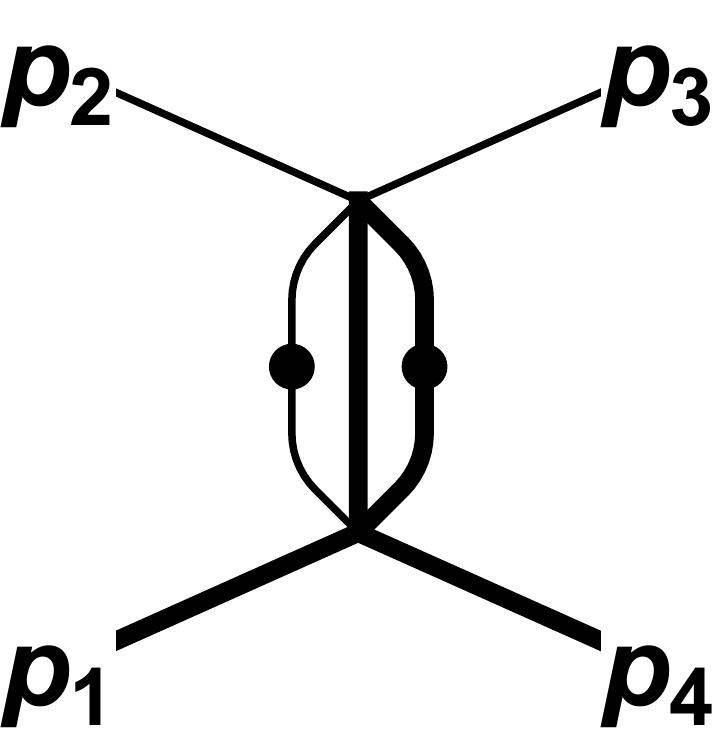}
  }
  \subfloat[$\mathcal{T}_6$]{%
    \includegraphics[width=0.14\textwidth]{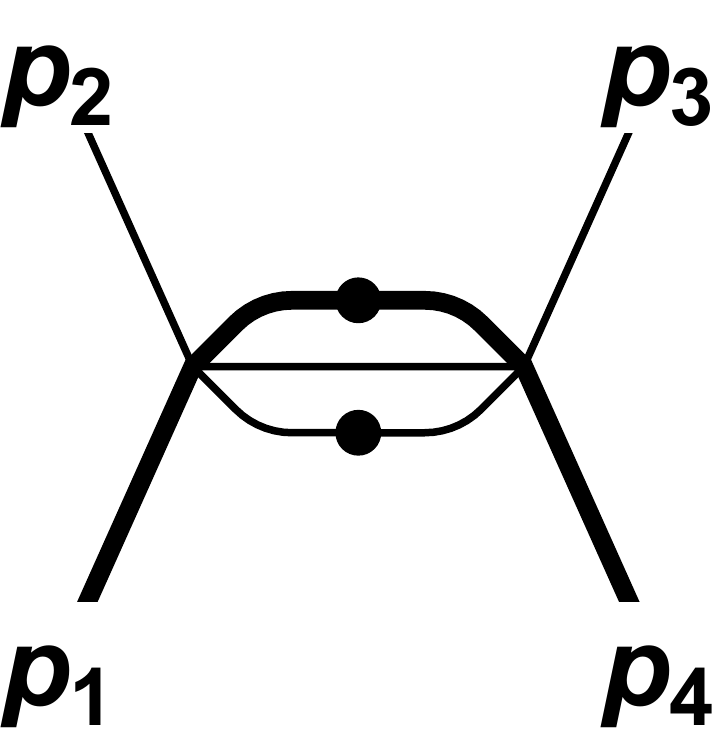}
  } \\
  \subfloat[$\mathcal{T}_7$]{%
    \includegraphics[width=0.14\textwidth]{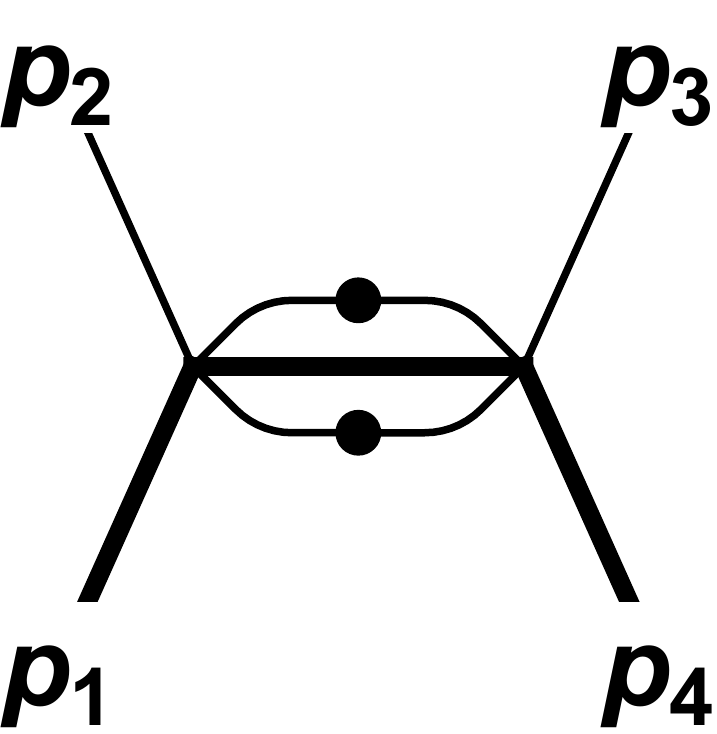}
  }
  \subfloat[$\mathcal{T}_8$]{%
    \includegraphics[width=0.14\textwidth]{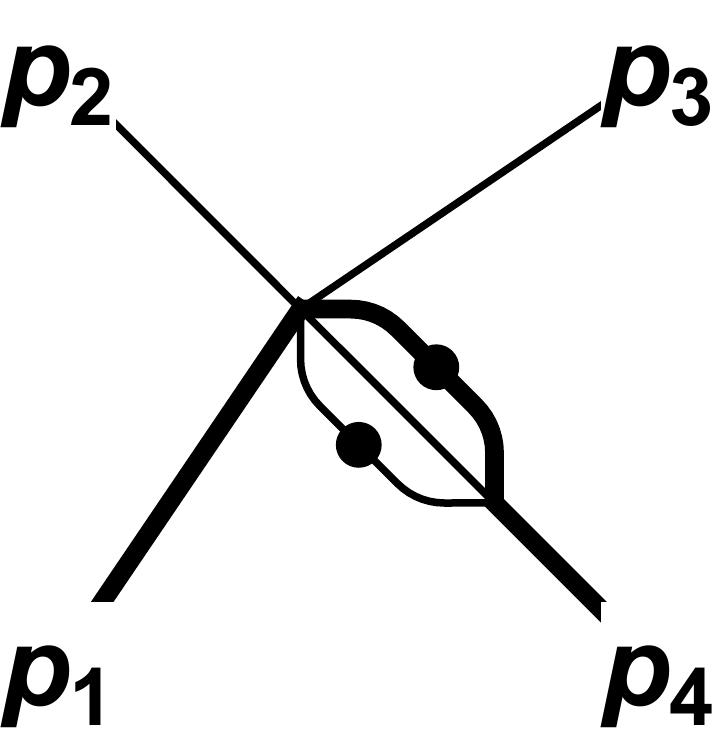}
  }
  \subfloat[$\mathcal{T}_9$]{%
    \includegraphics[width=0.14\textwidth]{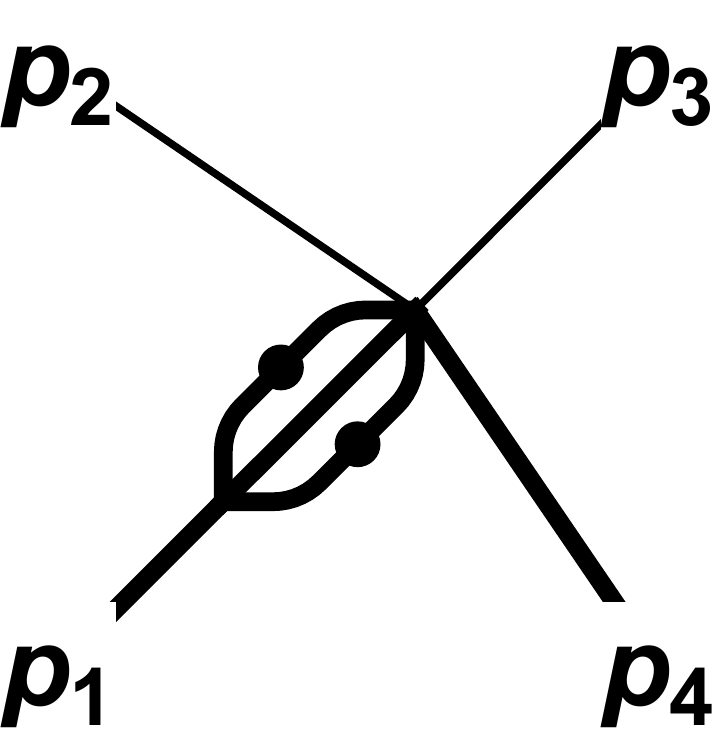}
  }
  \subfloat[$\mathcal{T}_{10}$]{%
    \includegraphics[width=0.14\textwidth]{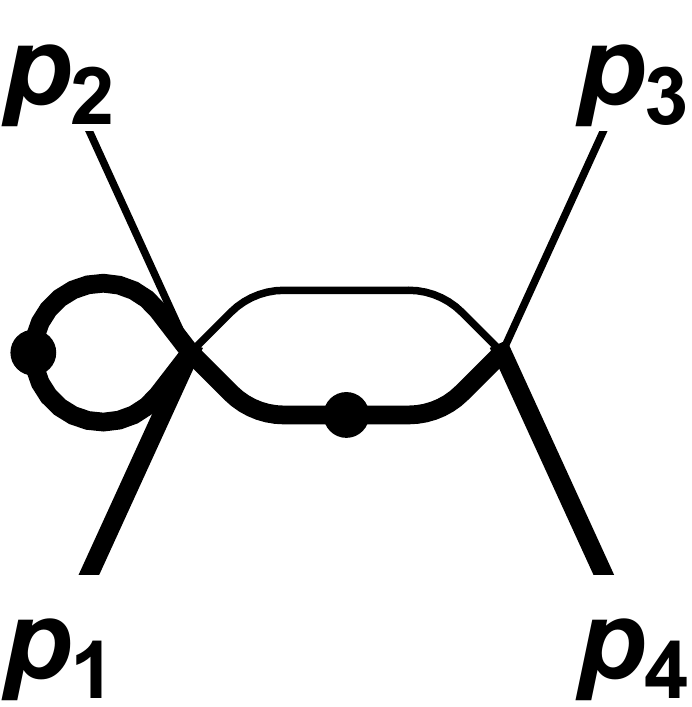}
  }
  \subfloat[$\mathcal{T}_{11}$]{%
    \includegraphics[width=0.14\textwidth]{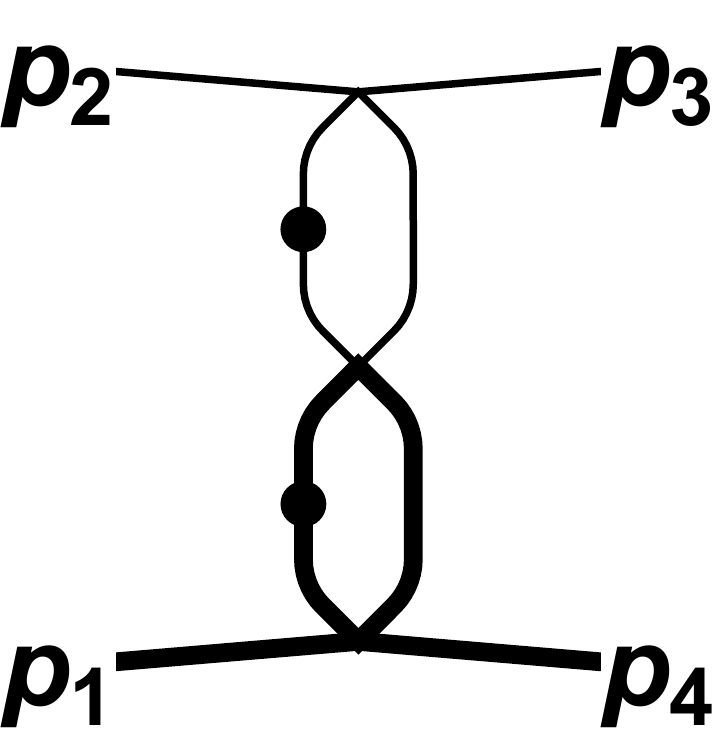}
  }
  \subfloat[$\mathcal{T}_{12}$]{%
    \includegraphics[width=0.14\textwidth]{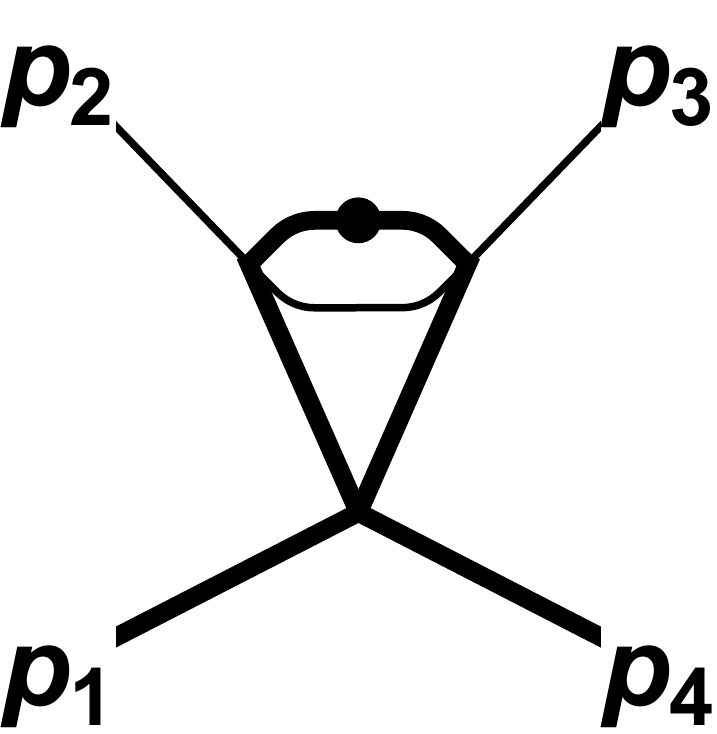}
  } \\
  \subfloat[$\mathcal{T}_{13}$]{%
    \includegraphics[width=0.14\textwidth]{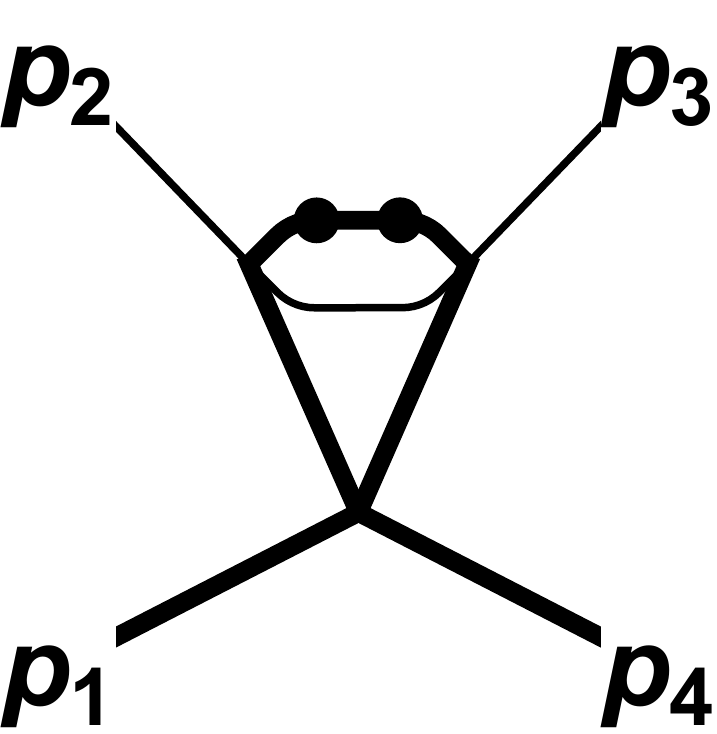}
  }
  \subfloat[$\mathcal{T}_{14}$]{%
    \includegraphics[width=0.14\textwidth]{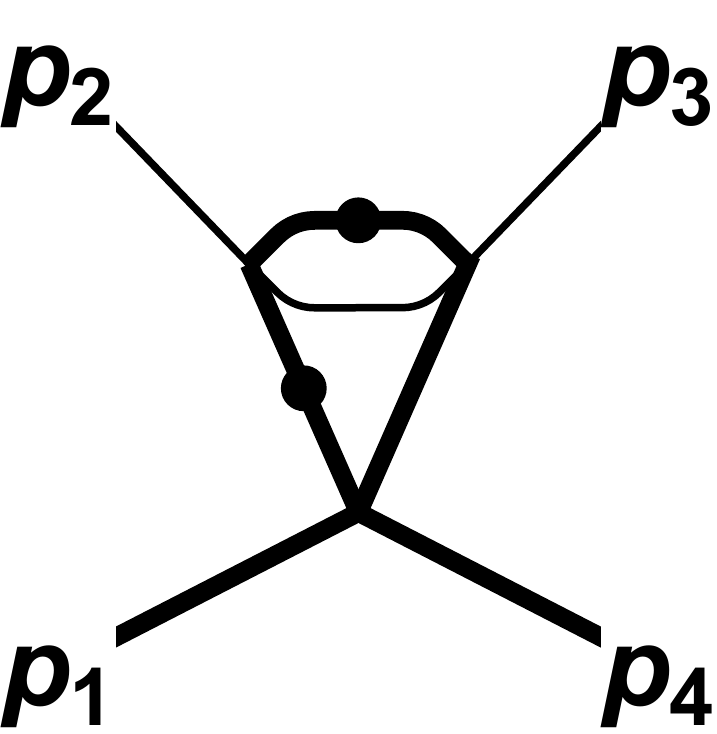}
  }
  \subfloat[$\mathcal{T}_{15}$]{%
    \includegraphics[width=0.14\textwidth]{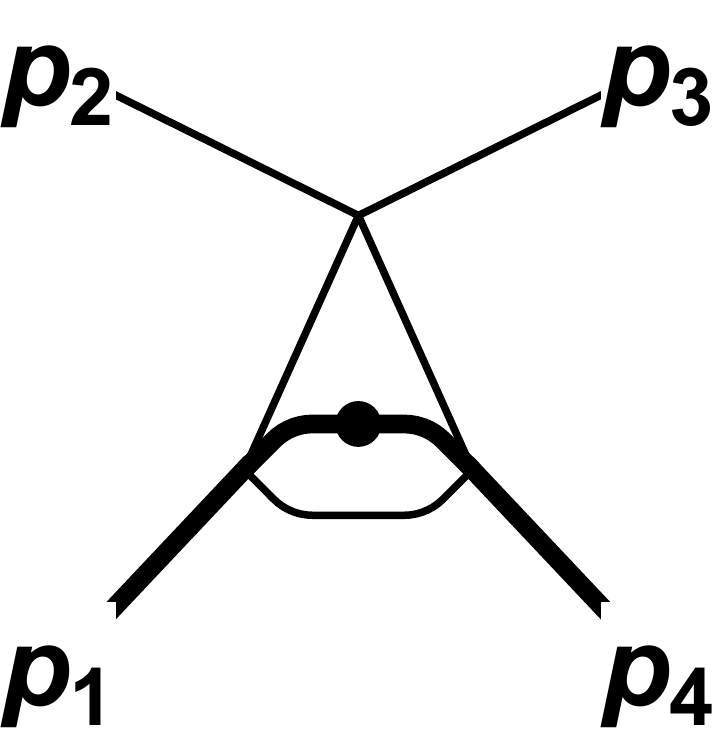}
  }
  \subfloat[$\mathcal{T}_{16}$]{%
    \includegraphics[width=0.14\textwidth]{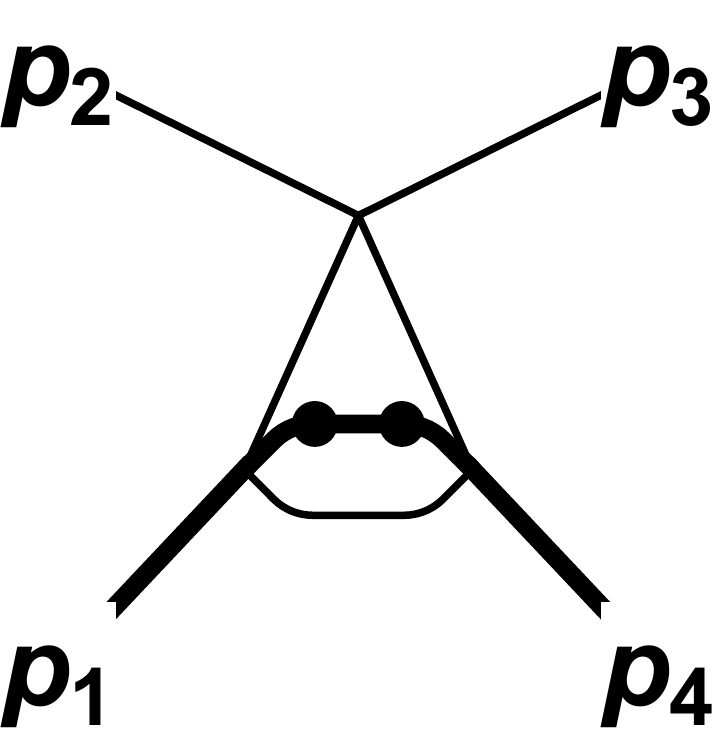}
  }
  \subfloat[$\mathcal{T}_{17}$]{%
    \includegraphics[width=0.14\textwidth]{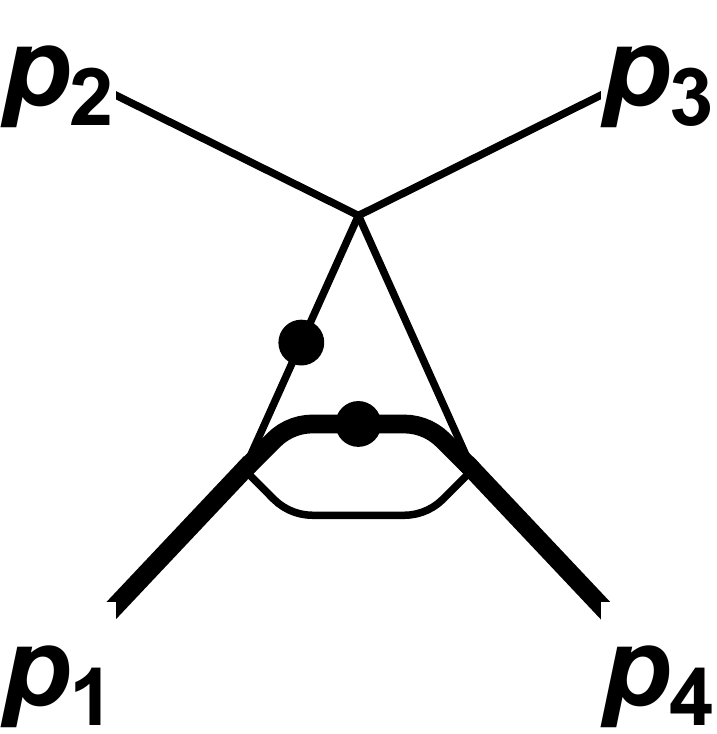}
  }
  \subfloat[$\mathcal{T}_{18}$]{%
    \includegraphics[width=0.14\textwidth]{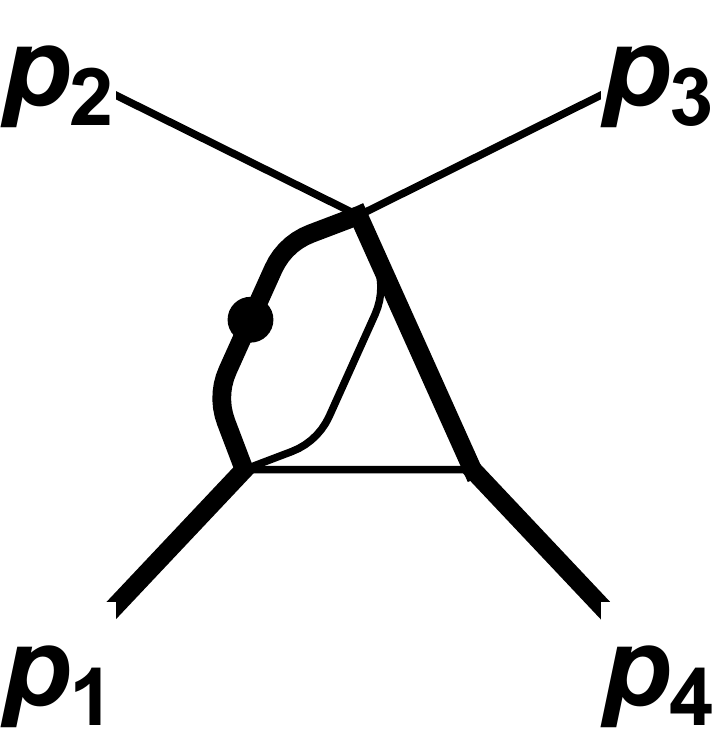}
  } \\
    \subfloat[$\mathcal{T}_{19}$]{%
    \includegraphics[width=0.14\textwidth]{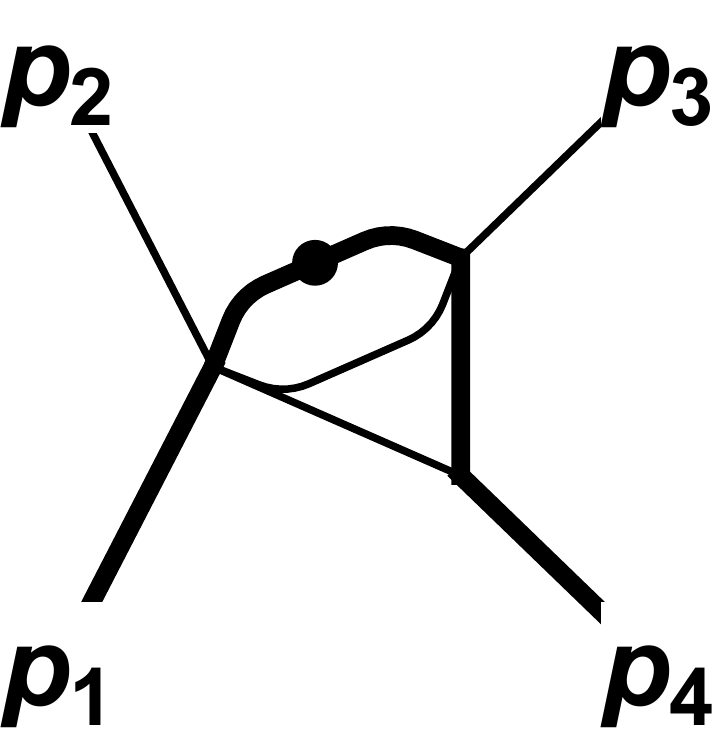}
  }
  \subfloat[$\mathcal{T}_{20}$]{%
    \includegraphics[width=0.14\textwidth]{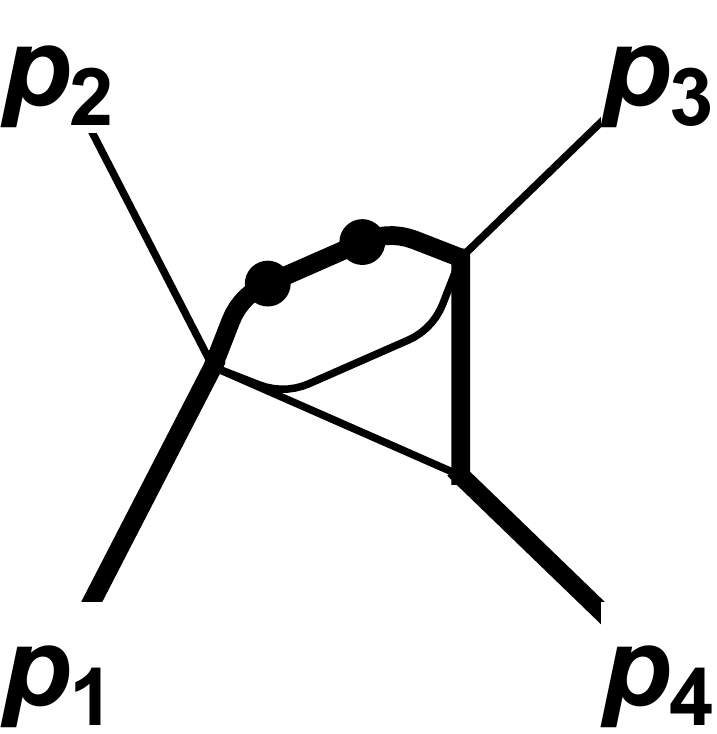}
  }
  \subfloat[$\mathcal{T}_{21}$]{%
    \includegraphics[width=0.14\textwidth]{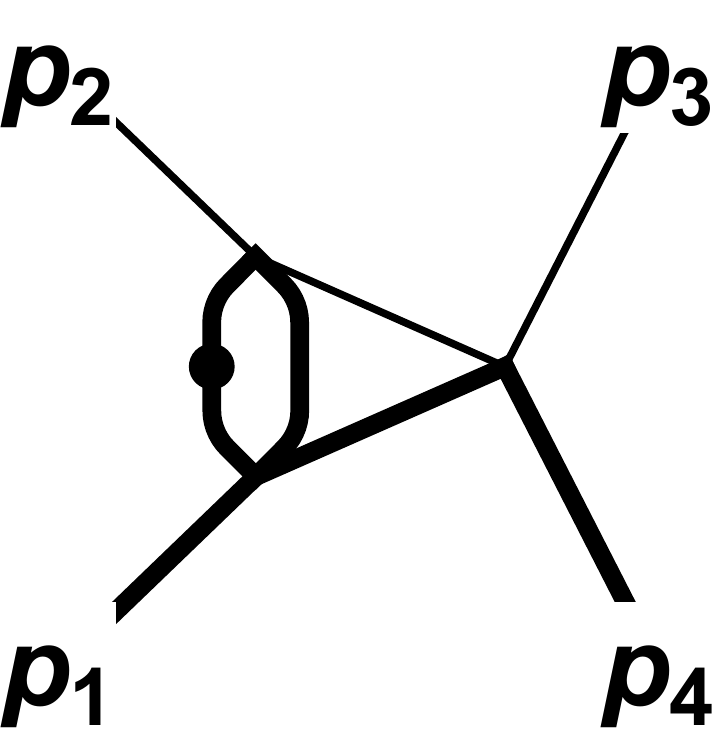}
  }
  \subfloat[$\mathcal{T}_{22}$]{%
    \includegraphics[width=0.14\textwidth]{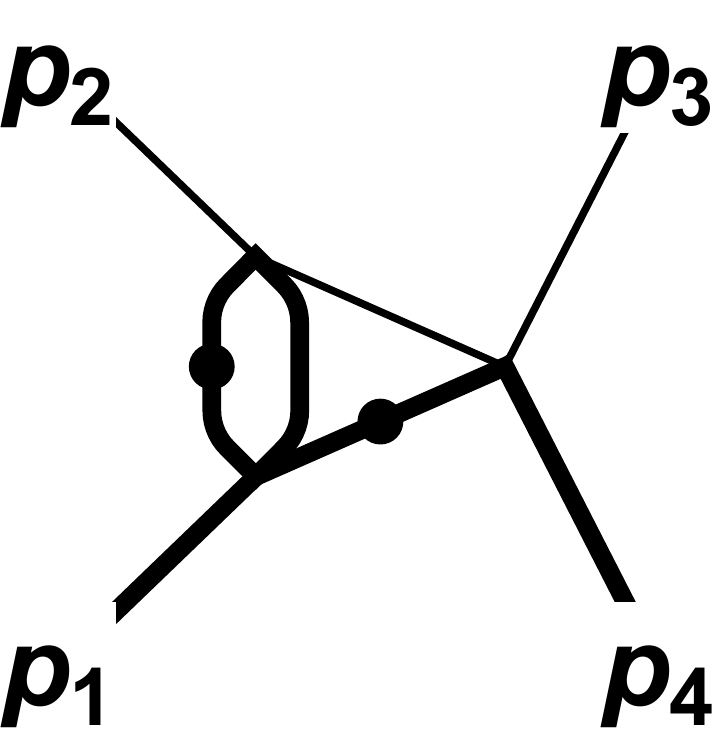}
  }
  \subfloat[$\mathcal{T}_{23}$]{%
    \includegraphics[width=0.14\textwidth]{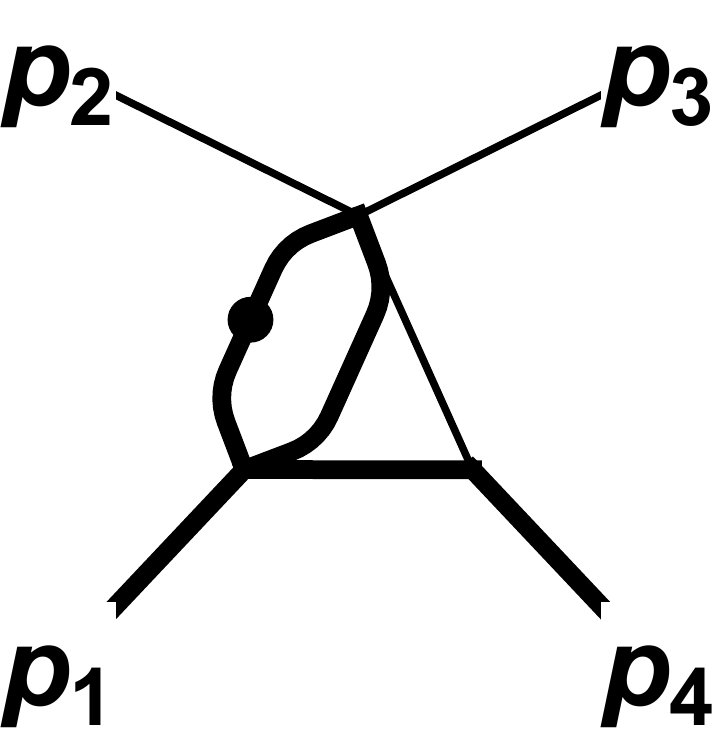}
  }
  \subfloat[$\mathcal{T}_{24}$]{%
    \includegraphics[width=0.14\textwidth]{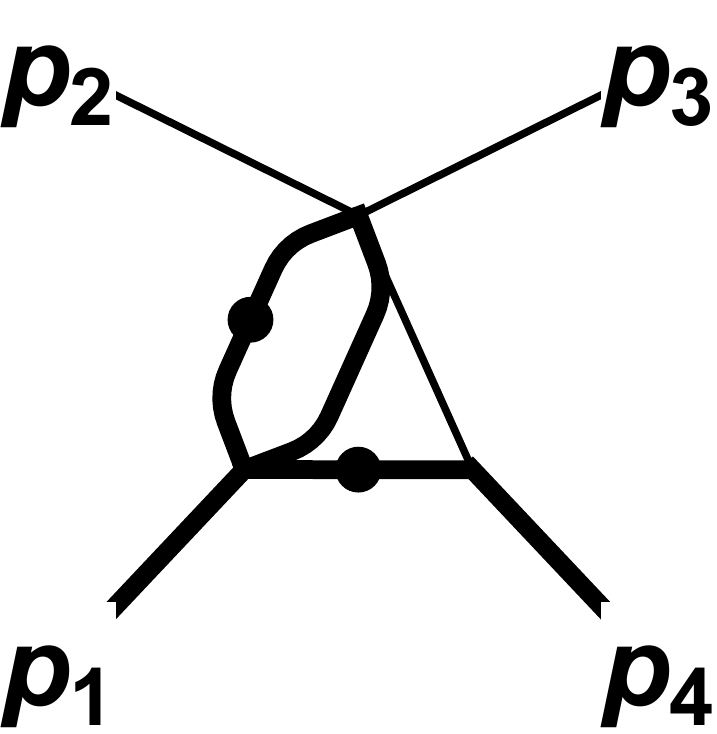}
  } \\
    \subfloat[$\mathcal{T}_{25}$]{%
    \includegraphics[width=0.14\textwidth]{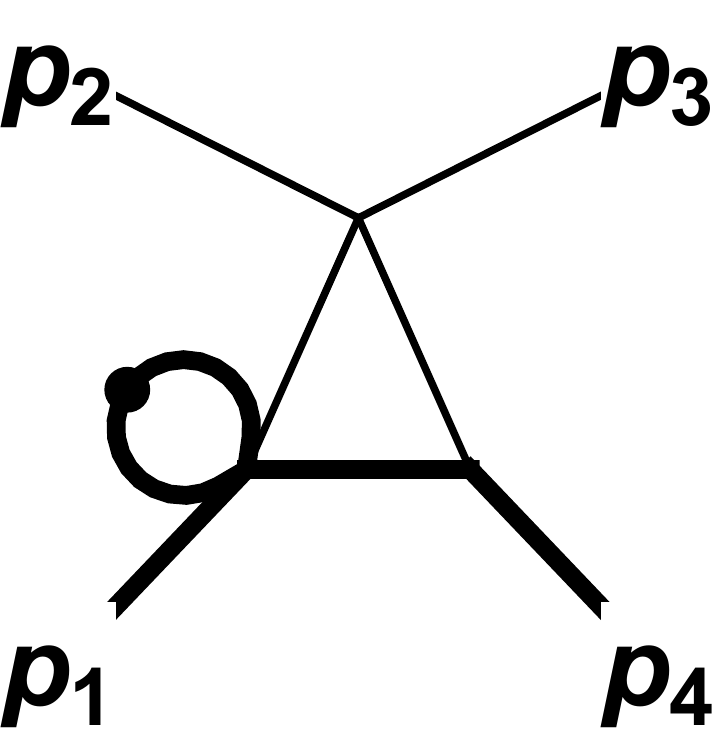}
  }
  \subfloat[$\mathcal{T}_{26}$]{%
    \includegraphics[width=0.14\textwidth]{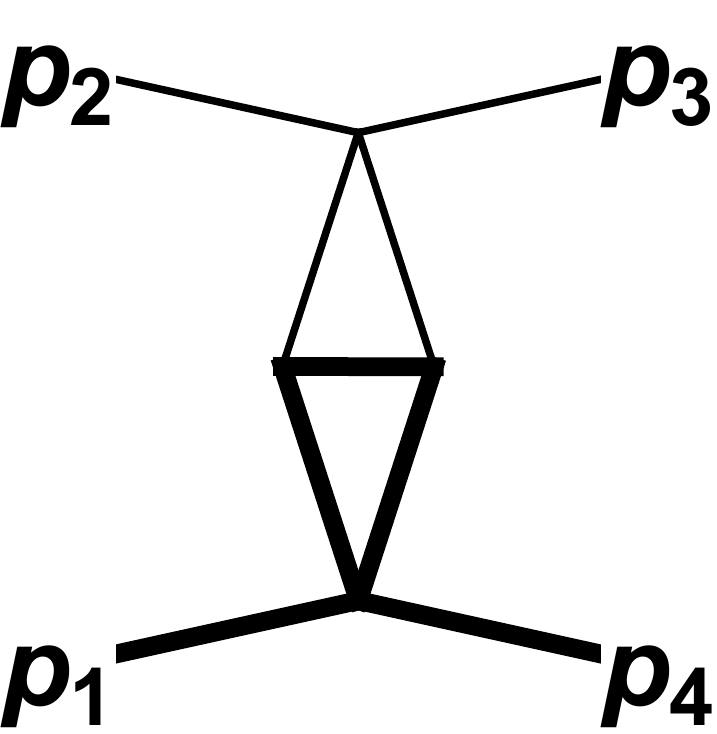}
  }
  \subfloat[$\mathcal{T}_{27}$]{%
    \includegraphics[width=0.14\textwidth]{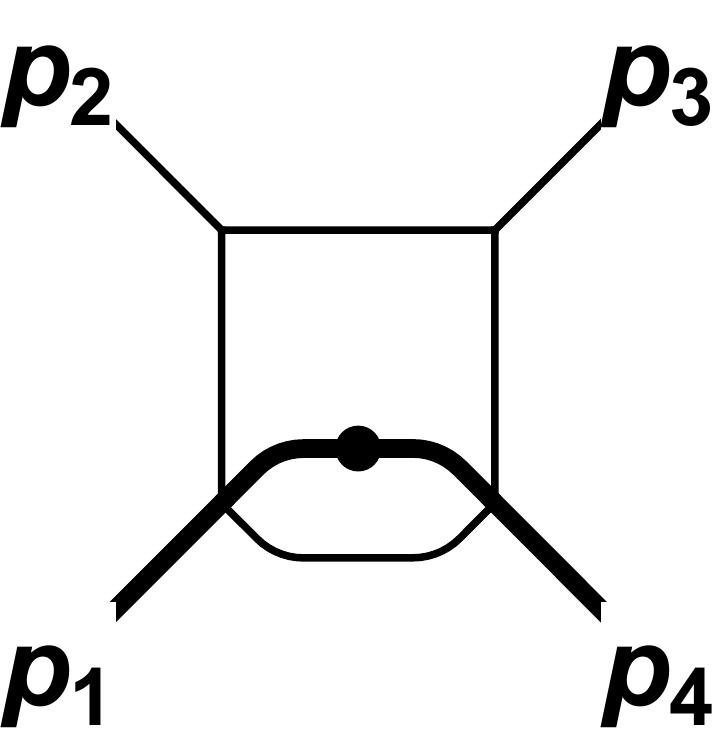}
  }
  \subfloat[$\mathcal{T}_{28}$]{%
    \includegraphics[width=0.14\textwidth]{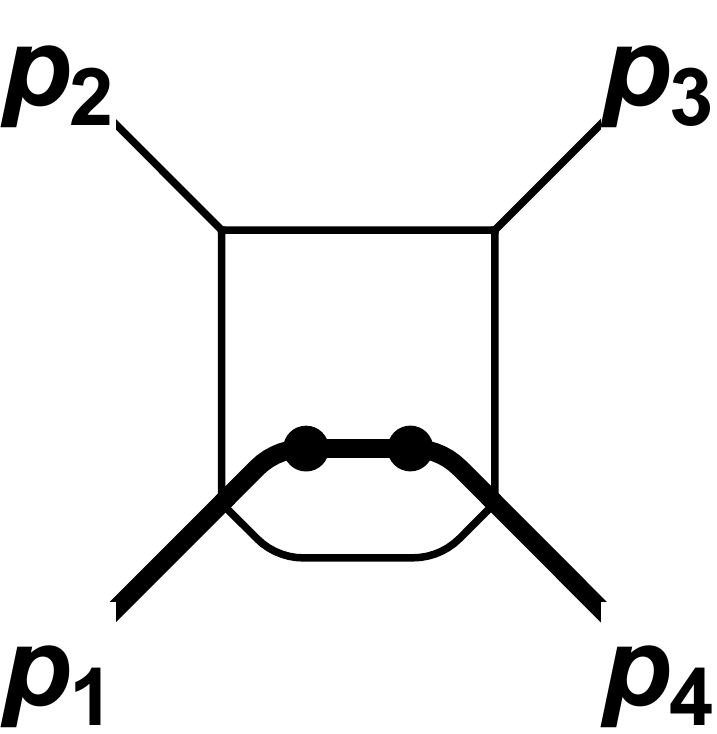}
  }
  \subfloat[$\mathcal{T}_{29}$]{%
    \includegraphics[width=0.14\textwidth]{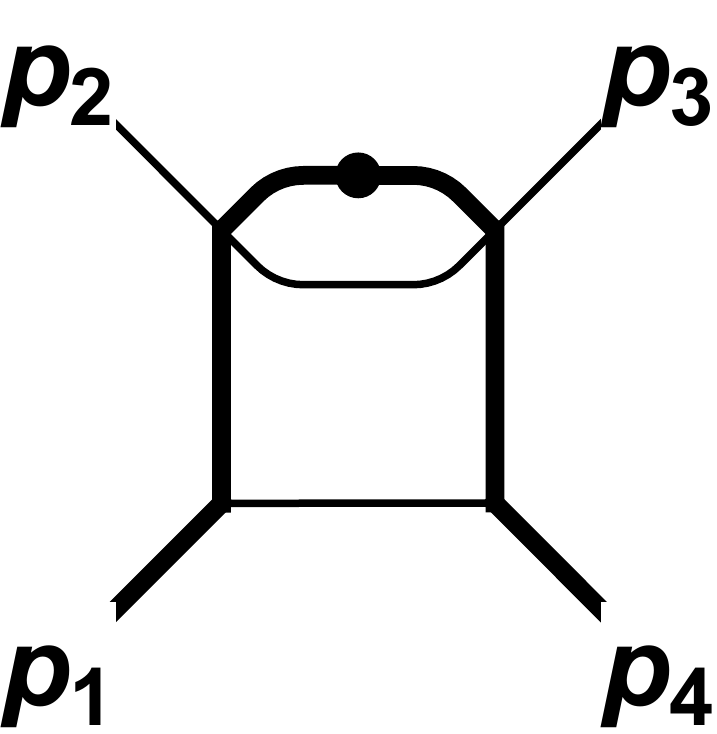}
  }
  \subfloat[$\mathcal{T}_{30}$]{%
    \includegraphics[width=0.14\textwidth]{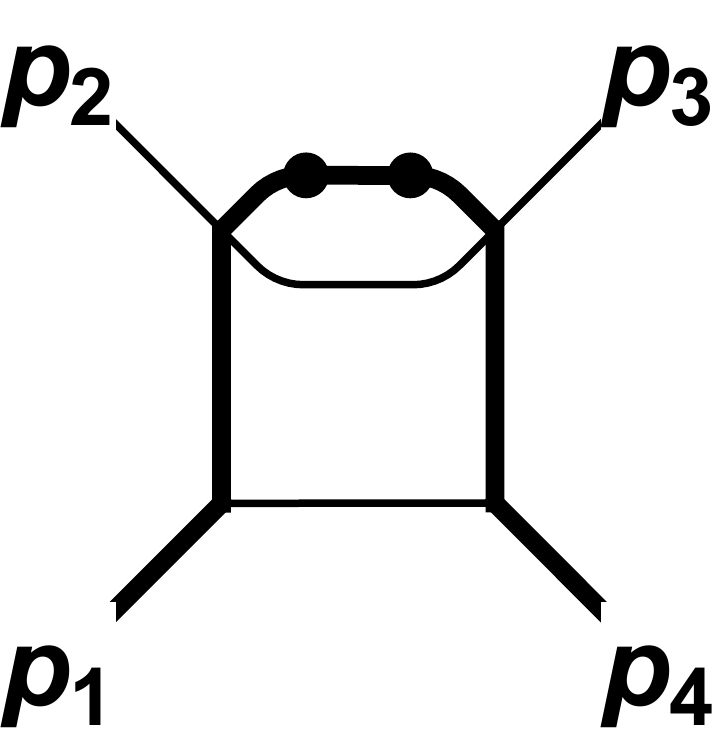}
  } \\
      \subfloat[$\mathcal{T}_{31}$]{%
    \includegraphics[width=0.14\textwidth]{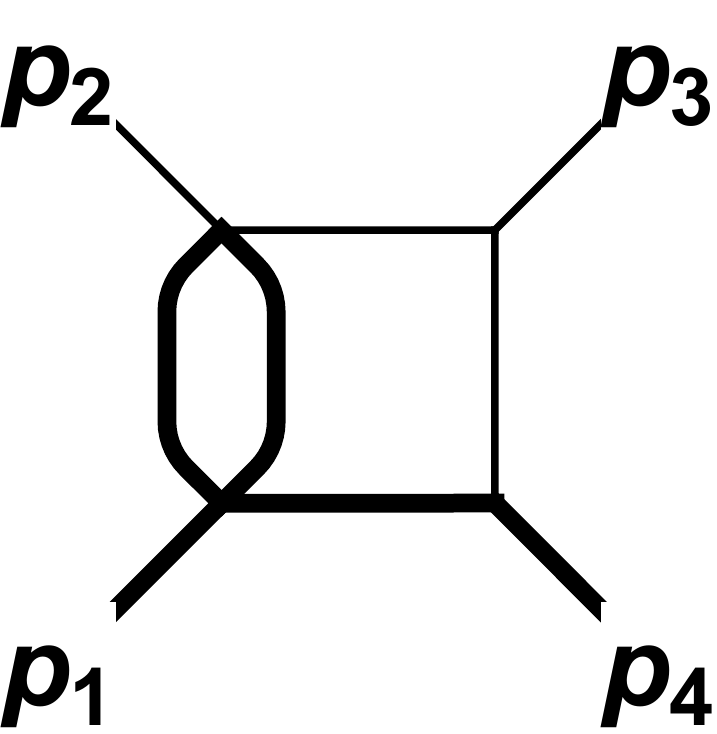}
  }
  \subfloat[$\mathcal{T}_{32}$]{%
    \includegraphics[width=0.14\textwidth]{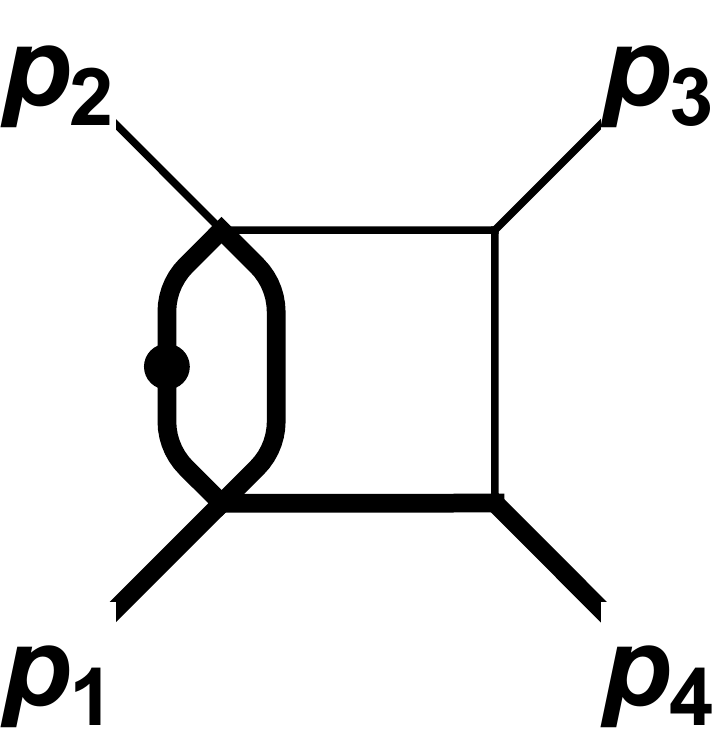}
  }
  \subfloat[$\mathcal{T}_{33}$]{%
    \includegraphics[width=0.14\textwidth]{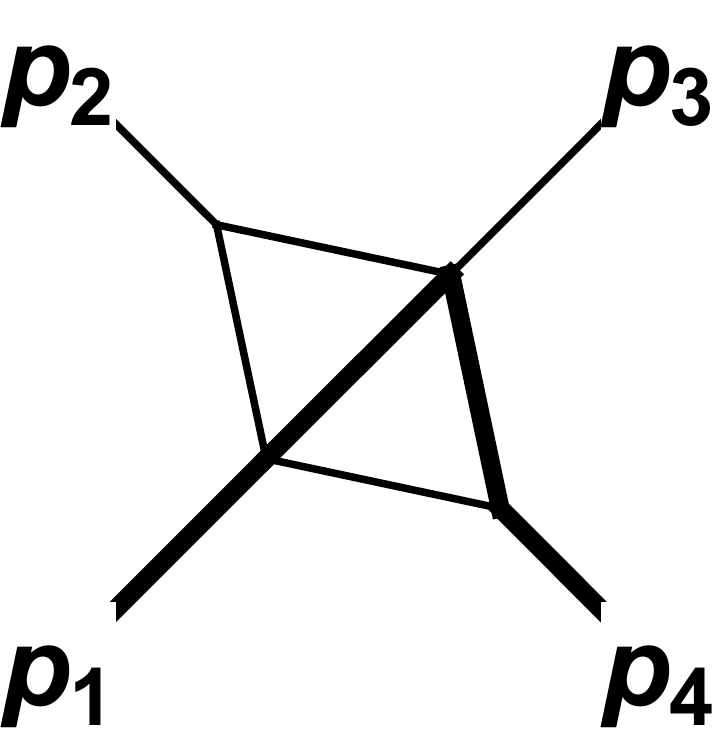}
  }
  \subfloat[$\mathcal{T}_{34}$]{%
    \includegraphics[width=0.14\textwidth]{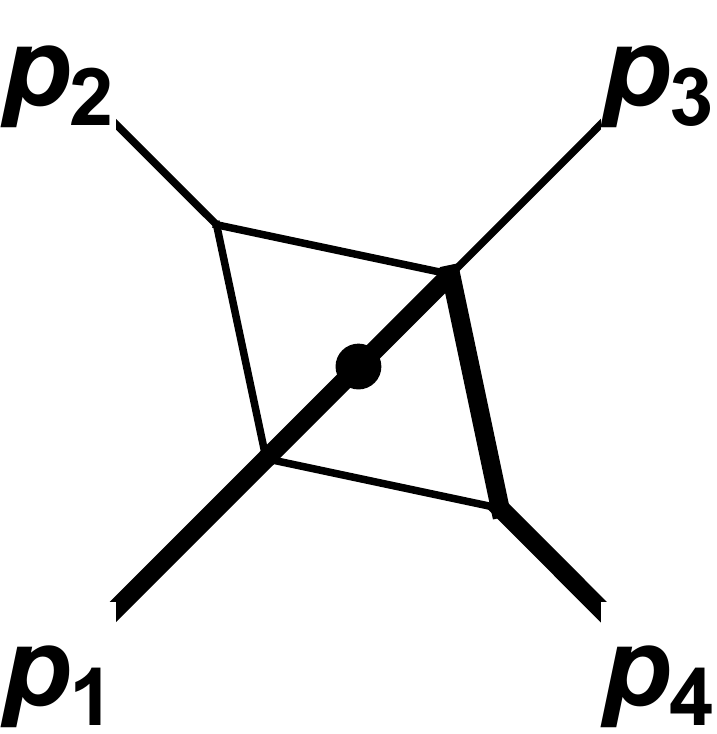}
  }
  \subfloat[$\mathcal{T}_{35}$]{%
    \includegraphics[width=0.14\textwidth]{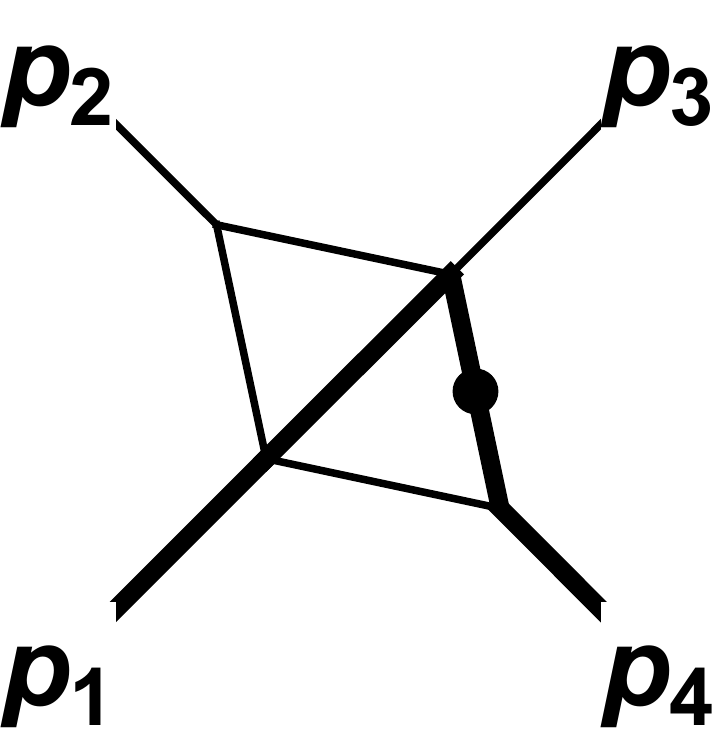}
  }
  \subfloat[$\mathcal{T}_{36}$]{%
    \includegraphics[width=0.14\textwidth]{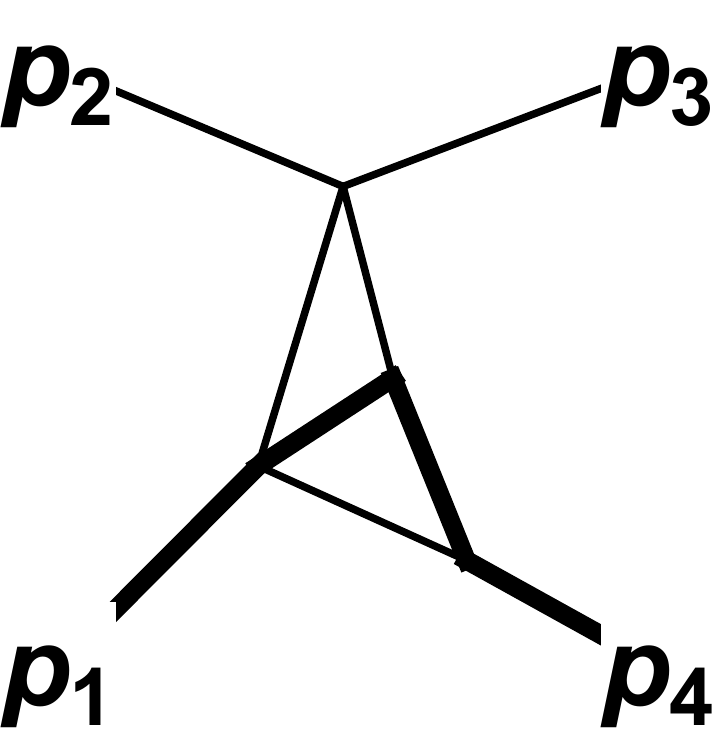}
  } \\
      \subfloat[$\mathcal{T}_{37}$]{%
    \includegraphics[width=0.14\textwidth]{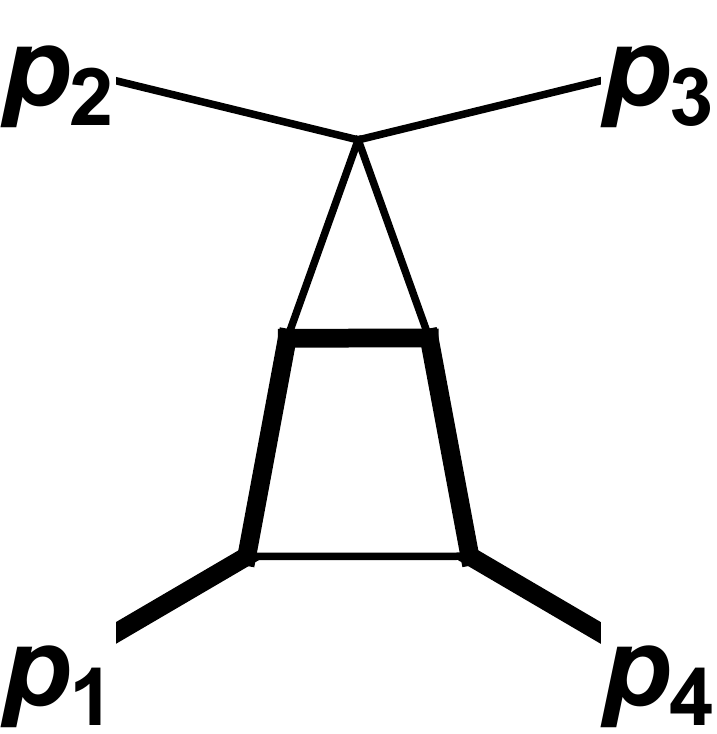}
  }
  \subfloat[$\mathcal{T}_{38}$]{%
    \includegraphics[width=0.14\textwidth]{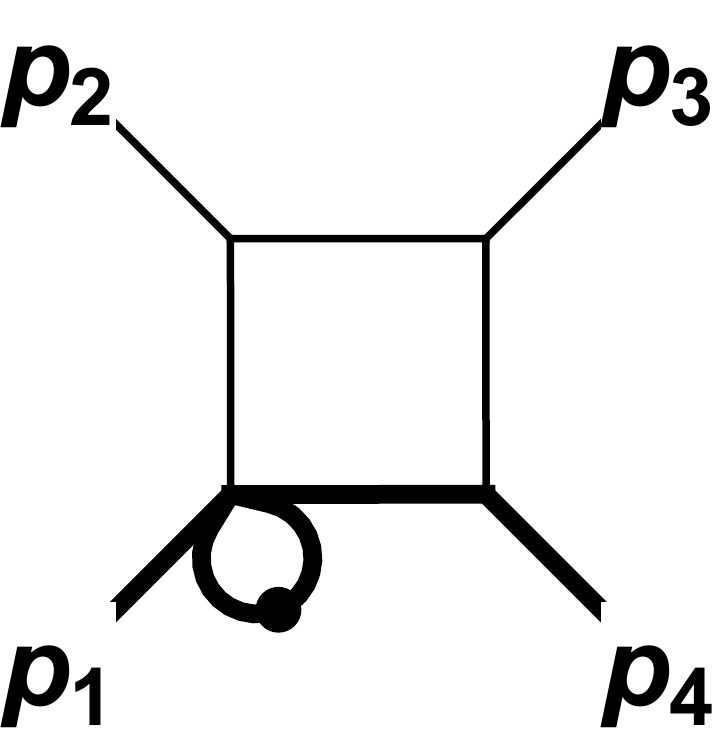}
  }
  \subfloat[$\mathcal{T}_{39}$]{%
    \includegraphics[width=0.14\textwidth]{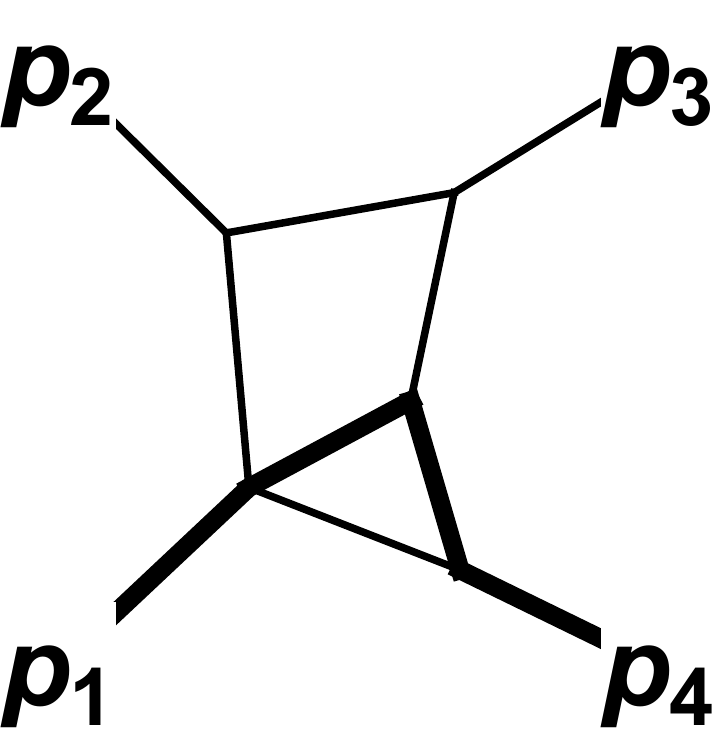}
  }
  \subfloat[$\mathcal{T}_{40}$]{%
    \includegraphics[width=0.14\textwidth]{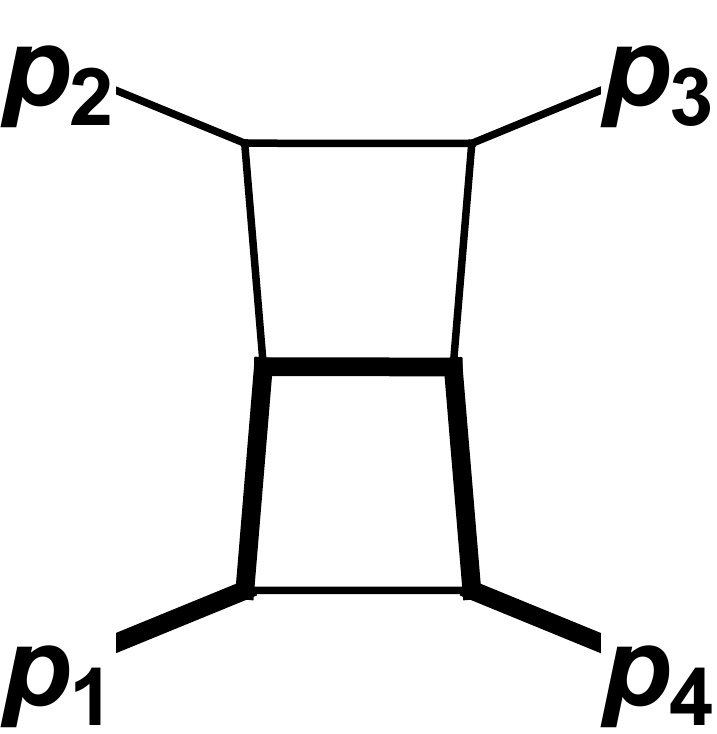}
  }
  \subfloat[$\mathcal{T}_{41}$]{%
    \includegraphics[width=0.14\textwidth]{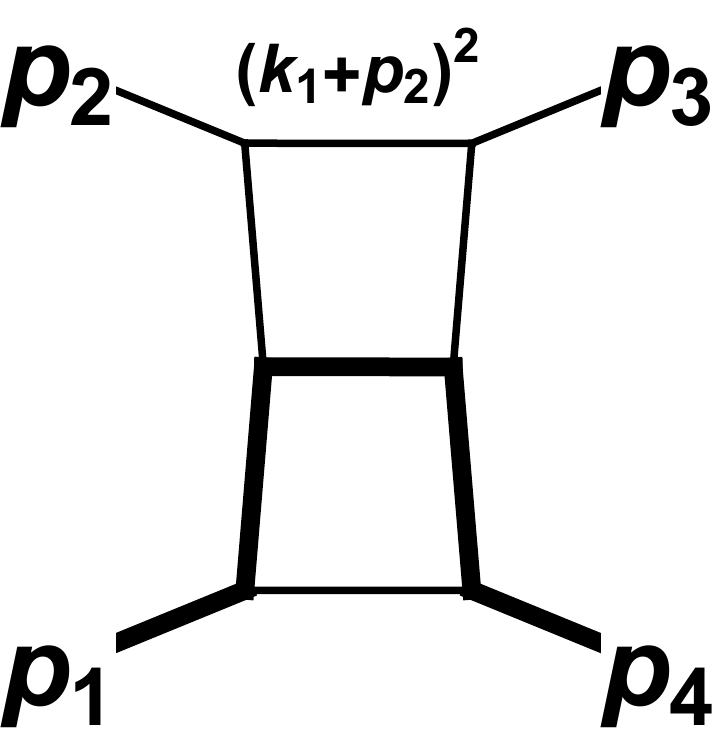}
  }
  \subfloat[$\mathcal{T}_{42}$]{%
    \includegraphics[width=0.14\textwidth]{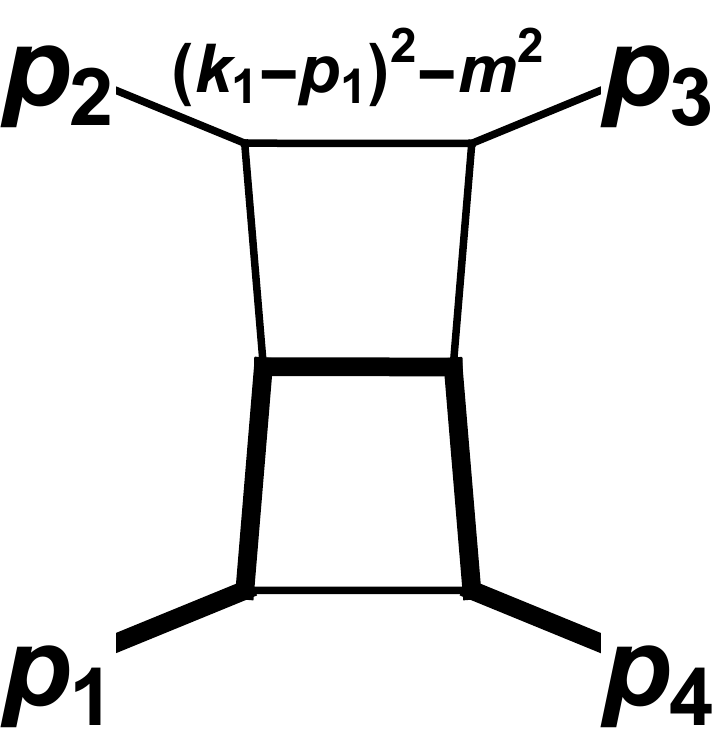}
  } \\
 \caption{Two-loop MIs $\mathcal{T}_{1,\ldots,42}$ for the second
   integral family. }
 \label{fig:MIsT4}
\end{figure}

For the two-loop integral family defined in~\eqref{eq:2Lfamily2}, we
identify 42 MIs obeying an $\eps$-linear system of DEQs:
\begin{align*}
\FF_{1}&=\eps^2 \, \top{1}\,,  &
\FF_{2}&=\eps^2 \, \top{2}\,,  &
\FF_{3}&=\eps^2 \, \top{3}\,,  \\
\FF_{4}&=\eps^2 \, \top{4}\,,  &
\FF_{5}&=\eps^2 \, \top{5}\,,  &
\FF_{6}&=\eps^2 \, \top{6}\,,  \\
\FF_{7}&=\eps^2 \, \top{7}\,,  &
\FF_{8}&=\eps^2 \, \top{8}\,,  &
\FF_{9}&=\eps^2 \, \top{9}\,,  \\
\FF_{10}&=\eps^2 \, \top{10}\,,  &
\FF_{11}&=\eps^2 \, \top{11}\,,  &
\FF_{12}&=\eps^3 \, \top{12}\,,  \\
\FF_{13}&=\eps^2 \, \top{13}\,,  &
\FF_{14}&=\eps^2 \, \top{14}\,,  &
\FF_{15}&=\eps^3 \, \top{15}\,,  \\
\FF_{16}&=\eps^2 \, \top{16}\,,  &
\FF_{17}&=\eps^2 \, \top{17}\,,  &
\FF_{18}&=\eps^3 \, \top{18}\,,  \\
\FF_{19}&=\eps^3 \, \top{19}\,,  &
\FF_{20}&=\eps^2 \, \top{20}\,,  &
\FF_{21}&=\eps^3 \, \top{21}\,,  \\
\FF_{22}&=\eps^2 \, \top{22}\,,  &
\FF_{23}&=\eps^3 \, \top{23}\,,  &
\FF_{24}&=\eps^2 \, \top{24}\,,  \\
\FF_{25}&=\eps^3 \, \top{25}\,,  &
\FF_{26}&=(1-2\eps)\eps^3 \, \top{26}\,,  &
\FF_{27}&=\eps^3 \, \top{27}\,,  \\
\FF_{28}&=\eps^2 \, \top{28}\,,  &
\FF_{29}&=\eps^3 \, \top{29}\,,  &
\FF_{30}&=\eps^2 \, \top{30}\,,  \\
\FF_{31}&=(1-2\eps)\eps^3 \, \top{31}\,,  &
\FF_{32}&=\eps^3 \, \top{32}\,,  &
\FF_{33}&=\eps^4 \, \top{33}\,,  \\
\FF_{34}&=\eps^3 \, \top{34}\,,  &
\FF_{35}&=\eps^3 \, \top{35}\,,  &
\FF_{36}&=\eps^4 \, \top{36}\,,  \\
\FF_{37}&=\eps^4 \, \top{37}\,,  &
\FF_{38}&=\eps^3 \, \top{38}\,,  &
\FF_{39}&=\eps^4 \, \top{39}\,,  \\
\FF_{40}&=\eps^4 \, \top{40}\,,  &
\FF_{41}&=\eps^4 \, \top{41}\,,  &
\FF_{42}&=\eps^4 \, \left(\top{26} +  \top{42} \right) \,,  \\
\stepcounter{equation}\tag{\theequation}
\label{def:LBasisT4}
\end{align*}
where the $\mathcal{T}_i$ are depicted in
figure~\ref{fig:MIsT4}. 
Through the Magnus exponential, we identify the corresponding canonical basis:
\begingroup
\allowdisplaybreaks[1]
\begin{align*}
  \GG_{1}&=   \FF_{1}\,, & \qquad
  \GG_{2}&= -t  \,  \FF_{2}\,,\nn
    \GG_{3}&=  \lambda_t  \FF_{3}\,, & \qquad
  \GG_{4}&= -t  \,  \FF_{4}\,,\nn
    \GG_{5}&= \frac{1}{2}(\lambda_t-t)\, \FF_4 -\lambda_t \,  \FF_{5}\,, & \qquad
  \GG_{6}&= -s  \,  \FF_{6}\,,\nn
    \GG_{7}&=  2 m^2 \, \FF_6 + (m^2-s) \,  \FF_{7}\,, & \qquad
  \GG_{8}&= m^2  \,  \FF_{8}\,,\nn
    \GG_{9}&= m^2  \FF_{9}\,, & \qquad
  \GG_{10}&= -s  \,  \FF_{10}\,,\nn
    \GG_{11}&= - t \, \lambda_t \, \FF_{11}\,, & \qquad
  \GG_{12}&= -t  \,  \FF_{12}\,,\nn
    \GG_{13}&=  - t \, m^2 \, \FF_{13}\,, & \qquad
  \GG_{14}&= -m^2( \lambda_t-t) \left( \frac{3}{2} \FF_{12} +\FF_{13}
            \right) + \nn & & & \quad - m^2  \, \lambda_t \, \FF_{14}\,,\nn
    \GG_{15}&= \lambda_t \,  \FF_{15}\,, & \qquad
  \GG_{16}&= m^2 \, \lambda_t  \,  \FF_{16}\,,\nn
    \GG_{17}&= m^2 (t- \lambda_t ) \left( \frac{3}{2} \FF_{15} + \FF_{16} \right) - m^2 \, t \,  \FF_{17} \,, & \qquad
  \GG_{18}&=  \lambda_t  \,  \FF_{18}\,,\nn
    \GG_{19}&= (m^2-s) \,   \FF_{19}\,, & \qquad
  \GG_{20}&= m^2 \, (m^2-s)  \,  \FF_{20}\,,\nn
    \GG_{21}&=(m^2-s) \,  \FF_{21}\,, & \qquad
  \GG_{22}&= -\frac{3}{2}s \, \FF_{9}+ (s^2-m^4)   \,  \FF_{22}\,,\nn
    \GG_{23}&=  \lambda_t \, \FF_{23}\,, & \qquad
  \GG_{24}&= \frac{1}{4} \left(4m^2-t+\lambda_t  \right) \left( \FF_4
            +2 \FF_5 \right)+ 
\nn & & & \quad + m^2 (4m^2-t) \, \FF_{24}  \,,\nn
    \GG_{25}&=   \lambda_t \,  \FF_{25}\,, & \qquad
  \GG_{26}&= -t  \,  \FF_{26}\,,\nn
    \GG_{27}&= s \, t \,  \FF_{27}\,, & \qquad
  \GG_{28}&= - m^4 \, t \, \FF_{27} -  m^2 (m^2-s ) \, t   \,  \FF_{28}\,,\nn
    \GG_{29}&= -s \, \lambda_t \,   \FF_{29}\,, & \qquad
  \GG_{30}&= m^4 \lambda_t \, \FF_{29} + m^2 \, (m^2 -s ) \, \lambda_t   \,  \FF_{30}\,,\nn
    \GG_{31}&=-(m^2-s) \FF_{31} -(m^2-s) \,(4m^2-t+\lambda_t) \, \FF_{31}   \,, & \qquad
  \GG_{32}&=(m^2-s) \lambda_t \FF_{32} \,,\nn
    \GG_{33}&=   (m^2-s-t) \,  \FF_{33}\,, & \qquad
  \GG_{34}&= (m^2-s) \, \lambda_t  \,  \FF_{34}\,,\nn
    \GG_{35}&= 2 \frac{m^4 (m^2-s)}{2m^2-t -\lambda_t}\, \FF_{34} +m^2 \,(m^2-s)   \FF_{35}\,, & \qquad
  \GG_{36}&=  \lambda_t  \,  \FF_{36}\,,\nn
    \GG_{37}&= -t \, (4m^2-t)\,  \FF_{37}\,, & \qquad
  \GG_{38}&= -(m^2 -s) \, t  \,  \FF_{38}\,,\nn
    \GG_{39}&= -(m^2-s) \, t  \FF_{39}\,, & \qquad
  \GG_{40}&= -(m^2-s) \, t \, \lambda_t  \,  \FF_{40}\,,\nn
    \GG_{41}&= t \lambda_t \left( \FF_{40} - \FF_{41} \right) \,, &
    \end{align*}
    \vspace{-0.599cm}
\begin{align*}
  \GG_{42}&= (m^2-t +\lambda_t) \times  \\ & \qquad \left( \frac{2}{3}
                                             \FF_{3} +\frac{1}{4}
                                             \FF_{4}+\frac{1}{2}\FF_{5}-
                                             \frac{1}{2} \, t \,
                                             \FF_{11} +\frac{5}{2}
                                             \FF_{12}+\frac{5}{3}m^2
                                             \FF_{13}+\frac{5}{3}m^2
                                             \FF_{14} \right. \\
                        & \qquad \qquad \qquad \qquad \qquad \qquad \qquad \qquad \left. +2 \FF_{36}-\frac{1}{2}(m^2+s)\FF_{40} + t
                                             \, \FF_{41} \right) + \\
&+ m^2 \left( \frac{1}{3} \FF_{3} -\frac{1}{2} t \FF_{11} + \frac{1}{2} \FF_{12} + \frac{1}{3}m^2 \FF_{13} + \frac{1}{3}m^2 \FF_{14} + \frac{1}{2} \FF_{18} -\frac{1}{2}\FF_{40}  \right) +
\\ &    -t\,(m^2-s) \, \FF_{11} - 2 \frac{m^4} {2m^2-t-\lambda_t} \FF_{15} 
                                                                       + t \, \FF_{26}
     + \frac{m^2 (m^2-s) (t+\lambda_t)}{2m^2-t-\lambda_t} \left(
                                                                       \frac{2}{3}
                                                                       \FF_{29}
                                                                       \!-\!
                                                                       \FF_{34} \!
                                                                       \right) \!+\!
  \\ & -\frac{2}{3} \frac{ m^2 \, s \, (t-\lambda_t)}{2m^2-t-\lambda_t} \FF_{29} +2 t \, \FF_{33} 
       + \frac{4}{3}t m^4 \frac{m^2-s}{\lambda_t+t} \FF_{30}    -t  \,  \FF_{42}\,, \qquad 
\label{def:CanonicalBasisT4}
\stepcounter{equation}\tag{\theequation}
 \end{align*}%
 \endgroup
which satisfies a system of DEQs of the form in
eq.(\ref{eq:canonicalDEQ}), whose corresponding $42 \times 42$
matrices are collected in appendix~\ref{dlog2ndIF}. 
We observe that $\GG_{1,2,6,7,8,10,15,16,17,27,28}$ correspond, respectively, to $\GG_{1,3,5,6,4,2,13,14,15,23,24}$ of integral family~\eqref{eq:2Lfamily1} previously discussed.
The boundary constants of the remaining integrals can be fixed in the following way:

\begin{itemize}
\item The integration constants of $\GG_{3,4,5,11,\dots,14,18,23,24,26,29\dots,35}$ by demanding regularity in the limit $t \rightarrow 0$.
\item The boundary values of $\GG_{9}$ can be obtained by direct integration and it is given by 
  \begin{align}
   \GG_{9}(\eps)= &-\frac{\zeta_2}{2}\eps^2+\frac{1}{4}  \left(12\zeta_2\log (2)-7 \zeta_3\right)\eps^3 \nn
   &+ \left(-12 \text{Li}_4\left(\frac{1}{2}\right)+\frac{31 }{40}\zeta_4-\frac{\log ^4(2)}{2}-6\zeta_2 \log
   ^2(2)\right)\eps^4+\mathcal{O}\left(\eps^5\right)
   \, .
   \end{align}
  
   \item The boundary constants of $\GG_{19,21}$ can be fixed by demanding regularity when $s \rightarrow 0$.
   
    \item The boundary constants of $\GG_{20,22,25}$ are determined by
      demanding regularity, respectively, when $s\rightarrow-
      \frac{1}{2}\left(2m^2-t\!-\!\lambda_t \right)$, $s \rightarrow
      -m^2$, and $t \rightarrow 4m^2$.
  
 \item Finally, the boundary constants of $\GG_{36 \dots 42}$ can be all determined by demanding regularity in the simultaneous limits $t \rightarrow \frac{9}{2}m^2$ and $s \rightarrow -2m^2$.
 \end{itemize}
 All results have been numerically checked with the help of the computer codes
\texttt{GiNaC} and \texttt{SecDec}, and the analytic expressions
of the MIs are given in electronic form in the ancillary files
attached to the \texttt{arXiv} version of this manuscript.
\newpage

\section{Towards the non-planar integrals}
 The complete computation of the NNLO virtual QED corrections to
 $\mu e$ scattering requires the evaluation of one last missing 
four-point topology, which corresponds to the non-planar diagram $T_{6}$
 of figure \ref{fig:Feyndiag}. In view of future studies dedicated to this last
 class of integrals, we hereby show how the previously adopted
 strategy, based on differential equations, Magnus
 exponential and regularity conditions, can be efficiently applied to
 compute the MIs of a simpler vertex integral belonging to same family.

\subsection{Master integrals for the non-planar vertex}

\begin{figure}[H]
\centering
\includegraphics[scale=0.25]{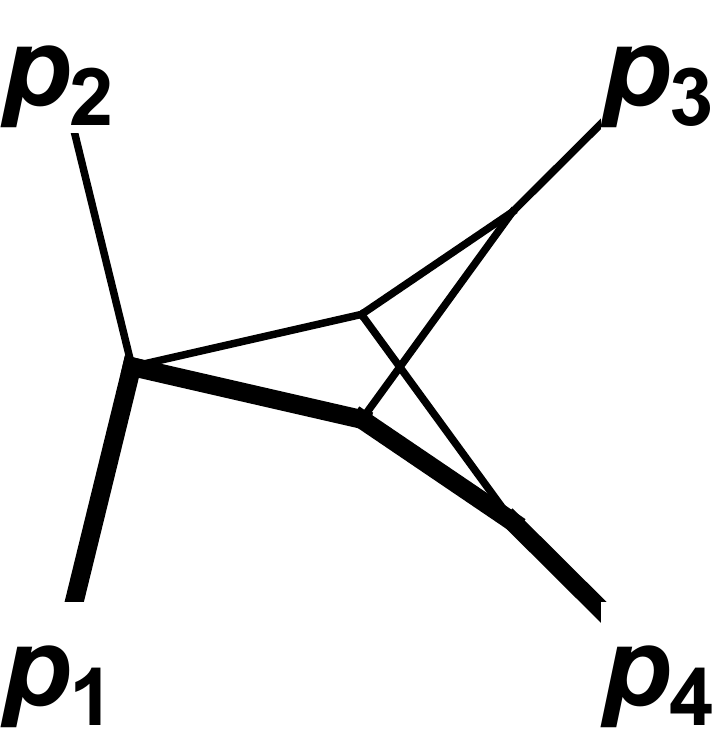}
\caption{Non-planar two-loop three-point topology}
\label{fig:NPfamily}
\end{figure}

We consider the non-planar vertex depicted in fig.~\ref{fig:NPfamily}, whose integral family is defined as
\begin{gather}
  \int \widetilde{\dd^d k_1}\widetilde{\dd^d k_2}\,
  \frac{\Den_{4}^{n_4}}{\Den_{1}^{n_1}\Den_{2}^{n_2}\Den_{3}^{n_3}\Den_{5}^{n_5}\Den_{6}^{n_6}\Den_{7}^{n_7}}\,,\quad n_i\geq0\,,
  \label{eq:familyNPtriangle}
\end{gather}
where the loop propagators are chosen to be
\begin{gather}
\Den_1 = k_1^2-m^2\,,\quad
\Den_2 = k_2^2-m^2\,,\quad
\Den_3 = (k_1+p_1+p_2)^2\,,\quad
\Den_4 = (k_2+p_1+p_2)^2\,, \nn
\Den_5 = (k_1-k_2+p_3)^2\,,\quad
\Den_6 = (k_2+p_4)^2\,,\quad
\Den_7 = (k_1-k_2)^2 \ . 
\end{gather}

\label{sec:Topo1}
The MIs belonging to this integral family, which will be part of the
full set of MIs needed for the computation of $T_6$, have been already
considered in the
literature~\cite{Bonciani:2008wf,Asatrian:2008uk,Beneke:2008ei,Bell:2008ws,Huber:2009se}. In
all previous computations, the determination of the boundary constants
resorted either to the fitting of numerical values to trascendental
constants~\cite{Bonciani:2008wf, Asatrian:2008uk, Beneke:2008ei} or to
Mellin-Barnes techniques~\cite{Huber:2009se}. With the present
calculation, we show that they can be fixed equivalently by imposing
the regularity of the solution at specific kinematic pseudo-thresholds
and by matching a particular linear combination of integrals
to their massless counterpart.
 
\begin{figure}[t]
  \centering
  \captionsetup[subfigure]{labelformat=empty}
  \subfloat[$\mathcal{T}_1$]{%
    \includegraphics[width=0.14\textwidth]{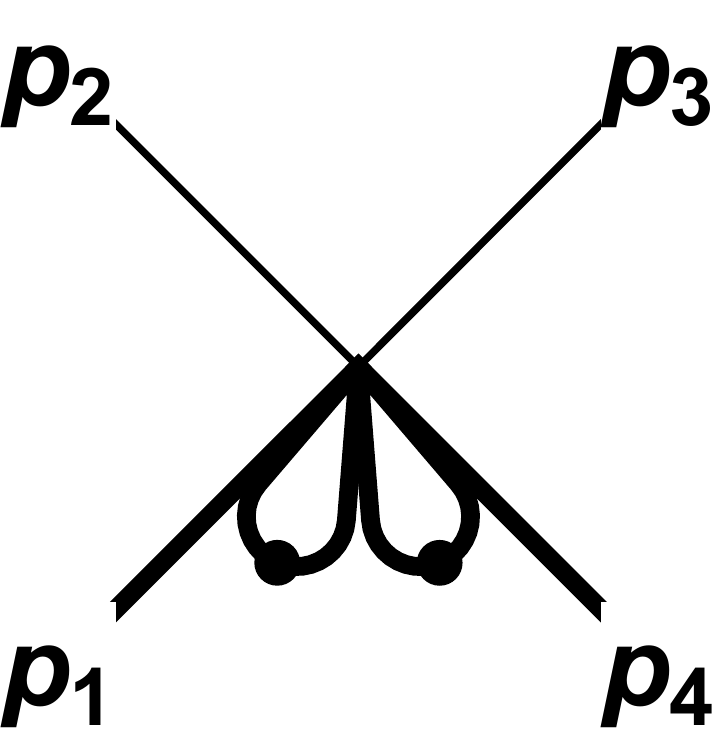}
  }
  \subfloat[$\mathcal{T}_2$]{%
    \includegraphics[width=0.14\textwidth]{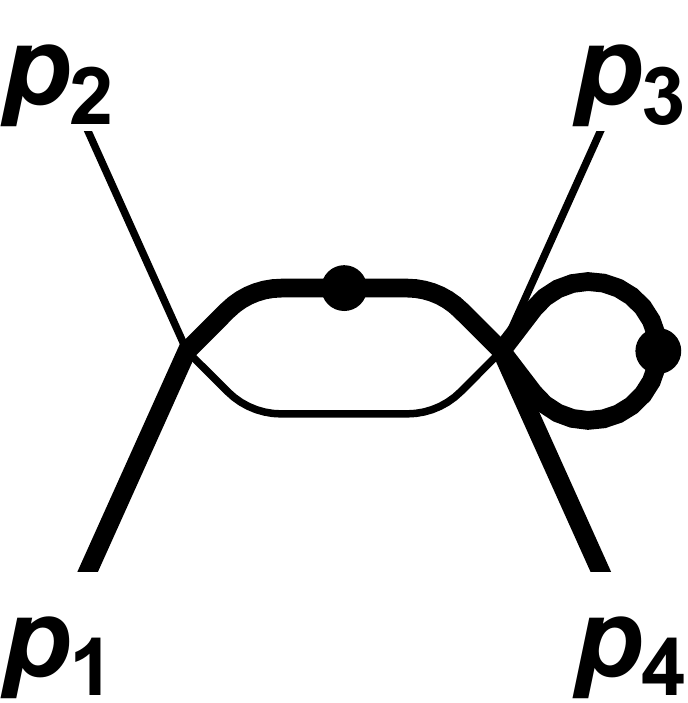}
  }
  \subfloat[$\mathcal{T}_3$]{%
    \includegraphics[width=0.14\textwidth]{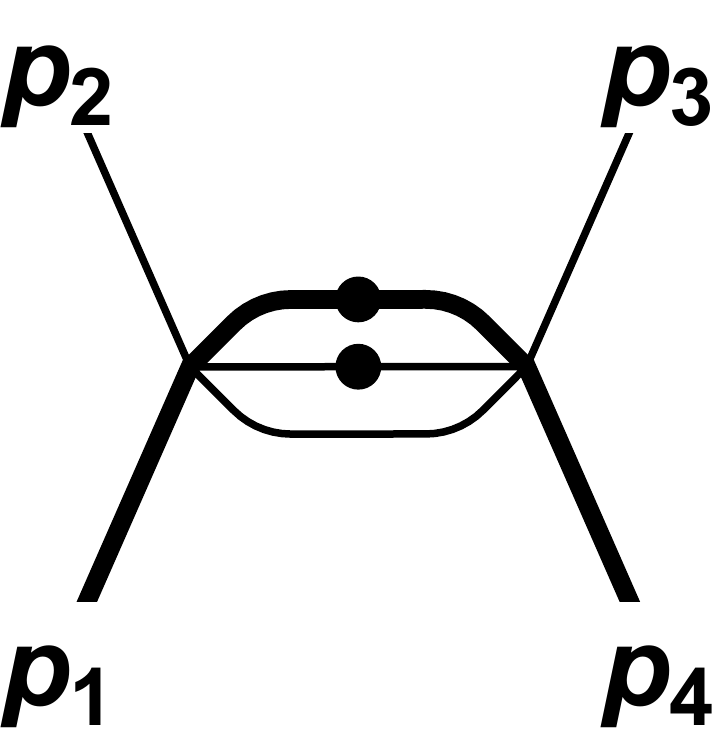}
  }
  \subfloat[$\mathcal{T}_4$]{%
    \includegraphics[width=0.14\textwidth]{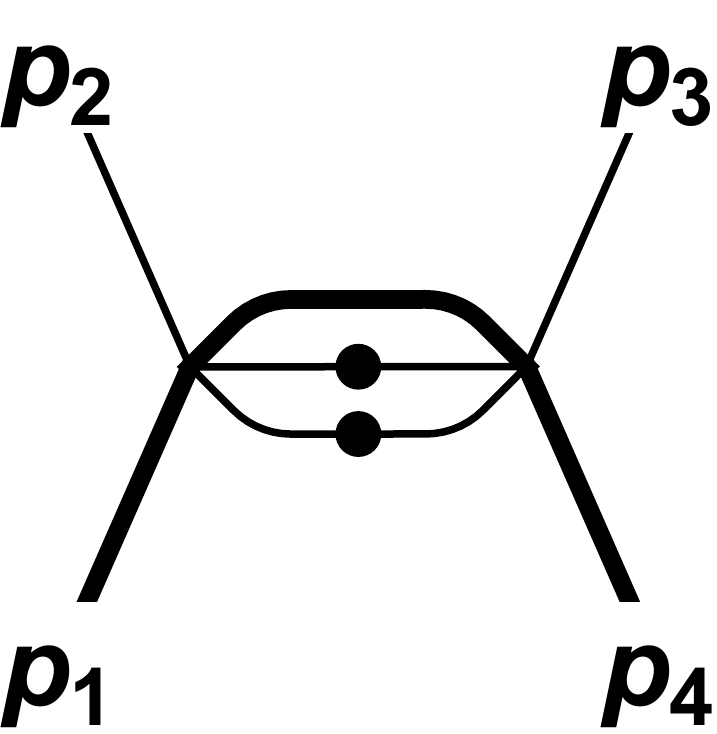}
  }
  \subfloat[$\mathcal{T}_5$]{%
    \includegraphics[width=0.14\textwidth]{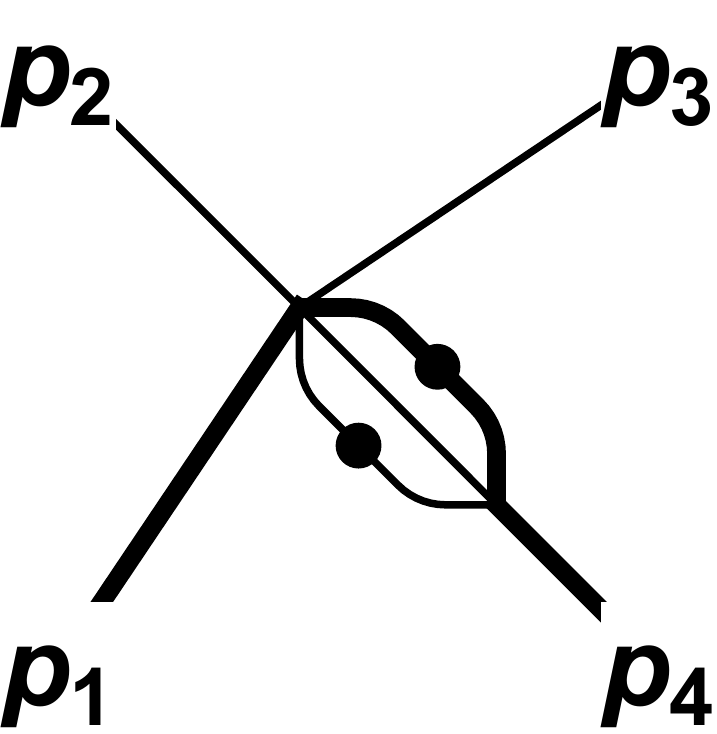}
  }
  \subfloat[$\mathcal{T}_6$]{%
    \includegraphics[width=0.14\textwidth]{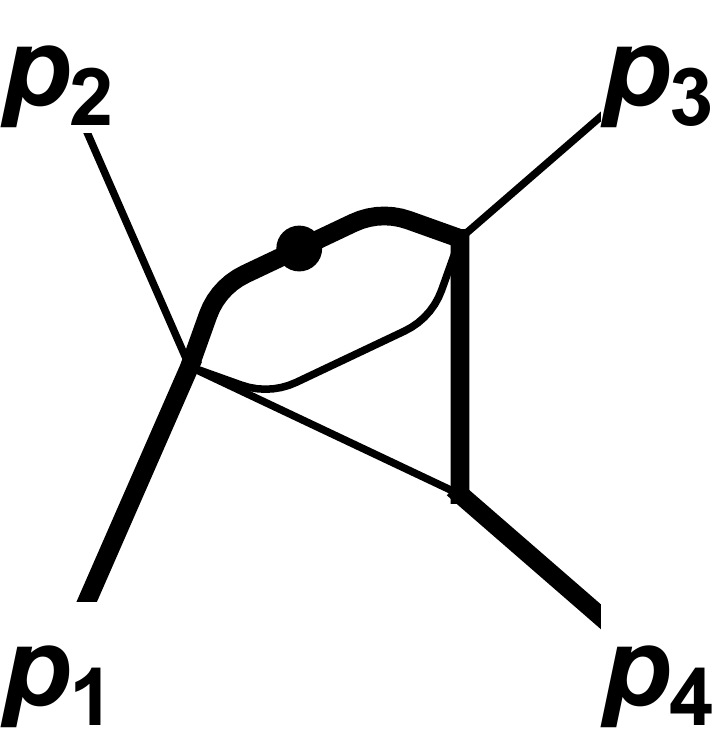}
  }
  \\
  \subfloat[$\mathcal{T}_7$]{%
    \includegraphics[width=0.14\textwidth]{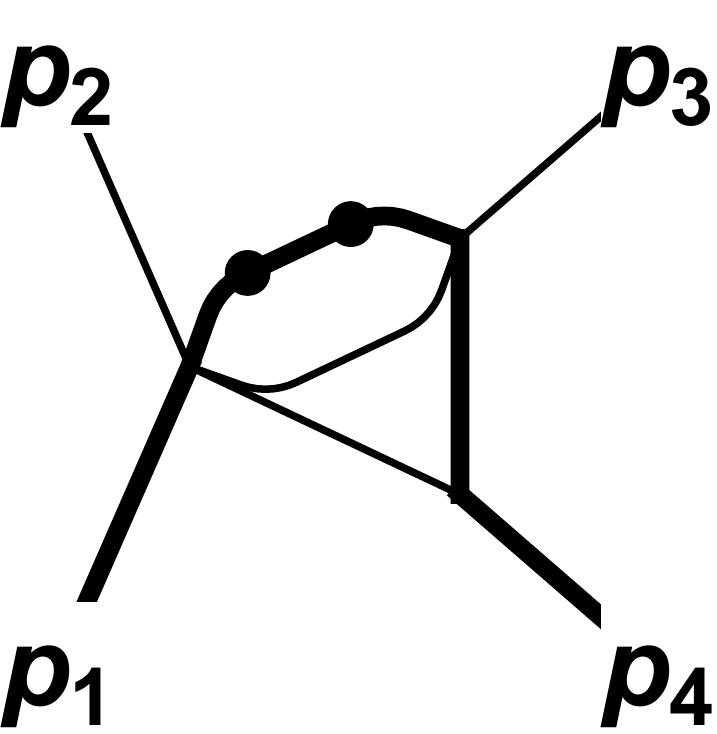}
  }
  \subfloat[$\mathcal{T}_8$]{%
    \includegraphics[width=0.14\textwidth]{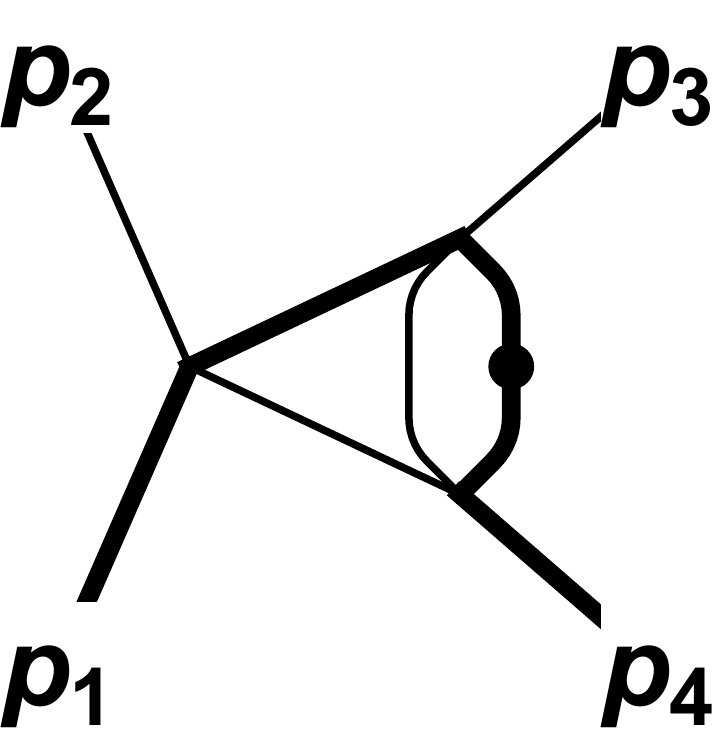}
  }
  \subfloat[$\mathcal{T}_9$]{%
    \includegraphics[width=0.14\textwidth]{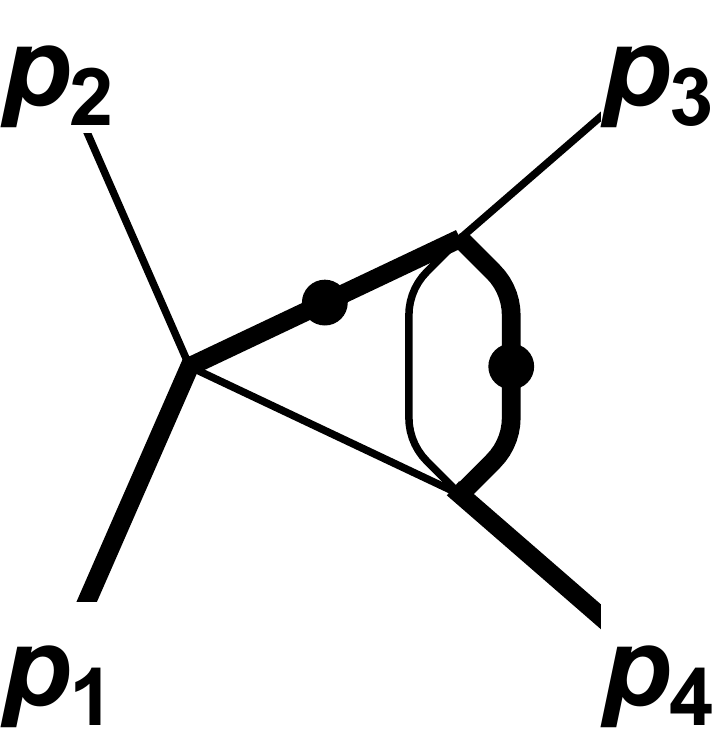}
  }
  \subfloat[$\mathcal{T}_{10}$]{%
    \includegraphics[width=0.14\textwidth]{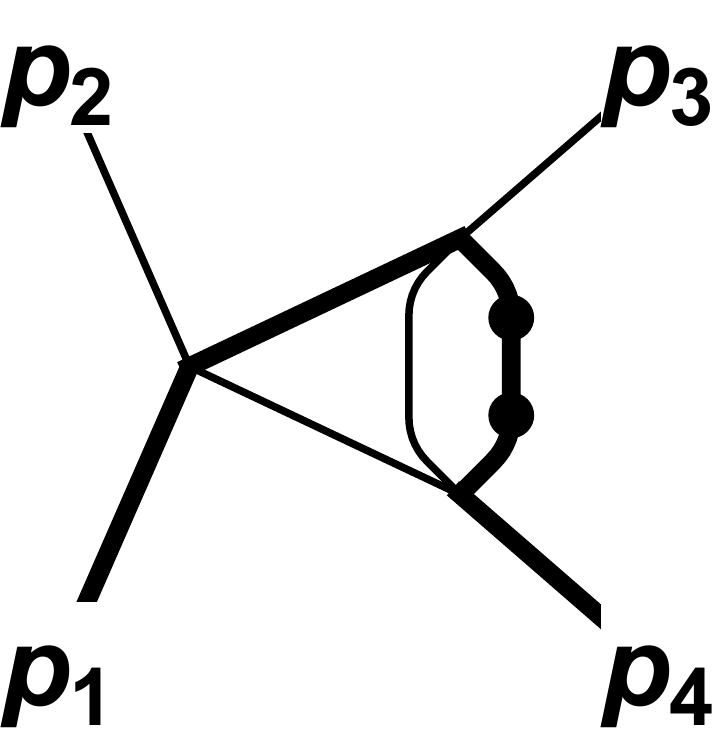}
  }
  \subfloat[$\mathcal{T}_{11}$]{%
    \includegraphics[width=0.14\textwidth]{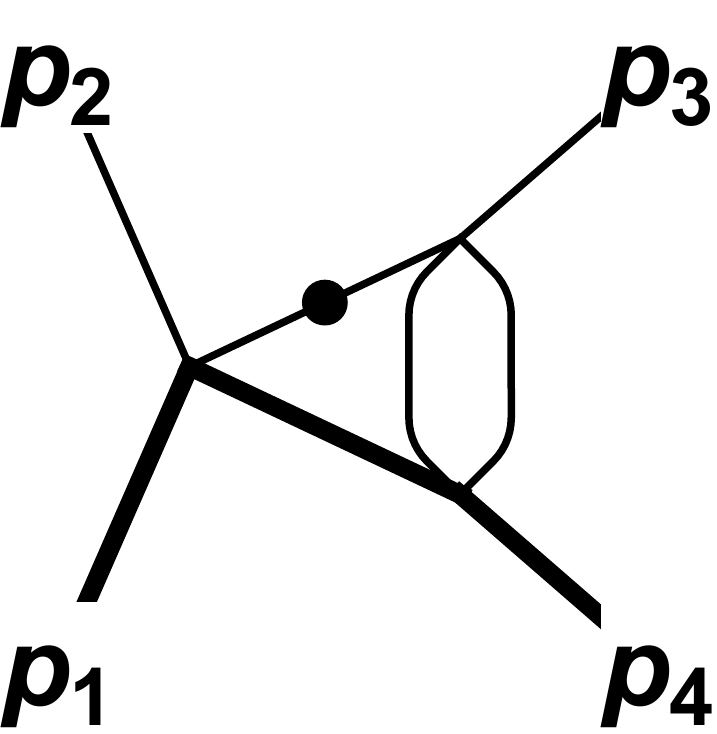}
  }
  \subfloat[$\mathcal{T}_{12}$]{%
    \includegraphics[width=0.14\textwidth]{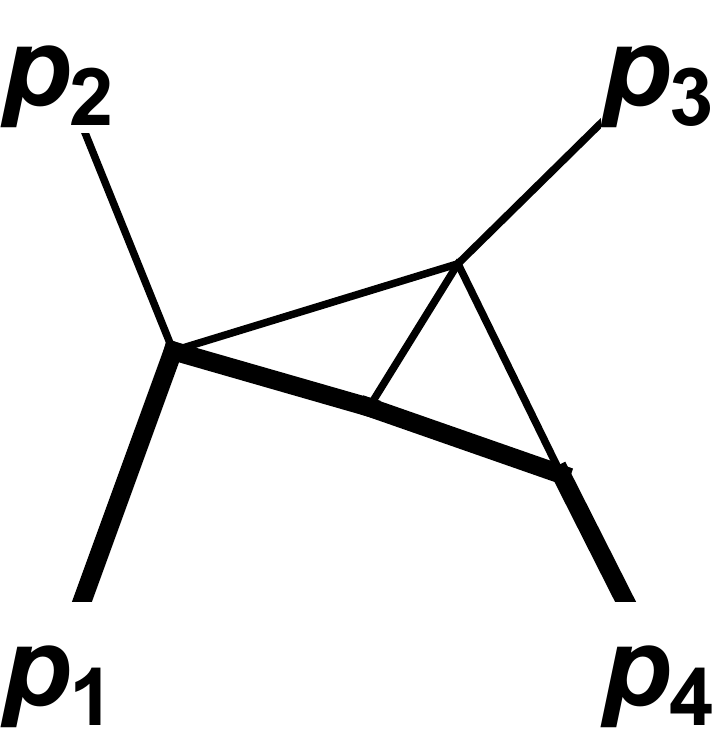}
  }
  \\
  \subfloat[$\mathcal{T}_{13}$]{%
    \includegraphics[width=0.14\textwidth]{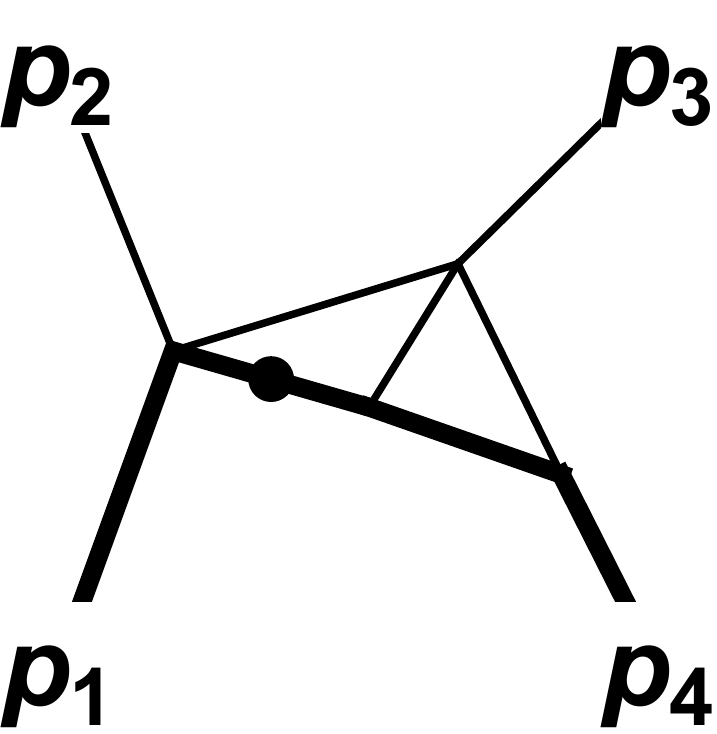}
  }
  \subfloat[$\mathcal{T}_{14}$]{%
    \includegraphics[width=0.14\textwidth]{figures/Topo6/T14-eps-converted-to.pdf}
  }
\caption{Two-loop MIs $\mathcal{T}_{1,\ldots,14}$ for the non-planar vertex.}
 \label{fig:MIsNPtriangle}
\end{figure}

In order to determine the MIs belonging to the integral
family~\eqref{eq:familyNPtriangle}, we derive their DEQ in the
dimensionless variable $x$.
We identify a set of 14 MIs which fulfills an $\eps$-linear system of DEQs,
\begin{align*}
\FF_{1}&=\eps^2 \, \top{1}\,,  &
\FF_{2}&=\eps^2 \, \top{2}\,,  &
\FF_{3}&=\eps^2 \, \top{3}\,,  \\
\FF_{4}&=\eps^2 \, \top{4}\,,  &
\FF_{5}&=\eps^2 \, \top{5}\,,  &
\FF_{6}&=\eps^3 \, \top{6}\,,  \\
\FF_{7}&=\eps^2 \, \top{7}\,,  &
\FF_{8}&=\eps^3 \, \top{8}\,,  &
\FF_{9}&=\eps^2 \, \top{9}\,,  \\
\FF_{10}&=\eps^2 \, \top{10}\,,  &
\FF_{11}&=\eps^2(2\eps-1) \, \top{11}\,,  &
\FF_{12}&=\eps^4 \, \top{12}\,,  \\
\FF_{13}&=\eps^3 \, \top{13}\,,  &
\FF_{14}&=\eps^4 \, \top{14}\,,  &
\stepcounter{equation}\tag{\theequation}
\label{def:LBasisT1}
\end{align*}
where the $\mathcal{T}_i$ are depicted in
figure~\ref{fig:MIsNPtriangle}. 
By making  use of the  Magnus exponential, we can transform these MIs into the canonical basis
\begin{align*}
  \begin{alignedat}{2}
  \GG_{1}&=   \FF_1\,, & \qquad
  \GG_{2}&= -s  \,  \FF_2\,,\nn
   \GG_{3}&= -s  \FF_3\,, & \qquad
  \GG_{4}&=2m^2\,\FF_3+(m^2-s) \,  \FF_4\,, \nn
  \GG_{5}&= m^2 \FF_5 \,, &\qquad
  \GG_{6}&= (m^2-s)\, \FF_6\,, \nn
  \GG_{7}&= m^2(m^2-s)\, \FF_7\,,  &\qquad
  \GG_{8}&=(m^2-s)\,\FF_8\,,  \nn
   \GG_{9}&= m^2\left(3\,\FF_8+(2m^2-s)\,\FF_9+2m^2\,\FF_{10}\right)\,, & \qquad
  \GG_{10}&=m^2(m^2-s)\FF_{10}\,,  \nn
  \GG_{11}&= \frac{1}{(s+m^2)}\left(-2m^4\,\FF_{5}+(s-m^2)s\,\FF_{11}\right)\,,& \qquad
  \GG_{12}&= (m^2-s)\, \FF_{12}\,,   \nn  
  \GG_{13}&= m^2(m^2-s)\, \FF_{13}\,,  & \qquad  
  \GG_{14}&=(m^2-s)^2 \, \FF_{14}\,,\nn
\label{def:CanonicalBasisNPtriangle}
   \end{alignedat}\stepcounter{equation}\tag{\theequation}
 \end{align*}
which satisfies a system of DEQ of the form,
\begin{equation}
d \GGvec = \eps \left(\MM_{1}  \dlog(x)+\MM_{2}  \dlog(1+x)+\MM_{3}  \dlog(2+x)\right) \GGvec \, ,
\label{eq:canonicalDEQx}
\end{equation}
where $\MM_i$ are the constant matrices 
\begin{align}
\MM_{1} = \scalemath{0.48}{
 \left(
\begin{array}{cccccccccccccc}
 0 &\pminus 0 &\pminus 0 &\pminus 0 &\pminus 0 &\pminus 0 &\pminus 0 &\pminus 0 &\pminus 0 &\pminus 0 &\pminus 0 &\pminus 0 &\pminus 0 &\pminus 0 \\
 0 &\pminus 1 &\pminus 0 &\pminus 0 &\pminus 0 &\pminus 0 &\pminus 0 &\pminus 0 &\pminus 0 &\pminus 0 &\pminus 0 &\pminus 0 &\pminus 0 &\pminus 0 \\
 0 &\pminus 0 &\pminus 1 &\pminus 0 &\pminus 0 &\pminus 0 &\pminus 0 &\pminus 0 &\pminus 0 &\pminus 0 &\pminus 0 &\pminus 0 &\pminus 0 &\pminus 0 \\
 0 &\pminus 0 &\pminus 4 &\pminus 0 &\pminus 0 &\pminus 0 &\pminus 0 &\pminus 0 &\pminus 0 &\pminus 0 &\pminus 0 &\pminus 0 &\pminus 0 &\pminus 0 \\
 0 &\pminus 0 &\pminus 0 &\pminus 0 &\pminus 0 &\pminus 0 &\pminus 0 &\pminus 0 &\pminus 0 &\pminus 0 &\pminus 0 &\pminus 0 &\pminus 0 &\pminus 0 \\
 \frac{1}{2} &\pminus 0 &\pminus 1 &\pminus \frac{1}{2} &\pminus 0 &\pminus 3 &\pminus 2 &\pminus 0 &\pminus 0 &\pminus 0 &\pminus 0 &\pminus 0 &\pminus 0 &\pminus 0 \\
 \frac{1}{2} &\pminus 0 &\pminus \frac{1}{2} &\pminus \frac{1}{2} &\pminus 0 &\pminus 3 &\pminus 2 &\pminus 0 &\pminus 0 &\pminus 0 &\pminus 0 &\pminus 0 &\pminus 0 &\pminus 0 \\
 \frac{1}{2} &\pminus 0 &\pminus 0 &\pminus 0 &\pminus 2 &\pminus 0 &\pminus 0 &\pminus 1 &\pminus 1 &\pminus 0 &\pminus 0 &\pminus 0 &\pminus 0 &\pminus 0 \\
 \frac{3}{2} &\pminus 1 &\pminus 0 &\pminus 0 &\pminus 6 &\pminus 0 &\pminus 0 &\pminus 3 &\pminus 3 &\pminus 0 &\pminus 0 &\pminus 0 &\pminus 0 &\pminus 0 \\
 0 &\pminus \frac{1}{2} &\pminus 0 &\pminus 0 &\pminus 0 &\pminus 0 &\pminus 0 &\pminus 0 &\pminus 0 &\pminus 0 &\pminus 0 &\pminus 0 &\pminus 0 &\pminus 0 \\
 0 &\pminus 0 &\pminus 0 &\pminus 0 &\pminus 4 &\pminus 0 &\pminus 0 &\pminus 0 &\pminus 0 &\pminus 0 &\pminus 2 &\pminus 0 &\pminus 0 &\pminus 0 \\
 \frac{1}{4} &\pminus 0 &\pminus 0 &\pminus \frac{1}{4} &\pminus 1 &\pminus 0 &\pminus 0 &\pminus 0 &\pminus 0 &\pminus 0 &\pminus \frac{1}{2} &\pminus 3 &\pminus 2 &\pminus 0 \\
 \frac{1}{2} &\pminus \frac{1}{2} &\pminus \frac{1}{2} &\pminus \frac{1}{2} &\pminus 0 &\pminus 0 &\pminus 0 &\pminus 0 &\pminus 0 &\pminus 0 &\pminus 0 &\pminus 6 &\pminus 4 &\pminus 0 \\
 0 &\pminus 0 &\pminus 0 &\pminus 0 &\pminus 0 &\pminus 0 &\pminus 0 &\pminus 0 &\pminus 0 &\pminus 0 &\pminus 0 &\pminus 0 &\pminus 0 &\pminus 0 \\
\end{array}
\right)
 }\,,\quad
 \MM_{2} = \scalemath{0.48}{
  \left(
\begin{array}{cccccccccccccc}
 0 &\pminus 0 &\pminus 0 &\pminus 0 &\pminus 0 &\pminus 0 &\pminus 0 &\pminus 0 &\pminus 0 &\pminus 0 &\pminus 0 &\pminus 0 &\pminus 0 &\pminus 0 \\
 1 &\pminus 2 &\pminus 0 &\pminus 0 &\pminus 0 &\pminus 0 &\pminus 0 &\pminus 0 &\pminus 0 &\pminus 0 &\pminus 0 &\pminus 0 &\pminus 0 &\pminus 0 \\
 0 &\pminus 0 &\pminus 2 &\pminus 1 &\pminus 0 &\pminus 0 &\pminus 0 &\pminus 0 &\pminus 0 &\pminus 0 &\pminus 0 &\pminus 0 &\pminus 0 &\pminus 0 \\
 0 &\pminus 0 &\pminus 4 &\pminus 2 &\pminus 0 &\pminus 0 &\pminus 0 &\pminus 0 &\pminus 0 &\pminus 0 &\pminus 0 &\pminus 0 &\pminus 0 &\pminus 0 \\
 0 &\pminus 0 &\pminus 0 &\pminus 0 &\pminus 0 &\pminus 0 &\pminus 0 &\pminus 0 &\pminus 0 &\pminus 0 &\pminus 0 &\pminus 0 &\pminus 0 &\pminus 0 \\
 0 &\pminus 0 &\pminus 0 &\pminus 0 &\pminus 0 &\pminus 2 &\pminus 0 &\pminus 0 &\pminus 0 &\pminus 0 &\pminus 0 &\pminus 0 &\pminus 0 &\pminus 0 \\
 0 &\pminus 0 &\pminus 0 &\pminus 0 &\pminus 0 &\pminus 3 &\pminus 2 &\pminus 0 &\pminus 0 &\pminus 0 &\pminus 0 &\pminus 0 &\pminus 0 &\pminus 0 \\
 0 &\pminus 0 &\pminus 0 &\pminus 0 &\pminus 0 &\pminus 0 &\pminus 0 &\pminus 2 &\pminus 0 &\pminus 0 &\pminus 0 &\pminus 0 &\pminus 0 &\pminus 0 \\
 \frac{3}{2} &\pminus 1 &\pminus 0 &\pminus 0 &\pminus 6 &\pminus 0 &\pminus 0 &\pminus 6 &\pminus 4 &\pminus 2 &\pminus 0 &\pminus 0 &\pminus 0 &\pminus 0 \\
 0 &\pminus 0 &\pminus 0 &\pminus 0 &\pminus 0 &\pminus 0 &\pminus 0 &\pminus 3 &\pminus 0 &\pminus 2 &\pminus 0 &\pminus 0 &\pminus 0 &\pminus 0 \\
 0 &\pminus 0 &\pminus 0 &\pminus 0 &\pminus 4 &\pminus 0 &\pminus 0 &\pminus 0 &\pminus 0 &\pminus 0 &\pminus 4 &\pminus 0 &\pminus 0 &\pminus 0 \\
 0 &\pminus 0 &\pminus 0 &\pminus 0 &\pminus 0 &\pminus 0 &\pminus 0 &\pminus 0 &\pminus 0 &\pminus 0 &\pminus 0 &\pminus 2 &\pminus 0 &\pminus 0 \\
 0 &\pminus 0 &\pminus 0 &\pminus 0 &\pminus 0 &\pminus 0 &\pminus 0 &\pminus 0 &\pminus 0 &\pminus 0 &\pminus 0 &\pminus 6 &\pminus 4 &\pminus 0 \\
 1 &\pminus 1 &\pminus \frac{4}{3} &\pminus \frac{2}{3} &\pminus 4 &\pminus 3 &\pminus 4 &\pminus 0 &\pminus 2 &\pminus 2 &\pminus 1 &\pminus 2 &\pminus 2 &\pminus 2 \\
\end{array}
\right)
  }
  \,,\quad
  \MM_{3} = \scalemath{0.48}{
  \left(
\begin{array}{cccccccccccccc}
 0 &\pminus 0 &\pminus 0 &\pminus 0 &\pminus 0 &\pminus 0 &\pminus 0 &\pminus 0 &\pminus 0 &\pminus 0 &\pminus 0 &\pminus 0 &\pminus 0 &\pminus 0 \\
 0 &\pminus 0 &\pminus 0 &\pminus 0 &\pminus 0 &\pminus 0 &\pminus 0 &\pminus 0 &\pminus 0 &\pminus 0 &\pminus 0 &\pminus 0 &\pminus 0 &\pminus 0 \\
 0 &\pminus 0 &\pminus 0 &\pminus 0 &\pminus 0 &\pminus 0 &\pminus 0 &\pminus 0 &\pminus 0 &\pminus 0 &\pminus 0 &\pminus 0 &\pminus 0 &\pminus 0 \\
 0 &\pminus 0 &\pminus 0 &\pminus 0 &\pminus 0 &\pminus 0 &\pminus 0 &\pminus 0 &\pminus 0 &\pminus 0 &\pminus 0 &\pminus 0 &\pminus 0 &\pminus 0 \\
 0 &\pminus 0 &\pminus 0 &\pminus 0 &\pminus 0 &\pminus 0 &\pminus 0 &\pminus 0 &\pminus 0 &\pminus 0 &\pminus 0 &\pminus 0 &\pminus 0 &\pminus 0 \\
 0 &\pminus 0 &\pminus 0 &\pminus 0 &\pminus 0 &\pminus 0 &\pminus 0 &\pminus 0 &\pminus 0 &\pminus 0 &\pminus 0 &\pminus 0 &\pminus 0 &\pminus 0 \\
 0 &\pminus 0 &\pminus 0 &\pminus 0 &\pminus 0 &\pminus 0 &\pminus 0 &\pminus 0 &\pminus 0 &\pminus 0 &\pminus 0 &\pminus 0 &\pminus 0 &\pminus 0 \\
 0 &\pminus 0 &\pminus 0 &\pminus 0 &\pminus 0 &\pminus 0 &\pminus 0 &\pminus 3 &\pminus 1 &\pminus 2 &\pminus 0 &\pminus 0 &\pminus 0 &\pminus 0 \\
 0 &\pminus 0 &\pminus 0 &\pminus 0 &\pminus 0 &\pminus 0 &\pminus 0 &\pminus 3 &\pminus 1 &\pminus 2 &\pminus 0 &\pminus 0 &\pminus 0 &\pminus 0 \\
 0 &\pminus 0 &\pminus 0 &\pminus 0 &\pminus 0 &\pminus 0 &\pminus 0 &\pminus 3 &\pminus 1 &\pminus 2 &\pminus 0 &\pminus 0 &\pminus 0 &\pminus 0 \\
 0 &\pminus 0 &\pminus 0 &\pminus 0 &\pminus 0 &\pminus 0 &\pminus 0 &\pminus 0 &\pminus 0 &\pminus 0 &\pminus 0 &\pminus 0 &\pminus 0 &\pminus 0 \\
 0 &\pminus 0 &\pminus 0 &\pminus 0 &\pminus 0 &\pminus 0 &\pminus 0 &\pminus 0 &\pminus 0 &\pminus 0 &\pminus 0 &\pminus 0 &\pminus 0 &\pminus 0 \\
 0 &\pminus 0 &\pminus 0 &\pminus 0 &\pminus 0 &\pminus 0 &\pminus 0 &\pminus 0 &\pminus 0 &\pminus 0 &\pminus 0 &\pminus 0 &\pminus 0 &\pminus 0 \\
 0 &\pminus 0 &\pminus 0 &\pminus 0 &\pminus 0 &\pminus 0 &\pminus 0 &\pminus 0 &\pminus 0 &\pminus 0 &\pminus 0 &\pminus 0 &\pminus 0 &\pminus 0 \\
\end{array}
\right)}\,.\nn
  \end{align}
The 3 letters are real and positive in the range $x>0$, which
corresponds to the Euclidean region $s<0$. The general solution of the system can be written in terms of one-dimensional GPLs. In order to completely determine the solution of the DEQ, we fix the boundary constants as follows:
\begin{itemize}
\item  $\GG_{1,2,3,4,5}$ correspond, respectively, to
  $\GG_{1,2,5,6,4}$ of the first integral family, defined in eq.~\eqref{eq:2Lfamily1}.
\item  $\GG_{6,7}$ correspond, respectively, to $\GG_{19,20}$ of the
  second integral family, defined in eq.~\eqref{eq:2Lfamily2}.
\item  The regularity at $s\to 0$ of $\GG_{8,9,10}$ can be used to fix
  the boundary constants of one single master integrals, which we
  choose to be $\GG_{9}$. The boundary values $\GG_{8}(\eps, 0)$ and
  $\GG_{10}(\eps, 0)$ can be obtained in the limit $p_4^2\to m^2$ of
  similar vertex integrals with off-shell momentum $p_4^2$ and $s\equiv (p_1+p_2)^2=p_3^2=0$,
   \begin{align}
 \GG_{8}(\eps, 0)=\eps^3m^2\lim_{p_4^2\to m^2} \raisebox{-38pt}{\includegraphics[scale=0.22]{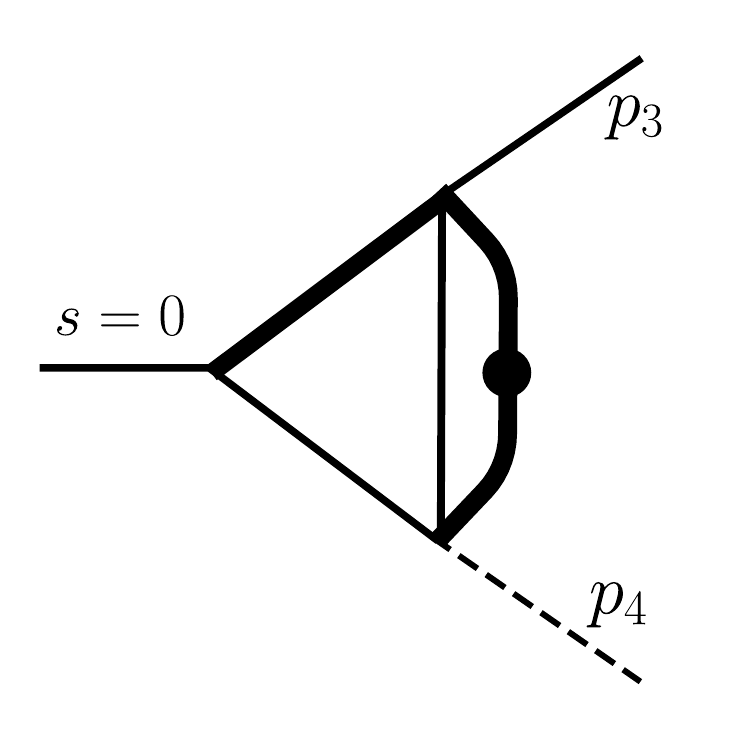}}\,,\quad \GG_{10}(\eps, 0)=\eps^2m^4\lim_{p_4^2\to m^2} \raisebox{-38pt}{\includegraphics[scale=0.22]{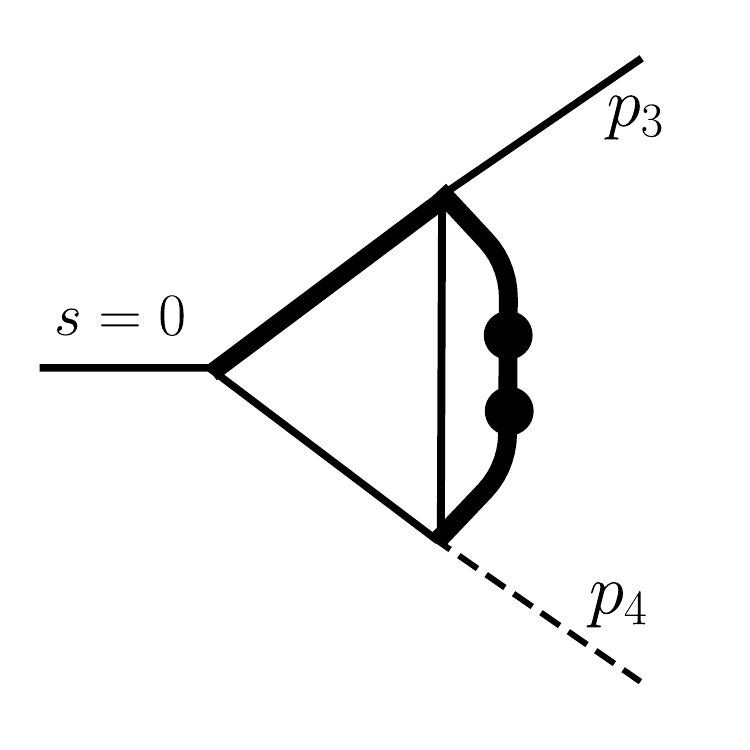}}\,.
 \label{eq:limit_i810}
 \end{align}
The computation of the auxiliary vertex integrals, which is discussed in appendix~\ref{sec:AuxVert}, leads to 
 \begin{align}
  \GG_{8}(\eps, 0)=&\left(\frac{5 \zeta_3}{4}-3\zeta_2\log (2)\right)\eps^3+\left(8 \text{Li}_4\left(\frac{1}{2}\right)-\frac{33}{8}\zeta_4+\frac{\log ^4(2)}{3}-2\zeta_2 \log ^2(2)\right)\eps^4 \,,\nn
  \GG_{10}(\eps, 0)=&\frac{\pi ^2}{12}\eps^2+
   \left(\frac{\zeta_3}{4}+3\zeta_2 \log (2)\right)\eps^3\nn
   &+\left(-8 \text{Li}_4\left(\frac{1}{2}\right)+\frac{65}{4}\zeta_4-\frac{\log ^4(2)}{3}+2\zeta_2 \log ^2(2)\right)\eps^4 +\mathcal{O}\left(\eps^5\right)\,.
  \label{eq:bc_i810}
  \end{align}
\item  The boundary constant of $\GG_{11}$ is determined by imposing regularity when $s\to 0$.
\item  $\GG_{12}$ corresponds to $\GG_{17}$ of the integral family~\eqref{eq:2Lfamily1} and the boundary constant of $\GG_{13}$ can be  fixed by demanding regularity when $s\to 0$.

\item The boundary condition for $\GG_{14}$ is determined from the
  $m \to 0$, or equivalently $s \to \infty$ behaviour of the
  solutions, where all the internal lines of the diagrams become {\it
    massless}. In this regime, we search for a combination of
  integrals behaving as,
\begin{gather}
\lim_{z \rightarrow 0} \sum_i c_i \GG_i = z^{a \eps}F(\eps) \ , \quad
a \in \mathbb{Z} \ , 
\end{gather} 
where $F(\eps)$ is finite as $z \rightarrow$ 0.

Following the ideas outlined in \cite{Dulat:2014mda},  we begin by performing a change of variables $z=1/x=(-m^2/s)$,
  yielding a total differential equation of the form,
\begin{equation}
d \GGvec = \eps \Big(\MM_{1}  \dlog(z)+\MM_{2}  \dlog(1+z)+\MM_{3}  \dlog(1+2z)\Big) \GGvec \, ,
\label{eq:canonicalDEQx}
\end{equation}
where $\MM_i$ are the constant matrices 
\begin{align}
\MM_{1} = \scalemath{0.41}{
 \left(
\begin{array}{cccccccccccccc}
 \pminus 0 & \pminus 0 & \pminus 0 & \pminus 0 & \pminus 0 & \pminus \
0 & \pminus 0 & \pminus 0 & \pminus 0 & \pminus 0 & \pminus 0 & \
\pminus 0 & \pminus 0 & \pminus 0 \\
 \pminus 1 & \pminus 1 & \pminus 0 & \pminus 0 & \pminus 0 & \pminus \
0 & \pminus 0 & \pminus 0 & \pminus 0 & \pminus 0 & \pminus 0 & \
\pminus 0 & \pminus 0 & \pminus 0 \\
 \pminus 0 & \pminus 0 & \pminus 1 & \pminus 1 & \pminus 0 & \pminus \
0 & \pminus 0 & \pminus 0 & \pminus 0 & \pminus 0 & \pminus 0 & \
\pminus 0 & \pminus 0 & \pminus 0 \\
 \pminus 0 & \pminus 0 & \pminus 0 & \pminus 2 & \pminus 0 & \pminus \
0 & \pminus 0 & \pminus 0 & \pminus 0 & \pminus 0 & \pminus 0 & \
\pminus 0 & \pminus 0 & \pminus 0 \\
 \pminus 0 & \pminus 0 & \pminus 0 & \pminus 0 & \pminus 0 & \pminus \
0 & \pminus 0 & \pminus 0 & \pminus 0 & \pminus 0 & \pminus 0 & \
\pminus 0 & \pminus 0 & \pminus 0 \\
 \pminus \frac{1}{2} & \pminus 0 & \minus  1 & \pminus \frac{1}{2} & \
\pminus 0 & \pminus 1 & \pminus 2 & \pminus 0 & \pminus 0 & \pminus 0 \
& \pminus 0 & \pminus 0 & \pminus 0 & \pminus 0 \\
 \minus  \frac{1}{2} & \pminus 0 & \pminus \frac{1}{2} & \minus  \
\frac{1}{2} & \pminus 0 & \pminus 0 & \pminus 0 & \pminus 0 & \pminus \
0 & \pminus 0 & \pminus 0 & \pminus 0 & \pminus 0 & \pminus 0 \\
 \pminus \frac{1}{2} & \pminus 0 & \pminus 0 & \pminus 0 & \pminus 2 \
& \pminus 0 & \pminus 0 & \pminus 2 & \pminus 2 & \pminus 2 & \pminus \
0 & \pminus 0 & \pminus 0 & \pminus 0 \\
 \pminus 0 & \pminus 0 & \pminus 0 & \pminus 0 & \pminus 0 & \pminus \
0 & \pminus 0 & \pminus 0 & \pminus 0 & \pminus 0 & \pminus 0 & \
\pminus 0 & \pminus 0 & \pminus 0 \\
 \pminus 0 & \pminus \frac{1}{2} & \pminus 0 & \pminus 0 & \pminus 0 \
& \pminus 0 & \pminus 0 & \pminus 0 & \minus  1 & \pminus 0 & \pminus \
0 & \pminus 0 & \pminus 0 & \pminus 0 \\
 \pminus 0 & \pminus 0 & \pminus 0 & \pminus 0 & \pminus 0 & \pminus \
0 & \pminus 0 & \pminus 0 & \pminus 0 & \pminus 0 & \pminus 2 & \
\pminus 0 & \pminus 0 & \pminus 0 \\
 \minus  \frac{1}{4} & \pminus 0 & \pminus 0 & \minus  \frac{1}{4} & \
\minus  1 & \pminus 0 & \pminus 0 & \pminus 0 & \pminus 0 & \pminus 0 \
& \minus  \frac{1}{2} & \pminus 1 & \pminus 2 & \pminus 0 \\
 \pminus \frac{1}{2} & \pminus \frac{1}{2} & \minus  \frac{1}{2} & \
\pminus \frac{1}{2} & \pminus 0 & \pminus 0 & \pminus 0 & \pminus 0 & \
\pminus 0 & \pminus 0 & \pminus 0 & \pminus 0 & \pminus 0 & \pminus 0 \
\\
 \minus  1 & \minus  1 & \minus  \frac{4}{3} & \minus  \frac{2}{3} & \
\minus  4 & \minus  3 & \minus  4 & \pminus 0 & \minus  2 & \pminus 2 \
& \minus  1 & \minus  2 & \minus  2 & \pminus 2 \\
\end{array}
\right)
 }\,,\quad
 \MM_{2} = \scalemath{0.41}{
  \left(
\begin{array}{cccccccccccccc}
 \pminus 0 & \pminus 0 & \pminus 0 & \pminus 0 & \pminus 0 & \pminus \
0 & \pminus 0 & \pminus 0 & \pminus 0 & \pminus 0 & \pminus 0 & \
\pminus 0 & \pminus 0 & \pminus 0 \\
 \minus  1 & \minus  2 & \pminus 0 & \pminus 0 & \pminus 0 & \pminus \
0 & \pminus 0 & \pminus 0 & \pminus 0 & \pminus 0 & \pminus 0 & \
\pminus 0 & \pminus 0 & \pminus 0 \\
 \pminus 0 & \pminus 0 & \minus  2 & \minus  1 & \pminus 0 & \pminus \
0 & \pminus 0 & \pminus 0 & \pminus 0 & \pminus 0 & \pminus 0 & \
\pminus 0 & \pminus 0 & \pminus 0 \\
 \pminus 0 & \pminus 0 & \minus  4 & \minus  2 & \pminus 0 & \pminus \
0 & \pminus 0 & \pminus 0 & \pminus 0 & \pminus 0 & \pminus 0 & \
\pminus 0 & \pminus 0 & \pminus 0 \\
 \pminus 0 & \pminus 0 & \pminus 0 & \pminus 0 & \pminus 0 & \pminus \
0 & \pminus 0 & \pminus 0 & \pminus 0 & \pminus 0 & \pminus 0 & \
\pminus 0 & \pminus 0 & \pminus 0 \\
 \pminus 0 & \pminus 0 & \pminus 0 & \pminus 0 & \pminus 0 & \pminus \
2 & \pminus 0 & \pminus 0 & \pminus 0 & \pminus 0 & \pminus 0 & \
\pminus 0 & \pminus 0 & \pminus 0 \\
 \pminus 0 & \pminus 0 & \pminus 0 & \pminus 0 & \pminus 0 & \minus  \
3 & \minus  2 & \pminus 0 & \pminus 0 & \pminus 0 & \pminus 0 & \
\pminus 0 & \pminus 0 & \pminus 0 \\
 \pminus 0 & \pminus 0 & \pminus 0 & \pminus 0 & \pminus 0 & \pminus \
0 & \pminus 0 & \pminus 2 & \pminus 0 & \pminus 0 & \pminus 0 & \
\pminus 0 & \pminus 0 & \pminus 0 \\
 \minus  \frac{3}{2} & \minus  1 & \pminus 0 & \pminus 0 & \minus  6 \
& \pminus 0 & \pminus 0 & \minus  6 & \minus  4 & \minus  2 & \pminus \
0 & \pminus 0 & \pminus 0 & \pminus 0 \\
 \pminus 0 & \pminus 0 & \pminus 0 & \pminus 0 & \pminus 0 & \pminus \
0 & \pminus 0 & \minus  3 & \pminus 0 & \minus  2 & \pminus 0 & \
\pminus 0 & \pminus 0 & \pminus 0 \\
 \pminus 0 & \pminus 0 & \pminus 0 & \pminus 0 & \minus  4 & \pminus \
0 & \pminus 0 & \pminus 0 & \pminus 0 & \pminus 0 & \minus  4 & \
\pminus 0 & \pminus 0 & \pminus 0 \\
 \pminus 0 & \pminus 0 & \pminus 0 & \pminus 0 & \pminus 0 & \pminus \
0 & \pminus 0 & \pminus 0 & \pminus 0 & \pminus 0 & \pminus 0 & \
\pminus 2 & \pminus 0 & \pminus 0 \\
 \pminus 0 & \pminus 0 & \pminus 0 & \pminus 0 & \pminus 0 & \pminus \
0 & \pminus 0 & \pminus 0 & \pminus 0 & \pminus 0 & \pminus 0 & \
\minus  6 & \minus  4 & \pminus 0 \\
 \pminus 1 & \pminus 1 & \pminus \frac{4}{3} & \pminus \frac{2}{3} & \
\pminus 4 & \pminus 3 & \pminus 4 & \pminus 0 & \pminus 2 & \minus  2 \
& \pminus 1 & \pminus 2 & \pminus 2 & \minus  2 \\
\end{array}
\right)
  }
  \,,\quad
  \MM_{3} = \scalemath{0.41}{
  \left(
\begin{array}{cccccccccccccc}
 \pminus 0 & \pminus 0 & \pminus 0 & \pminus 0 & \pminus 0 & \pminus \
0 & \pminus 0 & \pminus 0 & \pminus 0 & \pminus 0 & \pminus 0 & \
\pminus 0 & \pminus 0 & \pminus 0 \\
 \pminus 0 & \pminus 0 & \pminus 0 & \pminus 0 & \pminus 0 & \pminus \
0 & \pminus 0 & \pminus 0 & \pminus 0 & \pminus 0 & \pminus 0 & \
\pminus 0 & \pminus 0 & \pminus 0 \\
 \pminus 0 & \pminus 0 & \pminus 0 & \pminus 0 & \pminus 0 & \pminus \
0 & \pminus 0 & \pminus 0 & \pminus 0 & \pminus 0 & \pminus 0 & \
\pminus 0 & \pminus 0 & \pminus 0 \\
 \pminus 0 & \pminus 0 & \pminus 0 & \pminus 0 & \pminus 0 & \pminus \
0 & \pminus 0 & \pminus 0 & \pminus 0 & \pminus 0 & \pminus 0 & \
\pminus 0 & \pminus 0 & \pminus 0 \\
 \pminus 0 & \pminus 0 & \pminus 0 & \pminus 0 & \pminus 0 & \pminus \
0 & \pminus 0 & \pminus 0 & \pminus 0 & \pminus 0 & \pminus 0 & \
\pminus 0 & \pminus 0 & \pminus 0 \\
 \pminus 0 & \pminus 0 & \pminus 0 & \pminus 0 & \pminus 0 & \pminus \
0 & \pminus 0 & \pminus 0 & \pminus 0 & \pminus 0 & \pminus 0 & \
\pminus 0 & \pminus 0 & \pminus 0 \\
 \pminus 0 & \pminus 0 & \pminus 0 & \pminus 0 & \pminus 0 & \pminus \
0 & \pminus 0 & \pminus 0 & \pminus 0 & \pminus 0 & \pminus 0 & \
\pminus 0 & \pminus 0 & \pminus 0 \\
 \pminus 0 & \pminus 0 & \pminus 0 & \pminus 0 & \pminus 0 & \pminus \
0 & \pminus 0 & \minus  3 & \minus  1 & \minus  2 & \pminus 0 & \
\pminus 0 & \pminus 0 & \pminus 0 \\
 \pminus 0 & \pminus 0 & \pminus 0 & \pminus 0 & \pminus 0 & \pminus \
0 & \pminus 0 & \pminus 3 & \pminus 1 & \pminus 2 & \pminus 0 & \
\pminus 0 & \pminus 0 & \pminus 0 \\
 \pminus 0 & \pminus 0 & \pminus 0 & \pminus 0 & \pminus 0 & \pminus \
0 & \pminus 0 & \pminus 3 & \pminus 1 & \pminus 2 & \pminus 0 & \
\pminus 0 & \pminus 0 & \pminus 0 \\
 \pminus 0 & \pminus 0 & \pminus 0 & \pminus 0 & \pminus 0 & \pminus \
0 & \pminus 0 & \pminus 0 & \pminus 0 & \pminus 0 & \pminus 0 & \
\pminus 0 & \pminus 0 & \pminus 0 \\
 \pminus 0 & \pminus 0 & \pminus 0 & \pminus 0 & \pminus 0 & \pminus \
0 & \pminus 0 & \pminus 0 & \pminus 0 & \pminus 0 & \pminus 0 & \
\pminus 0 & \pminus 0 & \pminus 0 \\
 \pminus 0 & \pminus 0 & \pminus 0 & \pminus 0 & \pminus 0 & \pminus \
0 & \pminus 0 & \pminus 0 & \pminus 0 & \pminus 0 & \pminus 0 & \
\pminus 0 & \pminus 0 & \pminus 0 \\
 \pminus 0 & \pminus 0 & \pminus 0 & \pminus 0 & \pminus 0 & \pminus \
0 & \pminus 0 & \pminus 0 & \pminus 0 & \pminus 0 & \pminus 0 & \
\pminus 0 & \pminus 0 & \pminus 0 \\
\end{array}
\right)}\,.\nn
  \end{align}
Around the $z=0$ singularity, the system reduces to
\bea
d \GGvec \approx \eps \ \MM_{1}  \dlog(z) \GGvec \, .
\eea 
We perform a Jordan decomposition of $\MM_{1}$, identifying the
matrices ${\mathbb J}$ and ${ \mathbb S}$, related by
${\mathbb J} \equiv {\mathbb S} \MM_{1} {\mathbb S}^{-1}$, 
\begin{align}
{\mathbb J}=\scalemath{0.5}{\left(
\begin{array}{cccccccccccccc}
 0 &\pminus 1 &\pminus 0 &\pminus 0 &\pminus 0 &\pminus 0 &\pminus 0 &\pminus 0 &\pminus 0 &\pminus 0 &\pminus 0 &\pminus 0 &\pminus 0 &\pminus 0 \\
 0 &\pminus 0 &\pminus 0 &\pminus 0 &\pminus 0 &\pminus 0 &\pminus 0 &\pminus 0 &\pminus 0 &\pminus 0 &\pminus 0 &\pminus 0 &\pminus 0 &\pminus 0 \\
 0 &\pminus 0 &\pminus 0 &\pminus 0 &\pminus 0 &\pminus 0 &\pminus 0 &\pminus 0 &\pminus 0 &\pminus 0 &\pminus 0 &\pminus 0 &\pminus 0 &\pminus 0 \\
 0 &\pminus 0 &\pminus 0 &\pminus 0 &\pminus 0 &\pminus 0 &\pminus 0 &\pminus 0 &\pminus 0 &\pminus 0 &\pminus 0 &\pminus 0 &\pminus 0 &\pminus 0 \\
 0 &\pminus 0 &\pminus 0 &\pminus 0 &\pminus 0 &\pminus 1 &\pminus 0 &\pminus 0 &\pminus 0 &\pminus 0 &\pminus 0 &\pminus 0 &\pminus 0 &\pminus 0 \\
 0 &\pminus 0 &\pminus 0 &\pminus 0 &\pminus 0 &\pminus 0 &\pminus 0 &\pminus 0 &\pminus 0 &\pminus 0 &\pminus 0 &\pminus 0 &\pminus 0 &\pminus 0 \\
 0 &\pminus 0 &\pminus 0 &\pminus 0 &\pminus 0 &\pminus 0 &\pminus 1 &\pminus 0 &\pminus 0 &\pminus 0 &\pminus 0 &\pminus 0 &\pminus 0 &\pminus 0 \\
 0 &\pminus 0 &\pminus 0 &\pminus 0 &\pminus 0 &\pminus 0 &\pminus 0 &\pminus 1 &\pminus 0 &\pminus 0 &\pminus 0 &\pminus 0 &\pminus 0 &\pminus 0 \\
 0 &\pminus 0 &\pminus 0 &\pminus 0 &\pminus 0 &\pminus 0 &\pminus 0 &\pminus 0 &\pminus 1 &\pminus 1 &\pminus 0 &\pminus 0 &\pminus 0 &\pminus 0 \\
 0 &\pminus 0 &\pminus 0 &\pminus 0 &\pminus 0 &\pminus 0 &\pminus 0 &\pminus 0 &\pminus 0 &\pminus 1 &\pminus 0 &\pminus 0 &\pminus 0 &\pminus 0 \\
 0 &\pminus 0 &\pminus 0 &\pminus 0 &\pminus 0 &\pminus 0 &\pminus 0 &\pminus 0 &\pminus 0 &\pminus 0 &\pminus 2 &\pminus 0 &\pminus 0 &\pminus 0 \\
 0 &\pminus 0 &\pminus 0 &\pminus 0 &\pminus 0 &\pminus 0 &\pminus 0 &\pminus 0 &\pminus 0 &\pminus 0 &\pminus 0 &\pminus 2 &\pminus 0 &\pminus 0 \\
 0 &\pminus 0 &\pminus 0 &\pminus 0 &\pminus 0 &\pminus 0 &\pminus 0 &\pminus 0 &\pminus 0 &\pminus 0 &\pminus 0 &\pminus 0 &\pminus 2 &\pminus 0 \\
 0 &\pminus 0 &\pminus 0 &\pminus 0 &\pminus 0 &\pminus 0 &\pminus 0 &\pminus 0 &\pminus 0 &\pminus 0 &\pminus 0 &\pminus 0 &\pminus 0 &\pminus 2 \\
\end{array}
\right)
} \ ,
\qquad 
{\mathbb S}=\scalemath{0.48}{
\left(
\begin{array}{cccccccccccccc}
 \pminus \frac{5}{2} & \pminus 1 & \pminus 0 & \pminus 0 & \pminus 3 \
& \pminus 0 & \minus  1 & \pminus 0 & \pminus \frac{3}{2} & \minus  1 \
& \pminus 0 & \pminus 0 & \minus  1 & \pminus 0 \\
 \pminus 1 & \pminus 0 & \pminus 0 & \pminus 0 & \pminus 0 & \pminus \
0 & \pminus 0 & \pminus 0 & \pminus 1 & \pminus 0 & \pminus 0 & \
\pminus 0 & \pminus 0 & \pminus 0 \\
 \minus  \frac{1}{2} & \minus  \frac{1}{2} & \pminus \frac{1}{2} & \
\minus  \frac{1}{2} & \pminus 0 & \pminus 0 & \pminus 0 & \pminus 0 & \
\pminus 0 & \pminus 0 & \pminus 0 & \pminus 0 & \pminus 1 & \pminus 0 \
\\
 \pminus \frac{5}{4} & \pminus 1 & \minus  1 & \pminus 1 & \pminus 1 \
& \pminus 0 & \pminus 0 & \pminus 0 & \pminus 0 & \pminus 0 & \pminus \
0 & \pminus 0 & \minus  2 & \pminus 0 \\
 \minus  \frac{1}{2} & \minus  \frac{1}{2} & \pminus 0 & \pminus 0 & \
\pminus 0 & \pminus 0 & \pminus 0 & \pminus 0 & \pminus 0 & \pminus 1 \
& \pminus 0 & \pminus 0 & \pminus 0 & \pminus 0 \\
 \minus  \frac{1}{2} & \pminus 0 & \pminus 0 & \pminus 0 & \pminus 0 \
& \pminus 0 & \pminus 0 & \pminus 0 & \minus  1 & \pminus 0 & \pminus \
0 & \pminus 0 & \pminus 0 & \pminus 0 \\
 \pminus \frac{1}{2} & \pminus \frac{1}{2} & \pminus 0 & \pminus 0 & \
\pminus 0 & \pminus 0 & \pminus 0 & \pminus 0 & \pminus 0 & \pminus 0 \
& \pminus 0 & \pminus 0 & \pminus 0 & \pminus 0 \\
 \pminus \frac{3}{2} & \pminus 3 & \minus  \frac{8}{3} & \pminus \
\frac{25}{6} & \pminus 0 & \pminus 3 & \pminus 6 & \pminus 0 & \
\pminus 0 & \pminus 0 & \pminus 0 & \pminus 0 & \pminus 0 & \pminus 0 \
\\
 \minus  \frac{1}{4} & \pminus \frac{1}{4} & \minus  \frac{1}{6} & \
\pminus \frac{2}{3} & \minus  \frac{1}{2} & \pminus \frac{3}{4} & \
\pminus \frac{3}{2} & \pminus 0 & \pminus 0 & \pminus 0 & \pminus \
\frac{1}{4} & \pminus \frac{1}{2} & \pminus 1 & \pminus 0 \\
 \pminus \frac{1}{2} & \pminus \frac{1}{2} & \minus  \frac{1}{2} & \
\pminus \frac{1}{2} & \pminus 0 & \pminus 0 & \pminus 0 & \pminus 0 & \
\pminus 0 & \pminus 0 & \pminus 0 & \pminus 0 & \pminus 0 & \pminus 0 \
\\
 \minus  \frac{3}{2} & \minus  2 & \pminus \frac{2}{3} & \minus  \
\frac{8}{3} & \minus  1 & \minus  3 & \minus  5 & \pminus 0 & \minus  \
\frac{3}{2} & \pminus 1 & \minus  1 & \minus  2 & \minus  3 & \pminus \
1 \\
 \pminus 0 & \pminus 0 & \pminus 0 & \minus  \frac{1}{4} & \pminus 0 \
& \pminus 0 & \pminus 0 & \pminus 0 & \pminus 0 & \pminus 0 & \minus  \
\frac{1}{2} & \pminus 0 & \pminus 0 & \pminus 0 \\
 \pminus 0 & \pminus 0 & \pminus 0 & \pminus 0 & \pminus 0 & \pminus \
0 & \pminus 0 & \pminus 0 & \pminus 0 & \pminus 0 & \pminus 1 & \
\pminus 0 & \pminus 0 & \pminus 0 \\
 \pminus \frac{1}{2} & \pminus \frac{1}{2} & \pminus 0 & \pminus 0 & \
\pminus 1 & \pminus 0 & \pminus 0 & \pminus 1 & \pminus \frac{1}{2} & \
\pminus 1 & \pminus 0 & \pminus 0 & \pminus 0 & \pminus 0 \\
\end{array}
\right)
 
} \ .
\end{align} 
The latter can be used to define a change of the integral basis, 
${\bf H} \equiv {\mathbb S} \GGvec$, which, by construction, obeys 
a system of differential equations in Jordan form,
\bea
d {\bf H} = \eps \ {\mathbb J}  \dlog(z) {\bf H} \ .
\eea
In particular, the differential equation of 
the truly diagonal elements ${\bf H}_{i}$, say $i=2,3,4,6,10,11,12,13,14$ (not belonging to any
block-diagonal sector), obey a trivial first-order differential equation of the form,
\bea
{d {\bf H}_i \over d z} = \eps {{\mathbb J}_{ii} \over z}  {\bf H}_i \ ;
\eea
therefore, their expression is of the type,
\bea
{\bf H}_i = z^{{\mathbb J}_{ii} \eps} \, {\bf H}_{i,0} \ ,
\eea
where ${\bf H}_{i,0}$ is a boundary constant which may still depend on $\eps$.
Among the possible choices of $i$, we look at the element $i=11$,
\bea
{\bf H}_{11} &=& {\mathbb S}_{11,j} \GGvec_j  =  z^{2\eps} \ {\bf H}_{11,0} \ ,
\label{zlimit} 
\eea
from which we infer the behaviour around $z=0$ of the following
combination of canonical integrals,
\bea
{\bf H}_{11,0} &=& \lim_{z \rightarrow 0}  z^{-2\eps} \left( -\frac{3}{2} \GG_{1} -2
  \GG_2+\frac{2}{3}\GG_3 -\frac{8}{3}\GG_4-\GG_{5}
-3\GG_6-5\GG_7
+ \right. \nn & & \qquad \quad \quad \quad
\left. -\frac{3}{2}\GG_9
+\GG_{10}-\GG_{11}-2\GG_{12}-3\GG_{13}+\GG_{14}   
\right)  \ ,
\label{integralcomb}
\eea
involving the integral $\GG_{14}$. 
On the other side, ${\bf H}_{11}$ can be computed by taking the limit $z \to 0$ on the r.h.s. of
eq.~\eqref{integralcomb} directly at the {\it integrand level}, for the
integrals ${\bf I}$ must be evaluated in the limit $m \to 0$ (or alternatively
$s \to \infty$).
To this aim, we need to pull out the prefactor 
$m^{4 \eps}$ coming from the integration measure defined in eq.~\eqref{eq:intmeasure1},
and to consider the definition of the canonical integrals ${\bf I}$ 
in terms of the linear-$\eps$ basis ${\bf F}$ given in eq.~\eqref{def:CanonicalBasisNPtriangle},
\begin{gather}
 {\bf H}_{11,0} = (-s)^{2 \eps} \ \Big(2  s \FF_3+ 3  s \FF_6-  s \FF_{11} +2  s \FF_{12}+ s^2 \FF_{14}\Big)\Big|_{m=0} \, .
\label{boundarylhs} 
\end{gather} 
In the latter equation, we  took into account the vanishing of the massless tadpole in
dimensional regularisation and the symmetries arising from the
massless limit of the ${\bf F}$ integrals. 
After applying the IBPs to the massless integrals, the contributions due to all subtopologies cancel and
the contribution of the {\it massless non-planar vertex} $\FF_{14}\big|_{m=0}$~\cite{Gehrmann:2001ck}
is the only one left,
\bea
 {\bf H}_{11,0} &=& (-s)^{2+2 \eps} \ \FF_{14} \big|_{m=0} \,,
 \eea
 where
 \bea
 \FF_{14} \big|_{m=0}=  \ (-s)^{-2-2\eps} \ {\cal F}(\eps)
\eea
with
\bea
 {\cal F}(\eps) 
\equiv  
1-7 \, \zeta_2\,\eps^2-27 \,  \zeta_3\,\eps^3-\frac{57}{2}\, \zeta_4\, \eps^4 +
\mathcal{O}(\eps^5) \,.
\eea
Therefore,
\bea
{\bf H}_{11,0}= {\cal F}(\eps) \, .
\label{boundaryrhs}
\eea
Finally the boundary constant of integral $\GG_{14}$ can be determined by demanding the equality of eq.~\eqref{integralcomb} and eq.~\eqref{boundaryrhs}.
 \end{itemize}
 All results have been numerically checked with the help of the computer codes
\texttt{GiNaC} and \texttt{SecDec}, and the analytic expressions
of the MIs are given in electronic form in the ancillary files
attached to the \texttt{arXiv} version of this manuscript.

\section{Conclusions}

The scattering of high-energy muons on atomic electrons has been recently proposed as an ideal framework to determine the leading hadronic contribution to the anomalous magnetic moment of the muon. The ambitious experimental goal of measuring the differential cross section of the $\mu e \to \mu e$ process with an accuracy of 10ppm requires, on the theoretical side, the knowledge of the QED corrections at NNLO.
In this work, after calculating the QED corrections at NLO, which were found to be in agreement with the latest results in the literature, we investigated the feasibility of the evaluation of the corrections at NNLO. In particular, we began by considering the two-loop planar box-diagrams contributing to this process. We employed the method of differential equations and of the Magnus exponential series to identify a canonical set of master integrals. Boundary conditions were derived from the regularity requirements at pseudothresholds, or from the knowledge of the integrals at special kinematic points, evaluated by means of auxiliary, simpler systems of differential equations.

The considered master integrals were expressed as a Taylor series around four space-time dimensions, whose coefficients are written as a combination of generalised polylogarithms. We worked in the massless electron approximation, while keeping full dependence on the muon mass. Besides $\mu e$ scattering, our results are relevant also for crossing-related processes such as muon-pair production at $e^+ e^-$-colliders, as well as for the QCD corrections to $top$-pair production at hadron colliders.

The evaluation of the missing contributions due to non-planar box graphs will be the subject of a dedicated, future work -- we are confident that the techniques employed here can be systematically applied for that case as well.


\section*{Acknowledgments}

We thank Roberto Bonciani, Matteo Fael, Andrea Ferroglia and Stefano Laporta for useful discussions.  
We are happy to acknowledge the
stimulating discussions which took place during the workshops {\it
  Muon-electron scattering:~Theory kick-off meeting} (4--6 September
2017, Padova, Italy) and {\it Flavor Changing and Conserving Processes
  2017} (7--9 September 2017, Anacapri, Capri Island, Italy). We wish
to thank also Federico Gasparotto, for checks on the one-loop MIs.

This research was supported in part by Perimeter Institute for Theoretical Physics. Research at Perimeter Institute is supported by the Government of Canada through the Department of Innovation, Science and Economic Development and by the Province of Ontario through the Ministry of Research, Innovation and Science. The work of U.~S.\ was performed in part at the Aspen Center for Physics, which is supported by the National Science Foundation grant PHY-1607611. U.~S.\ is supported by the DOE contract DE-AC02-06CH11357. A.~P.\ wishes to thank the Institute for Particle Physics (IFIC) of Valencia for hospitality during the final stages of this project. M.~P.~acknowledges partial support by FP10 ITN Elusives (H2020-MSCA-ITN-2015-674896) and Invisibles-Plus (H2020-MSCA-RISE-2015-690575).

\appendix
\clearpage
\section{Auxiliary integrals}
\label{sec:AuxVert}
In this appendix we briefly discuss the solution of the system of differential equations for the vertex integrals which has been used in eqs.~\eqref{eq:limit_i10} and \eqref{eq:limit_i810} as an input for the determination of the boundary constants of some of the MIs considered in this paper.
\subsection*{Auxiliary vertex integral for eq.~\eqref{eq:limit_i10}}
\label{sec:AuxVert1}
 We consider the integral family
\begin{gather}
  \int \widetilde{\dd^d k_1}\widetilde{\dd^d k_2}\,
  \frac{\Den_{6}^{n_6}\Den_{7}^{n_7}}{\Den_{1}^{n_1}\Den_{2}^{n_2}\Den_{3}^{n_3}\Den_{4}^{n_4}\Den_{5}^{n_5}}\,,\quad n_i\geq0\,,
\end{gather}
identified by the set of denominators
\begin{gather}
\Den_1 = k_1^2-m^2,\quad
\Den_2 = k_2^2-m^2,\quad
\Den_3 = (k_1+p_1)^2,\quad
\Den_4 = (k_2+p_1+p_2)^2, \nonumber\\
\Den_5 = (k_1-k_2)^2,\quad
\Den_6 = (k_2+p_1)^2,\quad
\Den_7 = (k_1+p_1+p_2)^2,
\end{gather} 
and by external momenta $p_1$, $p_2$ and $p_3$ satisfying
\begin{align}
 p_2^2=0\,,\quad p_3^2=(p_1+p_2)^2=0\,.
\end{align}
All integrals belonging to this family can be reduced to a set of 8 MIs, whose dependence on $p_1^2$ is parametrized in terms of the dimensionless variable
\begin{align}
x=-\frac{p_1^2}{m^2}\,.
\end{align}
The basis of integrals
  \begin{align}
 \GG_{1}=&\eps^2 \raisebox{-19pt}{\includegraphics[scale=0.23]{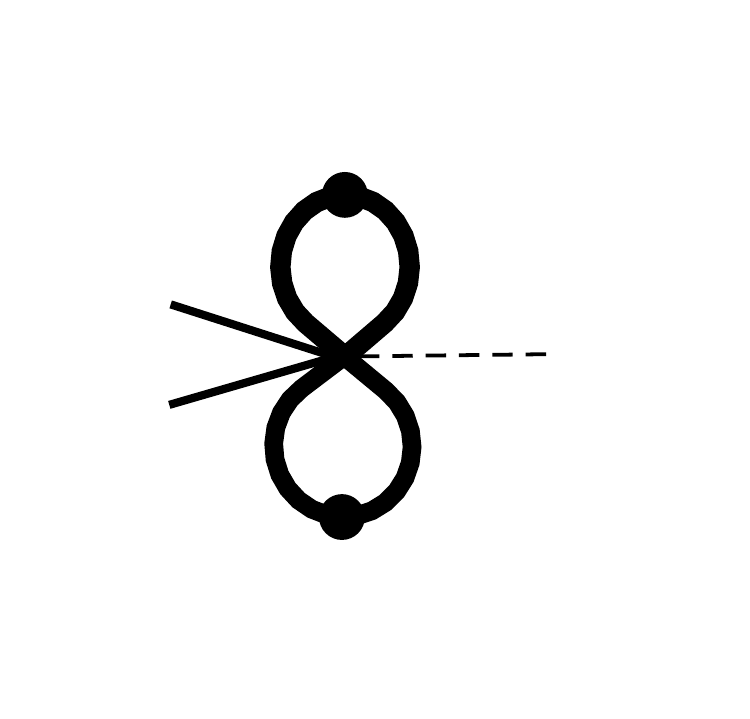}}\,,\quad \GG_{2}=-\eps^2p_1^2 \raisebox{-12pt}{\includegraphics[scale=0.23]{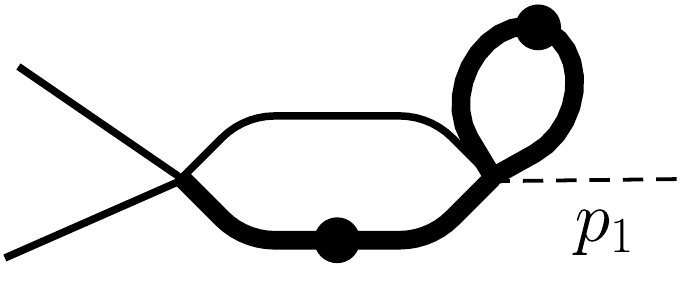}}\,,\quad \GG_{3}=-\eps^2p_1^2 \raisebox{-18pt}{\includegraphics[scale=0.23]{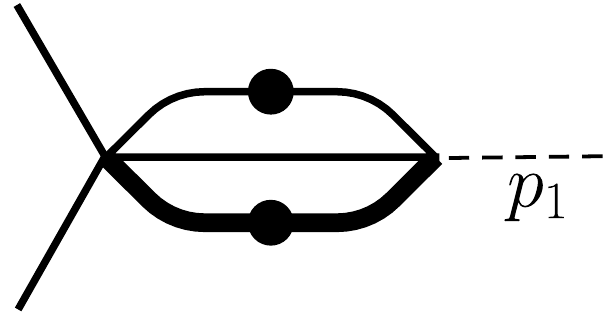}}\,,\nn
&\quad \GG_{4}=\eps^2\,2m^2\raisebox{-15pt}{\includegraphics[scale=0.23]{figures/AuxVert1011/Fig_Aux_Tri1011T3.pdf}}\;+\;\eps^2(m^2-p_1^2)\raisebox{-15pt}{\includegraphics[scale=0.23]{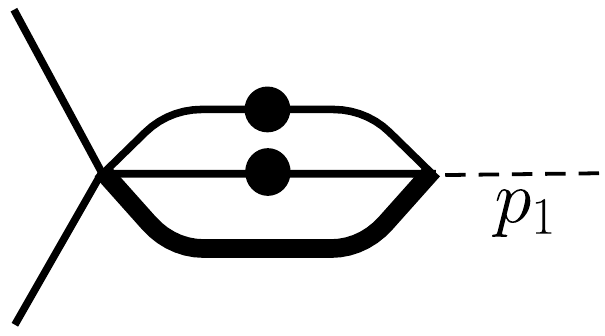}}\,,\nn
&\quad \GG_{5}=\eps(1-\eps)m^2 \raisebox{-9pt}{\includegraphics[scale=0.23]{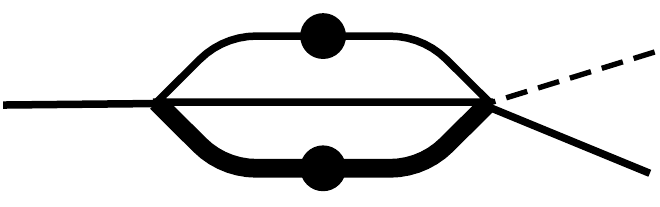}}\,,\quad
\GG_{6}=-\eps^3 p_1^2\raisebox{-25pt}{\includegraphics[scale=0.18]{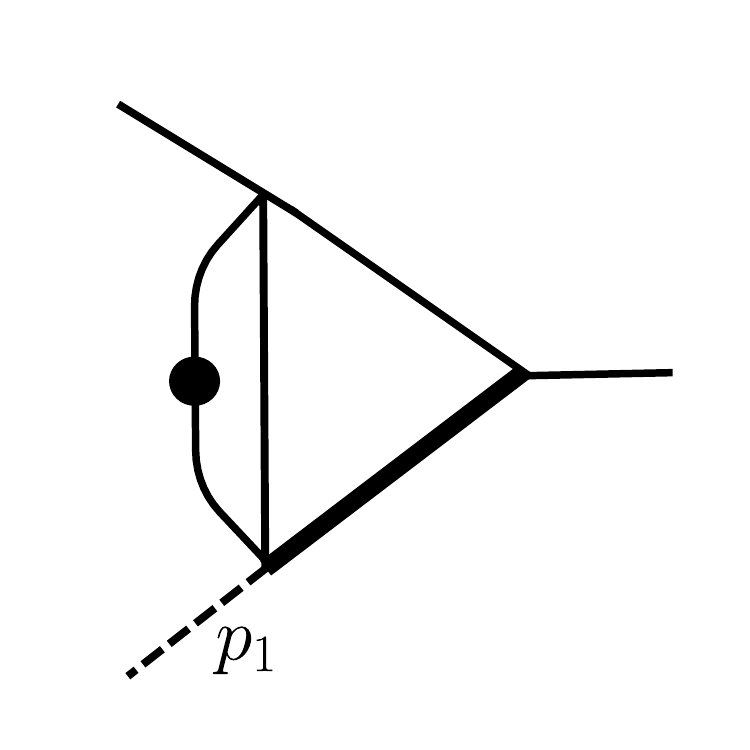}}\,,\nn
&\quad  \GG_{7}=-\eps^4\,p_1^2\raisebox{-26pt}{\includegraphics[scale=0.18]{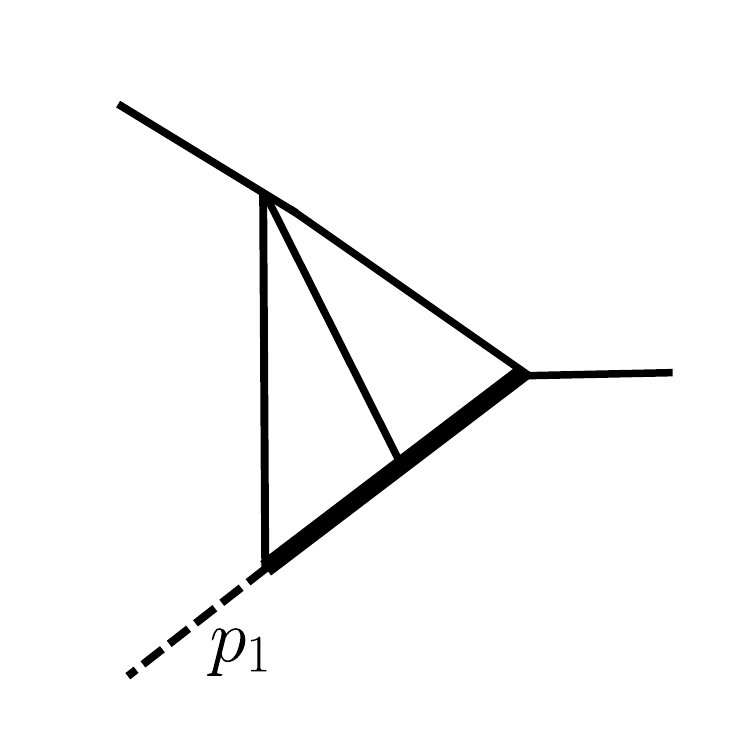}},\qquad   
 \GG_{8}=\eps^3 \,p_1^2(p_1^2-m^2)\raisebox{-26pt}{\includegraphics[scale=0.18]{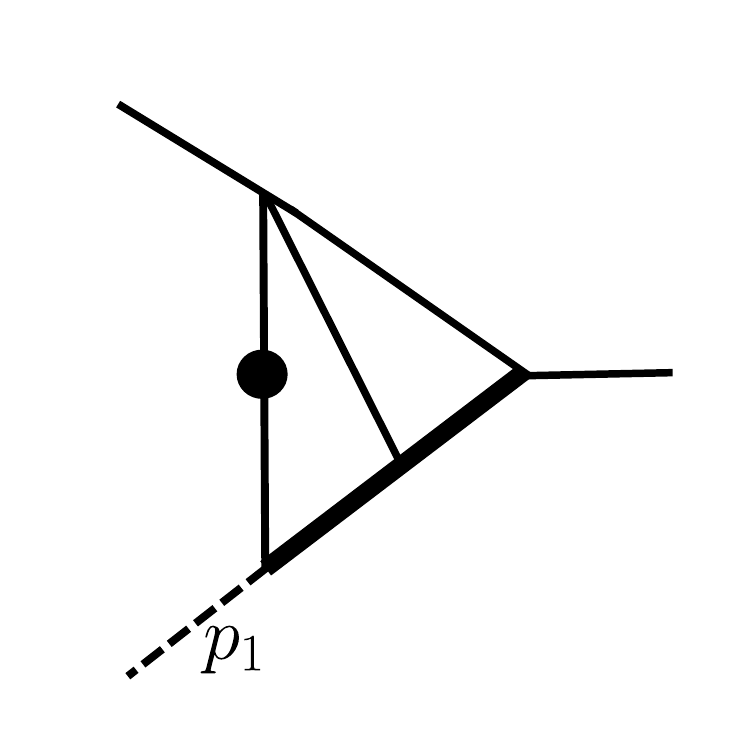}}
  \end{align}
  fulfills a canonical system of differential equations,
  \begin{align}
  d \GGvec = \eps \, \dA \, \GGvec \ ,
\end{align}
where 
\begin{align}
\dA =  \MM_1 \, \dlog x+\MM_{2} \, \dlog (1+x)\,,
\end{align}
with
\begin{align}
 \MM_1=\scalemath{0.65}{\left(
\begin{array}{cccccccc}
 0 &\pminus 0 &\pminus 0 &\pminus 0 &\pminus 0 &\pminus 0 &\pminus 0 &\pminus 0 \\
 0 &\pminus 1 &\pminus 0 &\pminus 0 &\pminus 0 &\pminus 0 &\pminus 0 &\pminus 0 \\
 0 &\pminus 0 &\pminus 1 &\pminus 0 &\pminus 0 &\pminus 0 &\pminus 0 &\pminus 0 \\
 0 &\pminus 0 &\pminus 4 &\pminus 0 &\pminus 0 &\pminus 0 &\pminus 0 &\pminus 0 \\
 0 &\pminus 0 &\pminus 0 &\pminus 0 &\pminus 0 &\pminus 0 &\pminus 0 &\pminus 0 \\
 0 &\pminus 0 &\pminus 0 &\pminus 0 &\pminus 0 &\pminus 1 &\pminus 0 &\pminus 0 \\
 0 &-\frac{1}{4} &-\frac{1}{2} &\pminus 0 &\pminus 0 &\pminus \frac{1}{4} &\pminus \frac{1}{2} &\pminus \frac{3}{4} \\
 0 &-\frac{1}{2} &-1 &\pminus 0 &\pminus 0 &\pminus \frac{1}{2} &-1 & \frac{5}{2} \\
\end{array}
\right)}
 \,,\quad
\MM_{2}=\scalemath{0.65}{
\left(
\begin{array}{cccccccc}
 0 &\pminus 0 &\pminus 0 &\pminus 0 &\pminus 0 &\pminus 0 &\pminus 0 &\pminus 0 \\
 -1 & -2 &\pminus 0 &\pminus 0 &\pminus 0 &\pminus 0 &\pminus 0 &\pminus 0 \\
 0 &\pminus 0 &-2 & -1 &\pminus 0 &\pminus 0 &\pminus 0 &\pminus 0 \\
 0 &\pminus 0 &-4 &\pminus -2 &\pminus 0 &\pminus 0 &\pminus 0 &\pminus 0 \\
 0 &\pminus 0 &\pminus 0 &\pminus 0 &\pminus 0 &\pminus 0 &\pminus 0 &\pminus 0 \\
 0 &\pminus 0 &\pminus 0 &\pminus 0 & -\frac{1}{2} &\pminus -3 &\pminus 0 &\pminus 0 \\
 0 &\pminus 0 &\pminus 0 &\pminus 0 &\pminus 0 &\pminus 0 &\pminus 0 &\pminus 0 \\
 1 &\pminus 2 &\pminus 0 &\pminus 0 & -\frac{1}{2} &\pminus -3 &\pminus 0 & -4 \\
\end{array}
\right)}\,.
\end{align}
In the Euclidean region $x>0$ the general solution of the system of
differential equations can be expressed in terms of 
harmonic polylogarithms (HPLs) \cite{Remiddi:1999ew},  and the boundary constants of all master integrals, with the only exception of $\GG_{1}=1$ and
\begin{align}
\GG_{5}(\eps)=1+2\zeta_2\eps^2-2\zeta_3 \eps^3 +9\zeta_4\eps^4+\mathcal{O}\left(\eps^5\right)\,,
\end{align}
can be fixed by demanding their regularity at $x\to 0$. In particular, for the $\GG_{7}(\eps,x)$ we obtain
\begin{align}
\GG_{7}(\eps,x)=&\left(-2\zeta_2 H(0,-1;x)-H(0,-1,-1,-1;x)+H(0,-1,0,-1;x)\right)\eps^4+\mathcal{O}(\eps^5).
 \end{align}
This expression, when it is analytically continued to the region $x<0$, has a smooth limit for $x\to -1$ ( $p_1^2=m^2$ ),
 \begin{align}
\GG_{7}(\eps,-1)=&\frac{27}{4}\zeta_4\eps^4+\mathcal{O}(\eps^5) \ , 
 \end{align}
 which has been used in eq.~\eqref{eq:bc_i10}. \\

 \subsection*{Auxiliary vertex integral for eq.~\eqref{eq:limit_i810}}
\label{sec:AuxVert2}
 We consider the integral family
\begin{gather}
  \int \widetilde{\dd^d k_1}\widetilde{\dd^d k_2}\,
  \frac{\Den_{5}^{n_5}\Den_{6}^{n_6}\Den_{7}^{n_7}}{\Den_{1}^{n_1}\Den_{2}^{n_2}\Den_{3}^{n_3}\Den_{4}^{n_4}}\,,\quad n_i\geq0\,,
\end{gather}
identified by the set of denominators
\begin{gather}
\Den_1 = k_1^2,\quad
\Den_2 = k_2^2-m^2,\quad
\Den_3 = (k_1+p_1)^2-m^2,\quad
\Den_4 = (k_1+k_2+p_1+p_2)^2, \nonumber\\
\Den_5 = (k_1+p_2)^2,\quad
\Den_6 = (k_2+p_1)^2,\quad
\Den_7 = (k_2+p_1+p_2)^2,
\end{gather} 
and by external momenta $p_1$, $p_2$ and $p_3$ satisfying
\begin{align}
p_1^2=p_2^2=0\,,\quad (p_1+p_2)^2=p_3^2\,.
\end{align}
All integrals belonging to this family can be reduced to a set of 5 MIs, whose dependence on $p_3^2$ is parametrized in terms of the dimensionless variable
\begin{align}
x=-\frac{p_3^2}{m^2}\,.
\end{align}
The basis of integrals
  \begin{align}
 \GG_{1}=&\quad\eps^2 \raisebox{-19pt}{\includegraphics[scale=0.23]{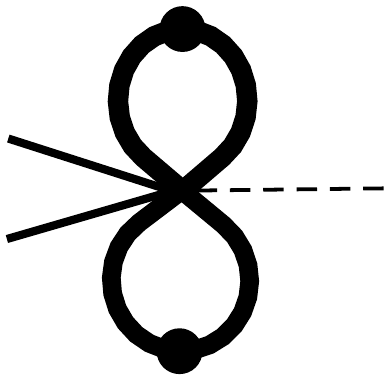}}\,,\qquad\qquad\qquad \GG_{2}=-\eps^2p_3^2 \raisebox{-10pt}{\includegraphics[scale=0.23]{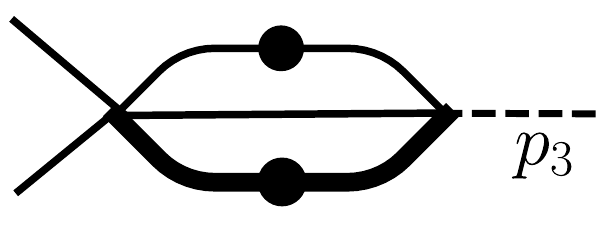}}\,,\nn
 \GG_{3}=&\qquad\eps^2\,2m^2\raisebox{-10pt}{\includegraphics[scale=0.23]{figures/AuxVert8910/Fig_Aux_Tri8910T2o.pdf}}\;+\;\eps^2(m^2-p_3^2)\raisebox{-10pt}{\includegraphics[scale=0.23]{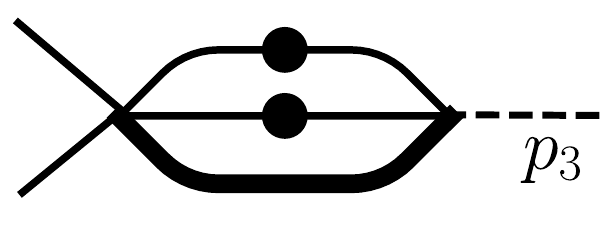}}\,,\nn
\GG_{4}=&-\eps^3\,p_3^2\raisebox{-35pt}{\includegraphics[scale=0.2]{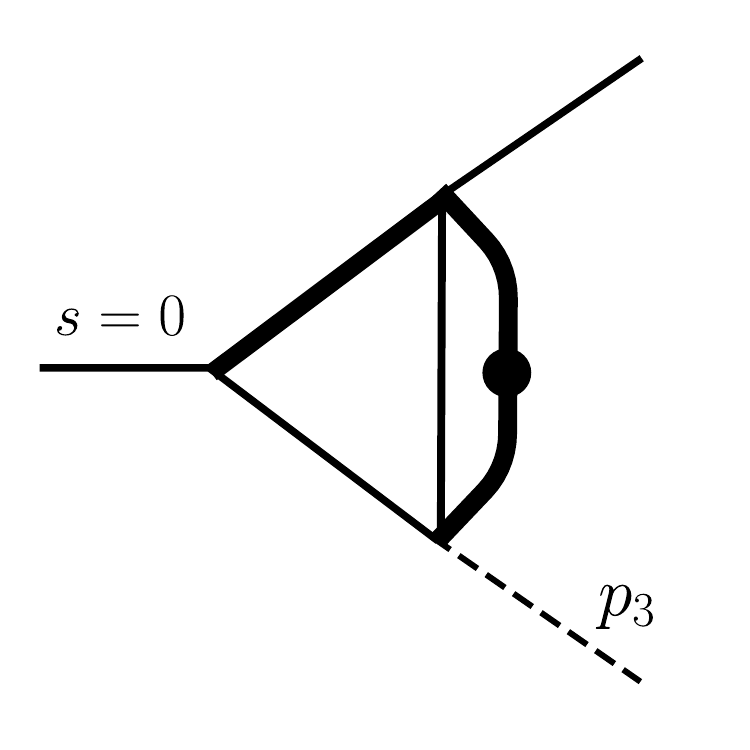}}\,,\quad \GG_{5}=-\eps^4\,p_3^2m^2 \raisebox{-35pt}{\includegraphics[scale=0.2]{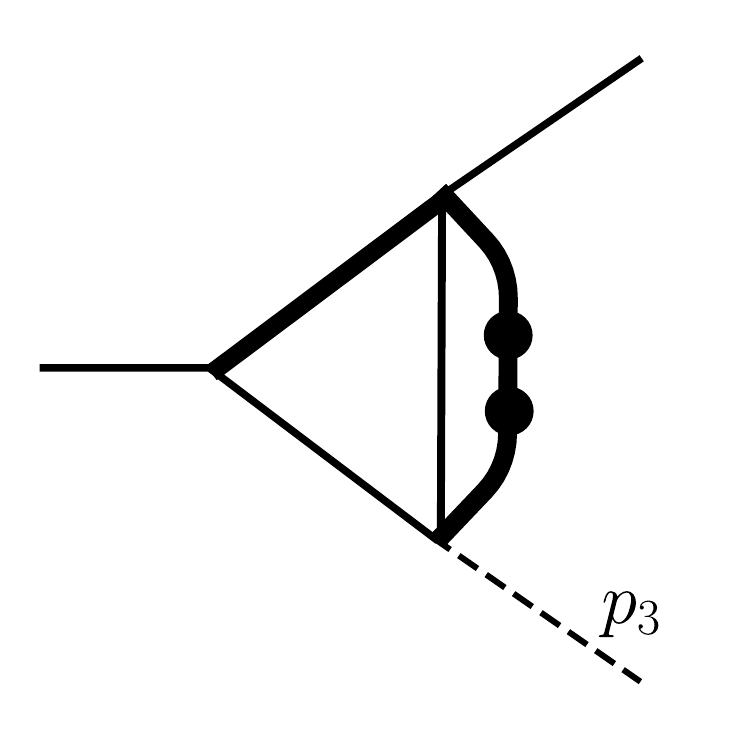}}\,,
  \end{align}
  fulfils a canonical system of differential equations,
  \begin{align}
  d \GGvec = \eps \, \dA \, \GGvec \ ,
\end{align}
where 
\begin{align}
\dA =  \MM_1 \, \dlog x+\MM_{2} \, \dlog (1+x)+\MM_{3} \, \dlog (1-x)\,,
\end{align}
with
\begin{align}
 \MM_1=\scalemath{0.65}{
 \left(
\begin{array}{ccccc}
 0 &\pminus 0 &\pminus 0 &\pminus 0 &\pminus 0 \\
 0 &\pminus 1 &\pminus 0 &\pminus 0 &\pminus 0 \\
 0 &\pminus 4 &\pminus 0 &\pminus 0 &\pminus 0 \\
 0 &\pminus 0 &\pminus 0 &\pminus 1 &\pminus 0 \\
 0 &\pminus \frac{1}{2} &\pminus 0 & -2 & -2 \\
\end{array}
\right)}\,,\quad
\MM_{2}=
\scalemath{0.65}{\left(
\begin{array}{ccccc}
 0 &\pminus 0 &\pminus 0 &\pminus 0 &\pminus 0 \\
 0 & -2 & -1 &\pminus 0 &\pminus 0 \\
 0 & -4 & -2 &\pminus 0 &\pminus 0 \\
 0 &\pminus 0 &\pminus 0 &\pminus 0 &\pminus 0 \\
 0 &\pminus 0 &\pminus 0 &\pminus 0 &\pminus 0 \\
\end{array}
\right)}
\,,\quad
\MM_{3}=\scalemath{0.65}{
\left(
\begin{array}{ccccc}
 0 &\pminus 0 &\pminus 0 &\pminus 0 &\pminus 0 \\
 0 &\pminus 0 &\pminus 0 &\pminus 0 &\pminus 0 \\
 0 &\pminus 0 &\pminus 0 &\pminus 0 &\pminus 0 \\
 -\frac{1}{2} &\pminus 1 & -\frac{1}{2} & -2 & -2 \\
 \frac{1}{2} & -1 &\pminus \frac{1}{2} &\pminus 2 &\pminus 2 \\
\end{array}
\right)}
\,.
\end{align}
In the Euclidean region $x>0$ the general solution of the system of differential equations can be expressed in terms of HPLs. The boundary constants $\GG_{4,5}$, which are the only MIs appearing for the first time in this computation, can be fixed by demanding their regularity at $x\to0$. In this way, we obtain
\begin{align}
\GG_{i}(\eps,x)=\sum_{k=2}^{4}\GG_{i}^{(k)}(x,y)\eps^k+\mathcal{O}(\eps^5)\,,
\end{align}
with
\begin{align}
\GG_{4}^{(2)}(x)=&\quad0\,,\nn
\GG_{4}^{(3)}(x)=&-\zeta_2 H(1;x)+2 H(1,0,-1;x)\,,\nn
\GG_{4}^{(4)}(x)=&\quad \zeta_3 H(1;x)-\zeta_2 H(0,1;x)+2 H(0,1,0,-1;x)-8
   H(1,0,-1,-1;x)\,,
\end{align}
and
\begin{align}
\GG_{5}^{(2)}(x)=&\quad\frac{1}{2} H(0,-1;x)\,,\nn
\GG_{5}^{(3)}(x)=&\quad\zeta_2 H(1;x)-2 H(0,-1,-1;x)-\frac{1}{2}H(0,0,-1;x) -2 H(1,0,-1;x))\,,\nn
\GG_{5}^{(4)}(x)=&-\zeta_3 H(1;x)+\zeta_2 H(0,-1;x)8 H(0,-1,-1,-1;x)-3 H(0,-1,0,-1;x)\nn
&+2 H(0,0,-1,-1;x)+\frac{3}{2} H(0,0,0,-1;x)+8H(1,0,-1,-1;x)\,.
\end{align}
The analytic continuation of these expressions to $x\to -1$ ( $p_3^2=m^2$ ) produces the smooth limits
  \begin{align}
  \GG_{4}(\eps, -1)=&-\left(\frac{5 \zeta_3}{4}-3\zeta_2 \log (2)\right)\eps^3-\left(8 \text{Li}_4\left(\frac{1}{2}\right)-\frac{33}{8}\zeta_4+\frac{\log ^4(2)}{3}-2\zeta_2 \log ^2(2)\right)\eps^4 \,,\nn
  \GG_{5}(\eps, -1)=&-\frac{\zeta_2}{2}\eps^2-
   \left(\frac{\zeta_3}{4}+3\zeta_2 \log (2)\right)\eps^3\nn
   &-\left(-8
     \text{Li}_4\left(\frac{1}{2}\right)+\frac{65}{4}\zeta_4-\frac{\log
     ^4(2)}{3}+2 \zeta_2 \log ^2(2)\right)\eps^4
     +\mathcal{O}\left(\eps^5\right)\, ,
  \end{align}
 which have been used in eq.~\eqref{eq:bc_i810}.

\section[dlog-forms]{$\dlog$-forms}
\label{sec:dlog-form}
In this appendix we collect the coefficient matrices of the
$\dlog$-forms
\begin{align}
  \dA = {} & \MM_1 \, \dlog(x) + \MM_2 \, \dlog(1+x) + \MM_3 \, \dlog(1-x) \nn
  &+ \MM_4 \, \dlog(y) + \MM_5 \, \dlog(1+y)  + \MM_6\, \dlog(1-y) \nn
  &+ \MM_7\, \dlog(x+y) + \MM_8 \, \dlog\left(1+x\,y\right) \nn
  &+ \MM_9\, \dlog\left(1-y(1-x-y)\right)\,,
\end{align}
for the master integrals in the first and second integral family,
respectively defined in eqs.~(\ref{eq:2Lfamily1},\ref{eq:2Lfamily2}).

\subsection{First integral family}
\label{dlog1stIF}
For the first integral family, given in eq.~\eqref{eq:2Lfamily1}, we have ($\MM_3$ is vanishing for this integral family): 
\begin{align}
\MM_1 = \scalemath{0.5}{
  \left(
\begin{array}{cccccccccccccccccccccccccccccccccc}
 \pminus 0 & \pminus 0 & \pminus 0 & \pminus 0 & \pminus 0 & \pminus 0 & \pminus 0 & \pminus 0 & \pminus 0 & \pminus 0 & \pminus 0 & \pminus 0 & \pminus 0 & \pminus 0 & \pminus 0 & \pminus 0 & \pminus 0 & \pminus 0 & \pminus 0 & \pminus 0 & \pminus 0 & \pminus 0 & \pminus 0 & \pminus 0 & \pminus 0 & \pminus 0 & \pminus 0 & \pminus 0 & \pminus 0 & \pminus 0 & \pminus 0 & \pminus 0 & \pminus 0 & \pminus 0 \\
 \pminus 0 & \pminus 1 & \pminus 0 & \pminus 0 & \pminus 0 & \pminus 0 & \pminus 0 & \pminus 0 & \pminus 0 & \pminus 0 & \pminus 0 & \pminus 0 & \pminus 0 & \pminus 0 & \pminus 0 & \pminus 0 & \pminus 0 & \pminus 0 & \pminus 0 & \pminus 0 & \pminus 0 & \pminus 0 & \pminus 0 & \pminus 0 & \pminus 0 & \pminus 0 & \pminus 0 & \pminus 0 & \pminus 0 & \pminus 0 & \pminus 0 & \pminus 0 & \pminus 0 & \pminus 0 \\
 \pminus 0 & \pminus 0 & \pminus 0 & \pminus 0 & \pminus 0 & \pminus 0 & \pminus 0 & \pminus 0 & \pminus 0 & \pminus 0 & \pminus 0 & \pminus 0 & \pminus 0 & \pminus 0 & \pminus 0 & \pminus 0 & \pminus 0 & \pminus 0 & \pminus 0 & \pminus 0 & \pminus 0 & \pminus 0 & \pminus 0 & \pminus 0 & \pminus 0 & \pminus 0 & \pminus 0 & \pminus 0 & \pminus 0 & \pminus 0 & \pminus 0 & \pminus 0 & \pminus 0 & \pminus 0 \\
 \pminus 0 & \pminus 0 & \pminus 0 & \pminus 0 & \pminus 0 & \pminus 0 & \pminus 0 & \pminus 0 & \pminus 0 & \pminus 0 & \pminus 0 & \pminus 0 & \pminus 0 & \pminus 0 & \pminus 0 & \pminus 0 & \pminus 0 & \pminus 0 & \pminus 0 & \pminus 0 & \pminus 0 & \pminus 0 & \pminus 0 & \pminus 0 & \pminus 0 & \pminus 0 & \pminus 0 & \pminus 0 & \pminus 0 & \pminus 0 & \pminus 0 & \pminus 0 & \pminus 0 & \pminus 0 \\
\end{array}
\right)
  }\,,
\end{align}

\begin{align}
\MM_2 = \scalemath{0.53}{
  \left(
\begin{array}{cccccccccccccccccccccccccccccccccc}
 \pminus 0 & \pminus 0 & \pminus 0 & \pminus 0 & \pminus 0 & \pminus 0 & \pminus 0 & \pminus 0 & \pminus 0 & \pminus 0 & \pminus 0 & \pminus 0 & \pminus 0 & \pminus 0 & \pminus 0 & \pminus 0 & \pminus 0 & \pminus 0 & \pminus 0 & \pminus 0 & \pminus 0 & \pminus 0 & \pminus 0 & \pminus 0 & \pminus 0 & \pminus 0 & \pminus 0 & \pminus 0 & \pminus 0 & \pminus 0 & \pminus 0 & \pminus 0 & \pminus 0 & \pminus 0 \\
 \minus  1 & \minus  2 & \pminus 0 & \pminus 0 & \pminus 0 & \pminus 0 & \pminus 0 & \pminus 0 & \pminus 0 & \pminus 0 & \pminus 0 & \pminus 0 & \pminus 0 & \pminus 0 & \pminus 0 & \pminus 0 & \pminus 0 & \pminus 0 & \pminus 0 & \pminus 0 & \pminus 0 & \pminus 0 & \pminus 0 & \pminus 0 & \pminus 0 & \pminus 0 & \pminus 0 & \pminus 0 & \pminus 0 & \pminus 0 & \pminus 0 & \pminus 0 & \pminus 0 & \pminus 0 \\
 \pminus 0 & \pminus 0 & \pminus 0 & \pminus 0 & \pminus 0 & \pminus 0 & \pminus 0 & \pminus 0 & \pminus 0 & \pminus 0 & \pminus 0 & \pminus 0 & \pminus 0 & \pminus 0 & \pminus 0 & \pminus 0 & \pminus 0 & \pminus 0 & \pminus 0 & \pminus 0 & \pminus 0 & \pminus 0 & \pminus 0 & \pminus 0 & \pminus 0 & \pminus 0 & \pminus 0 & \pminus 0 & \pminus 0 & \pminus 0 & \pminus 0 & \pminus 0 & \pminus 0 & \pminus 0 \\
 \pminus 0 & \pminus 0 & \pminus 0 & \pminus 0 & \pminus 0 & \pminus 0 & \pminus 0 & \pminus 0 & \pminus 0 & \pminus 0 & \pminus 0 & \pminus 0 & \pminus 0 & \pminus 0 & \pminus 0 & \pminus 0 & \pminus 0 & \pminus 0 & \pminus 0 & \pminus 0 & \pminus 0 & \pminus 0 & \pminus 0 & \pminus 0 & \pminus 0 & \pminus 0 & \pminus 0 & \pminus 0 & \pminus 0 & \pminus 0 & \pminus 0 & \pminus 0 & \pminus 0 & \pminus 0 \\
\end{array}
\right)
  }\,,
\end{align}

\begin{align}
\MM_4 = \scalemath{0.53}{
  \left(
\begin{array}{cccccccccccccccccccccccccccccccccc}
 \pminus 0 & \pminus 0 & \pminus 0 & \pminus 0 & \pminus 0 & \pminus 0 & \pminus 0 & \pminus 0 & \pminus 0 & \pminus 0 & \pminus 0 & \pminus 0 & \pminus 0 & \pminus 0 & \pminus 0 & \pminus 0 & \pminus 0 & \pminus 0 & \pminus 0 & \pminus 0 & \pminus 0 & \pminus 0 & \pminus 0 & \pminus 0 & \pminus 0 & \pminus 0 & \pminus 0 & \pminus 0 & \pminus 0 & \pminus 0 & \pminus 0 & \pminus 0 & \pminus 0 & \pminus 0 \\
 \pminus 0 & \pminus 0 & \pminus 0 & \pminus 0 & \pminus 0 & \pminus 0 & \pminus 0 & \pminus 0 & \pminus 0 & \pminus 0 & \pminus 0 & \pminus 0 & \pminus 0 & \pminus 0 & \pminus 0 & \pminus 0 & \pminus 0 & \pminus 0 & \pminus 0 & \pminus 0 & \pminus 0 & \pminus 0 & \pminus 0 & \pminus 0 & \pminus 0 & \pminus 0 & \pminus 0 & \pminus 0 & \pminus 0 & \pminus 0 & \pminus 0 & \pminus 0 & \pminus 0 & \pminus 0 \\
 \pminus 0 & \pminus 0 & \pminus 1 & \pminus 0 & \pminus 0 & \pminus 0 & \pminus 0 & \pminus 0 & \pminus 0 & \pminus 0 & \pminus 0 & \pminus 0 & \pminus 0 & \pminus 0 & \pminus 0 & \pminus 0 & \pminus 0 & \pminus 0 & \pminus 0 & \pminus 0 & \pminus 0 & \pminus 0 & \pminus 0 & \pminus 0 & \pminus 0 & \pminus 0 & \pminus 0 & \pminus 0 & \pminus 0 & \pminus 0 & \pminus 0 & \pminus 0 & \pminus 0 & \pminus 0 \\
 \pminus 0 & \pminus 0 & \pminus 0 & \pminus 0 & \pminus 0 & \pminus 0 & \pminus 0 & \pminus 0 & \pminus 0 & \pminus 0 & \pminus 0 & \pminus 0 & \pminus 0 & \pminus 0 & \pminus 0 & \pminus 0 & \pminus 0 & \pminus 0 & \pminus 0 & \pminus 0 & \pminus 0 & \pminus 0 & \pminus 0 & \pminus 0 & \pminus 0 & \pminus 0 & \pminus 0 & \pminus 0 & \pminus 0 & \pminus 0 & \pminus 0 & \pminus 0 & \pminus 0 & \pminus 0 \\
\end{array}
\right)
  }\,,
\end{align}

\begin{align}
\MM_5 = \scalemath{0.53}{
  \left(
\begin{array}{cccccccccccccccccccccccccccccccccc}
 \pminus 0 & \pminus 0 & \pminus 0 & \pminus 0 & \pminus 0 & \pminus 0 & \pminus 0 & \pminus 0 & \pminus 0 & \pminus 0 & \pminus 0 & \pminus 0 & \pminus 0 & \pminus 0 & \pminus 0 & \pminus 0 & \pminus 0 & \pminus 0 & \pminus 0 & \pminus 0 & \pminus 0 & \pminus 0 & \pminus 0 & \pminus 0 & \pminus 0 & \pminus 0 & \pminus 0 & \pminus 0 & \pminus 0 & \pminus 0 & \pminus 0 & \pminus 0 & \pminus 0 & \pminus 0 \\
 \pminus 0 & \pminus 0 & \pminus 0 & \pminus 0 & \pminus 0 & \pminus 0 & \pminus 0 & \pminus 0 & \pminus 0 & \pminus 0 & \pminus 0 & \pminus 0 & \pminus 0 & \pminus 0 & \pminus 0 & \pminus 0 & \pminus 0 & \pminus 0 & \pminus 0 & \pminus 0 & \pminus 0 & \pminus 0 & \pminus 0 & \pminus 0 & \pminus 0 & \pminus 0 & \pminus 0 & \pminus 0 & \pminus 0 & \pminus 0 & \pminus 0 & \pminus 0 & \pminus 0 & \pminus 0 \\
 \pminus 0 & \pminus 0 & \pminus 0 & \pminus 0 & \pminus 0 & \pminus 0 & \pminus 0 & \pminus 0 & \pminus 0 & \pminus 0 & \pminus 0 & \pminus 0 & \pminus 0 & \pminus 0 & \pminus 0 & \pminus 0 & \pminus 0 & \pminus 0 & \pminus 0 & \pminus 0 & \pminus 0 & \pminus 0 & \pminus 0 & \pminus 0 & \pminus 0 & \pminus 0 & \pminus 0 & \pminus 0 & \pminus 0 & \pminus 0 & \pminus 0 & \pminus 0 & \pminus 0 & \pminus 0 \\
 \pminus 0 & \pminus 0 & \pminus 0 & \pminus 0 & \pminus 0 & \pminus 0 & \pminus 0 & \pminus 0 & \pminus 0 & \pminus 0 & \pminus 0 & \pminus 0 & \pminus 0 & \pminus 0 & \pminus 0 & \pminus 0 & \pminus 0 & \pminus 0 & \pminus 0 & \pminus 0 & \pminus 0 & \pminus 0 & \pminus 0 & \pminus 0 & \pminus 0 & \pminus 0 & \pminus 0 & \pminus 0 & \pminus 0 & \pminus 0 & \pminus 0 & \pminus 0 & \pminus 0 & \pminus 0 \\
\end{array}
\right)
  }\,,
\end{align}

\begin{align}
\MM_6 = \scalemath{0.53}{
  \left(
\begin{array}{cccccccccccccccccccccccccccccccccc}
 \pminus 0 & \pminus 0 & \pminus 0 & \pminus 0 & \pminus 0 & \pminus 0 & \pminus 0 & \pminus 0 & \pminus 0 & \pminus 0 & \pminus 0 & \pminus 0 & \pminus 0 & \pminus 0 & \pminus 0 & \pminus 0 & \pminus 0 & \pminus 0 & \pminus 0 & \pminus 0 & \pminus 0 & \pminus 0 & \pminus 0 & \pminus 0 & \pminus 0 & \pminus 0 & \pminus 0 & \pminus 0 & \pminus 0 & \pminus 0 & \pminus 0 & \pminus 0 & \pminus 0 & \pminus 0 \\
 \pminus 0 & \pminus 0 & \pminus 0 & \pminus 0 & \pminus 0 & \pminus 0 & \pminus 0 & \pminus 0 & \pminus 0 & \pminus 0 & \pminus 0 & \pminus 0 & \pminus 0 & \pminus 0 & \pminus 0 & \pminus 0 & \pminus 0 & \pminus 0 & \pminus 0 & \pminus 0 & \pminus 0 & \pminus 0 & \pminus 0 & \pminus 0 & \pminus 0 & \pminus 0 & \pminus 0 & \pminus 0 & \pminus 0 & \pminus 0 & \pminus 0 & \pminus 0 & \pminus 0 & \pminus 0 \\
 \pminus 0 & \pminus 0 & \minus  2 & \pminus 0 & \pminus 0 & \pminus 0 & \pminus 0 & \pminus 0 & \pminus 0 & \pminus 0 & \pminus 0 & \pminus 0 & \pminus 0 & \pminus 0 & \pminus 0 & \pminus 0 & \pminus 0 & \pminus 0 & \pminus 0 & \pminus 0 & \pminus 0 & \pminus 0 & \pminus 0 & \pminus 0 & \pminus 0 & \pminus 0 & \pminus 0 & \pminus 0 & \pminus 0 & \pminus 0 & \pminus 0 & \pminus 0 & \pminus 0 & \pminus 0 \\
 \pminus 0 & \pminus 0 & \pminus 0 & \pminus 0 & \pminus 0 & \pminus 0 & \pminus 0 & \pminus 0 & \pminus 0 & \pminus 0 & \pminus 0 & \pminus 0 & \pminus 0 & \pminus 0 & \pminus 0 & \pminus 0 & \pminus 0 & \pminus 0 & \pminus 0 & \pminus 0 & \pminus 0 & \pminus 0 & \pminus 0 & \pminus 0 & \pminus 0 & \pminus 0 & \pminus 0 & \pminus 0 & \pminus 0 & \pminus 0 & \pminus 0 & \pminus 0 & \pminus 0 & \pminus 0 \\
\end{array}
\right)
  }\,,
\end{align}

\begin{align}
\MM_7 = \scalemath{0.53}{
	\left(
\begin{array}{cccccccccccccccccccccccccccccccccc}
 \pminus 0 & \pminus 0 & \pminus 0 & \pminus 0 & \pminus 0 & \pminus 0 & \pminus 0 & \pminus 0 & \pminus 0 & \pminus 0 & \pminus 0 & \pminus 0 & \pminus 0 & \pminus 0 & \pminus 0 & \pminus 0 & \pminus 0 & \pminus 0 & \pminus 0 & \pminus 0 & \pminus 0 & \pminus 0 & \pminus 0 & \pminus 0 & \pminus 0 & \pminus 0 & \pminus 0 & \pminus 0 & \pminus 0 & \pminus 0 & \pminus 0 & \pminus 0 & \pminus 0 & \pminus 0 \\
 \pminus 0 & \pminus 0 & \pminus 0 & \pminus 0 & \pminus 0 & \pminus 0 & \pminus 0 & \pminus 0 & \pminus 0 & \pminus 0 & \pminus 0 & \pminus 0 & \pminus 0 & \pminus 0 & \pminus 0 & \pminus 0 & \pminus 0 & \pminus 0 & \pminus 0 & \pminus 0 & \pminus 0 & \pminus 0 & \pminus 0 & \pminus 0 & \pminus 0 & \pminus 0 & \pminus 0 & \pminus 0 & \pminus 0 & \pminus 0 & \pminus 0 & \pminus 0 & \pminus 0 & \pminus 0 \\
 \pminus 0 & \pminus 0 & \pminus 0 & \pminus 0 & \pminus 0 & \pminus 0 & \pminus 0 & \pminus 0 & \pminus 0 & \pminus 0 & \pminus 0 & \pminus 0 & \pminus 0 & \pminus 0 & \pminus 0 & \pminus 0 & \pminus 0 & \pminus 0 & \pminus 0 & \pminus 0 & \pminus 0 & \pminus 0 & \pminus 0 & \pminus 0 & \pminus 0 & \pminus 0 & \pminus 0 & \pminus 0 & \pminus 0 & \pminus 0 & \pminus 0 & \pminus 0 & \pminus 0 & \pminus 0 \\
 \pminus 0 & \pminus 0 & \pminus 0 & \pminus 0 & \pminus 0 & \pminus 0 & \pminus 0 & \pminus 0 & \pminus 0 & \pminus 0 & \pminus 0 & \pminus 0 & \pminus 0 & \pminus 0 & \pminus 0 & \pminus 0 & \pminus 0 & \pminus 0 & \pminus 0 & \pminus 0 & \pminus 0 & \pminus 0 & \pminus 0 & \pminus 0 & \pminus 0 & \pminus 0 & \pminus 0 & \pminus 0 & \pminus 0 & \pminus 0 & \pminus 0 & \pminus 0 & \pminus 0 & \pminus 0 \\
\end{array}
\right)
}\,,
\end{align}

\begin{align}
\MM_8 = \scalemath{0.53}{
	\left(
\begin{array}{cccccccccccccccccccccccccccccccccc}
 \pminus 0 & \pminus 0 & \pminus 0 & \pminus 0 & \pminus 0 & \pminus 0 & \pminus 0 & \pminus 0 & \pminus 0 & \pminus 0 & \pminus 0 & \pminus 0 & \pminus 0 & \pminus 0 & \pminus 0 & \pminus 0 & \pminus 0 & \pminus 0 & \pminus 0 & \pminus 0 & \pminus 0 & \pminus 0 & \pminus 0 & \pminus 0 & \pminus 0 & \pminus 0 & \pminus 0 & \pminus 0 & \pminus 0 & \pminus 0 & \pminus 0 & \pminus 0 & \pminus 0 & \pminus 0 \\
 \pminus 0 & \pminus 0 & \pminus 0 & \pminus 0 & \pminus 0 & \pminus 0 & \pminus 0 & \pminus 0 & \pminus 0 & \pminus 0 & \pminus 0 & \pminus 0 & \pminus 0 & \pminus 0 & \pminus 0 & \pminus 0 & \pminus 0 & \pminus 0 & \pminus 0 & \pminus 0 & \pminus 0 & \pminus 0 & \pminus 0 & \pminus 0 & \pminus 0 & \pminus 0 & \pminus 0 & \pminus 0 & \pminus 0 & \pminus 0 & \pminus 0 & \pminus 0 & \pminus 0 & \pminus 0 \\
 \pminus 0 & \pminus 0 & \pminus 0 & \pminus 0 & \pminus 0 & \pminus 0 & \pminus 0 & \pminus 0 & \pminus 0 & \pminus 0 & \pminus 0 & \pminus 0 & \pminus 0 & \pminus 0 & \pminus 0 & \pminus 0 & \pminus 0 & \pminus 0 & \pminus 0 & \pminus 0 & \pminus 0 & \pminus 0 & \pminus 0 & \pminus 0 & \pminus 0 & \pminus 0 & \pminus 0 & \pminus 0 & \pminus 0 & \pminus 0 & \pminus 0 & \pminus 0 & \pminus 0 & \pminus 0 \\
 \pminus 0 & \pminus 0 & \pminus 0 & \pminus 0 & \pminus 0 & \pminus 0 & \pminus 0 & \pminus 0 & \pminus 0 & \pminus 0 & \pminus 0 & \pminus 0 & \pminus 0 & \pminus 0 & \pminus 0 & \pminus 0 & \pminus 0 & \pminus 0 & \pminus 0 & \pminus 0 & \pminus 0 & \pminus 0 & \pminus 0 & \pminus 0 & \pminus 0 & \pminus 0 & \pminus 0 & \pminus 0 & \pminus 0 & \pminus 0 & \pminus 0 & \pminus 0 & \pminus 0 & \pminus 0 \\
\end{array}
\right)
}\,,
\end{align}

\begin{align}
\MM_9 = \scalemath{0.53}{
	\left(
\begin{array}{cccccccccccccccccccccccccccccccccc}
 \pminus 0 & \pminus 0 & \pminus 0 & \pminus 0 & \pminus 0 & \pminus 0 & \pminus 0 & \pminus 0 & \pminus 0 & \pminus 0 & \pminus 0 & \pminus 0 & \pminus 0 & \pminus 0 & \pminus 0 & \pminus 0 & \pminus 0 & \pminus 0 & \pminus 0 & \pminus 0 & \pminus 0 & \pminus 0 & \pminus 0 & \pminus 0 & \pminus 0 & \pminus 0 & \pminus 0 & \pminus 0 & \pminus 0 & \pminus 0 & \pminus 0 & \pminus 0 & \pminus 0 & \pminus 0 \\
 \pminus 0 & \pminus 0 & \pminus 0 & \pminus 0 & \pminus 0 & \pminus 0 & \pminus 0 & \pminus 0 & \pminus 0 & \pminus 0 & \pminus 0 & \pminus 0 & \pminus 0 & \pminus 0 & \pminus 0 & \pminus 0 & \pminus 0 & \pminus 0 & \pminus 0 & \pminus 0 & \pminus 0 & \pminus 0 & \pminus 0 & \pminus 0 & \pminus 0 & \pminus 0 & \pminus 0 & \pminus 0 & \pminus 0 & \pminus 0 & \pminus 0 & \pminus 0 & \pminus 0 & \pminus 0 \\
 \pminus 0 & \pminus 0 & \pminus 0 & \pminus 0 & \pminus 0 & \pminus 0 & \pminus 0 & \pminus 0 & \pminus 0 & \pminus 0 & \pminus 0 & \pminus 0 & \pminus 0 & \pminus 0 & \pminus 0 & \pminus 0 & \pminus 0 & \pminus 0 & \pminus 0 & \pminus 0 & \pminus 0 & \pminus 0 & \pminus 0 & \pminus 0 & \pminus 0 & \pminus 0 & \pminus 0 & \pminus 0 & \pminus 0 & \pminus 0 & \pminus 0 & \pminus 0 & \pminus 0 & \pminus 0 \\
 \pminus 0 & \pminus 0 & \pminus 0 & \pminus 0 & \pminus 0 & \pminus 0 & \pminus 0 & \pminus 0 & \pminus 0 & \pminus 0 & \pminus 0 & \pminus 0 & \pminus 0 & \pminus 0 & \pminus 0 & \pminus 0 & \pminus 0 & \pminus 0 & \pminus 0 & \pminus 0 & \pminus 0 & \pminus 0 & \pminus 0 & \pminus 0 & \pminus 0 & \pminus 0 & \pminus 0 & \pminus 0 & \pminus 0 & \pminus 0 & \pminus 0 & \pminus 0 & \pminus 0 & \pminus 0 \\
\end{array}
\right)
}\,.
\end{align}

\subsection{Second integral family}
\label{dlog2ndIF}
For the second integral family, given in eq.~\eqref{eq:2Lfamily2},  we find: 

\begin{align}
\MM_1 = \scalemath{0.4}{
  \left(
\begin{array}{cccccccccccccccccccccccccccccccccccccccccc}
 \pminus 0 & \pminus 0 & \pminus 0 & \pminus 0 & \pminus 0 & \pminus 0 & \pminus 0 & \pminus 0 & \pminus 0 & \pminus 0 & \pminus 0 & \pminus 0 & \pminus 0 & \pminus 0 & \pminus 0 & \pminus 0 & \pminus 0 & \pminus 0 & \pminus 0 & \pminus 0 & \pminus 0 & \pminus 0 & \pminus 0 & \pminus 0 & \pminus 0 & \pminus 0 & \pminus 0 & \pminus 0 & \pminus 0 & \pminus 0 & \pminus 0 & \pminus 0 & \pminus 0 & \pminus 0 & \pminus 0 & \pminus 0 & \pminus 0 & \pminus 0 & \pminus 0 & \pminus 0 & \pminus 0 & \pminus 0 \\
 \pminus 0 & \pminus 0 & \pminus 0 & \pminus 0 & \pminus 0 & \pminus 0 & \pminus 0 & \pminus 0 & \pminus 0 & \pminus 0 & \pminus 0 & \pminus 0 & \pminus 0 & \pminus 0 & \pminus 0 & \pminus 0 & \pminus 0 & \pminus 0 & \pminus 0 & \pminus 0 & \pminus 0 & \pminus 0 & \pminus 0 & \pminus 0 & \pminus 0 & \pminus 0 & \pminus 0 & \pminus 0 & \pminus 0 & \pminus 0 & \pminus 0 & \pminus 0 & \pminus 0 & \pminus 0 & \pminus 0 & \pminus 0 & \pminus 0 & \pminus 0 & \pminus 0 & \pminus 0 & \pminus 0 & \pminus 0 \\
 \pminus 0 & \pminus 0 & \pminus 0 & \pminus 0 & \pminus 0 & \pminus 0 & \pminus 0 & \pminus 0 & \pminus 0 & \pminus 0 & \pminus 0 & \pminus 0 & \pminus 0 & \pminus 0 & \pminus 0 & \pminus 0 & \pminus 0 & \pminus 0 & \pminus 0 & \pminus 0 & \pminus 0 & \pminus 0 & \pminus 0 & \pminus 0 & \pminus 0 & \pminus 0 & \pminus 0 & \pminus 0 & \pminus 0 & \pminus 0 & \pminus 0 & \pminus 0 & \pminus 0 & \pminus 0 & \pminus 0 & \pminus 0 & \pminus 0 & \pminus 0 & \pminus 0 & \pminus 0 & \pminus 0 & \pminus 0 \\
\end{array}
\right)
  }\,,
\end{align}

\begin{align}
\MM_2 = \scalemath{0.4}{
  \left(
\begin{array}{cccccccccccccccccccccccccccccccccccccccccc}
 \pminus 0 & \pminus 0 & \pminus 0 & \pminus 0 & \pminus 0 & \pminus 0 & \pminus 0 & \pminus 0 & \pminus 0 & \pminus 0 & \pminus 0 & \pminus 0 & \pminus 0 & \pminus 0 & \pminus 0 & \pminus 0 & \pminus 0 & \pminus 0 & \pminus 0 & \pminus 0 & \pminus 0 & \pminus 0 & \pminus 0 & \pminus 0 & \pminus 0 & \pminus 0 & \pminus 0 & \pminus 0 & \pminus 0 & \pminus 0 & \pminus 0 & \pminus 0 & \pminus 0 & \pminus 0 & \pminus 0 & \pminus 0 & \pminus 0 & \pminus 0 & \pminus 0 & \pminus 0 & \pminus 0 & \pminus 0 \\
 \pminus 0 & \pminus 0 & \pminus 0 & \pminus 0 & \pminus 0 & \pminus 0 & \pminus 0 & \pminus 0 & \pminus 0 & \pminus 0 & \pminus 0 & \pminus 0 & \pminus 0 & \pminus 0 & \pminus 0 & \pminus 0 & \pminus 0 & \pminus 0 & \pminus 0 & \pminus 0 & \pminus 0 & \pminus 0 & \pminus 0 & \pminus 0 & \pminus 0 & \pminus 0 & \pminus 0 & \pminus 0 & \pminus 0 & \pminus 0 & \pminus 0 & \pminus 0 & \pminus 0 & \pminus 0 & \pminus 0 & \pminus 0 & \pminus 0 & \pminus 0 & \pminus 0 & \pminus 0 & \pminus 0 & \pminus 0 \\
 \pminus 0 & \pminus 0 & \pminus 0 & \pminus 0 & \pminus 0 & \pminus 0 & \pminus 0 & \pminus 0 & \pminus 0 & \pminus 0 & \pminus 0 & \pminus 0 & \pminus 0 & \pminus 0 & \pminus 0 & \pminus 0 & \pminus 0 & \pminus 0 & \pminus 0 & \pminus 0 & \pminus 0 & \pminus 0 & \pminus 0 & \pminus 0 & \pminus 0 & \pminus 0 & \pminus 0 & \pminus 0 & \pminus 0 & \pminus 0 & \pminus 0 & \pminus 0 & \pminus 0 & \pminus 0 & \pminus 0 & \pminus 0 & \pminus 0 & \pminus 0 & \pminus 0 & \pminus 0 & \pminus 0 & \pminus 0 \\
\end{array}
\right)
  }\,,
\end{align}

\begin{align}
\MM_3 = \scalemath{0.4}{
  \left(
\begin{array}{cccccccccccccccccccccccccccccccccccccccccc}
 \pminus 0 & \pminus 0 & \pminus 0 & \pminus 0 & \pminus 0 & \pminus 0 & \pminus 0 & \pminus 0 & \pminus 0 & \pminus 0 & \pminus 0 & \pminus 0 & \pminus 0 & \pminus 0 & \pminus 0 & \pminus 0 & \pminus 0 & \pminus 0 & \pminus 0 & \pminus 0 & \pminus 0 & \pminus 0 & \pminus 0 & \pminus 0 & \pminus 0 & \pminus 0 & \pminus 0 & \pminus 0 & \pminus 0 & \pminus 0 & \pminus 0 & \pminus 0 & \pminus 0 & \pminus 0 & \pminus 0 & \pminus 0 & \pminus 0 & \pminus 0 & \pminus 0 & \pminus 0 & \pminus 0 & \pminus 0 \\
 \pminus 0 & \pminus 0 & \pminus 0 & \pminus 0 & \pminus 0 & \pminus 0 & \pminus 0 & \pminus 0 & \pminus 0 & \pminus 0 & \pminus 0 & \pminus 0 & \pminus 0 & \pminus 0 & \pminus 0 & \pminus 0 & \pminus 0 & \pminus 0 & \pminus 0 & \pminus 0 & \pminus 0 & \pminus 0 & \pminus 0 & \pminus 0 & \pminus 0 & \pminus 0 & \pminus 0 & \pminus 0 & \pminus 0 & \pminus 0 & \pminus 0 & \pminus 0 & \pminus 0 & \pminus 0 & \pminus 0 & \pminus 0 & \pminus 0 & \pminus 0 & \pminus 0 & \pminus 0 & \pminus 0 & \pminus 0 \\
 \pminus 0 & \pminus 0 & \pminus 0 & \pminus 0 & \pminus 0 & \pminus 0 & \pminus 0 & \pminus 0 & \pminus 0 & \pminus 0 & \pminus 0 & \pminus 0 & \pminus 0 & \pminus 0 & \pminus 0 & \pminus 0 & \pminus 0 & \pminus 0 & \pminus 0 & \pminus 0 & \pminus 0 & \pminus 0 & \pminus 0 & \pminus 0 & \pminus 0 & \pminus 0 & \pminus 0 & \pminus 0 & \pminus 0 & \pminus 0 & \pminus 0 & \pminus 0 & \pminus 0 & \pminus 0 & \pminus 0 & \pminus 0 & \pminus 0 & \pminus 0 & \pminus 0 & \pminus 0 & \pminus 0 & \pminus 0 \\
\end{array}
\right)
  }\,,
\end{align}

\begin{align}
\MM_4 = \scalemath{0.4}{
  \left(
\begin{array}{cccccccccccccccccccccccccccccccccccccccccc}
 \pminus 0 & \pminus 0 & \pminus 0 & \pminus 0 & \pminus 0 & \pminus 0 & \pminus 0 & \pminus 0 & \pminus 0 & \pminus 0 & \pminus 0 & \pminus 0 & \pminus 0 & \pminus 0 & \pminus 0 & \pminus 0 & \pminus 0 & \pminus 0 & \pminus 0 & \pminus 0 & \pminus 0 & \pminus 0 & \pminus 0 & \pminus 0 & \pminus 0 & \pminus 0 & \pminus 0 & \pminus 0 & \pminus 0 & \pminus 0 & \pminus 0 & \pminus 0 & \pminus 0 & \pminus 0 & \pminus 0 & \pminus 0 & \pminus 0 & \pminus 0 & \pminus 0 & \pminus 0 & \pminus 0 & \pminus 0 \\
 \pminus 0 & \pminus 1 & \pminus 0 & \pminus 0 & \pminus 0 & \pminus 0 & \pminus 0 & \pminus 0 & \pminus 0 & \pminus 0 & \pminus 0 & \pminus 0 & \pminus 0 & \pminus 0 & \pminus 0 & \pminus 0 & \pminus 0 & \pminus 0 & \pminus 0 & \pminus 0 & \pminus 0 & \pminus 0 & \pminus 0 & \pminus 0 & \pminus 0 & \pminus 0 & \pminus 0 & \pminus 0 & \pminus 0 & \pminus 0 & \pminus 0 & \pminus 0 & \pminus 0 & \pminus 0 & \pminus 0 & \pminus 0 & \pminus 0 & \pminus 0 & \pminus 0 & \pminus 0 & \pminus 0 & \pminus 0 \\
 \pminus 1 & \pminus 0 & \pminus 1 & \pminus 0 & \pminus 0 & \pminus 0 & \pminus 0 & \pminus 0 & \pminus 0 & \pminus 0 & \pminus 0 & \pminus 0 & \pminus 0 & \pminus 0 & \pminus 0 & \pminus 0 & \pminus 0 & \pminus 0 & \pminus 0 & \pminus 0 & \pminus 0 & \pminus 0 & \pminus 0 & \pminus 0 & \pminus 0 & \pminus 0 & \pminus 0 & \pminus 0 & \pminus 0 & \pminus 0 & \pminus 0 & \pminus 0 & \pminus 0 & \pminus 0 & \pminus 0 & \pminus 0 & \pminus 0 & \pminus 0 & \pminus 0 & \pminus 0 & \pminus 0 & \pminus 0 \\
\end{array}
\right)
  }\,,
\end{align}

\begin{align}
\MM_5 = \scalemath{0.4}{
  \left(
\begin{array}{cccccccccccccccccccccccccccccccccccccccccc}
 \pminus 0 & \pminus 0 & \pminus 0 & \pminus 0 & \pminus 0 & \pminus 0 & \pminus 0 & \pminus 0 & \pminus 0 & \pminus 0 & \pminus 0 & \pminus 0 & \pminus 0 & \pminus 0 & \pminus 0 & \pminus 0 & \pminus 0 & \pminus 0 & \pminus 0 & \pminus 0 & \pminus 0 & \pminus 0 & \pminus 0 & \pminus 0 & \pminus 0 & \pminus 0 & \pminus 0 & \pminus 0 & \pminus 0 & \pminus 0 & \pminus 0 & \pminus 0 & \pminus 0 & \pminus 0 & \pminus 0 & \pminus 0 & \pminus 0 & \pminus 0 & \pminus 0 & \pminus 0 & \pminus 0 & \pminus 0 \\
 \pminus 0 & \pminus 0 & \pminus 0 & \pminus 0 & \pminus 0 & \pminus 0 & \pminus 0 & \pminus 0 & \pminus 0 & \pminus 0 & \pminus 0 & \pminus 0 & \pminus 0 & \pminus 0 & \pminus 0 & \pminus 0 & \pminus 0 & \pminus 0 & \pminus 0 & \pminus 0 & \pminus 0 & \pminus 0 & \pminus 0 & \pminus 0 & \pminus 0 & \pminus 0 & \pminus 0 & \pminus 0 & \pminus 0 & \pminus 0 & \pminus 0 & \pminus 0 & \pminus 0 & \pminus 0 & \pminus 0 & \pminus 0 & \pminus 0 & \pminus 0 & \pminus 0 & \pminus 0 & \pminus 0 & \pminus 0 \\
 \pminus 0 & \pminus 0 & \minus  2 & \pminus 0 & \pminus 0 & \pminus 0 & \pminus 0 & \pminus 0 & \pminus 0 & \pminus 0 & \pminus 0 & \pminus 0 & \pminus 0 & \pminus 0 & \pminus 0 & \pminus 0 & \pminus 0 & \pminus 0 & \pminus 0 & \pminus 0 & \pminus 0 & \pminus 0 & \pminus 0 & \pminus 0 & \pminus 0 & \pminus 0 & \pminus 0 & \pminus 0 & \pminus 0 & \pminus 0 & \pminus 0 & \pminus 0 & \pminus 0 & \pminus 0 & \pminus 0 & \pminus 0 & \pminus 0 & \pminus 0 & \pminus 0 & \pminus 0 & \pminus 0 & \pminus 0 \\
\end{array}
\right)
  }\,,
\end{align}

\begin{align}
\MM_6 = \scalemath{0.4}{
  \left(
\begin{array}{cccccccccccccccccccccccccccccccccccccccccc}
 \pminus 0 & \pminus 0 & \pminus 0 & \pminus 0 & \pminus 0 & \pminus 0 & \pminus 0 & \pminus 0 & \pminus 0 & \pminus 0 & \pminus 0 & \pminus 0 & \pminus 0 & \pminus 0 & \pminus 0 & \pminus 0 & \pminus 0 & \pminus 0 & \pminus 0 & \pminus 0 & \pminus 0 & \pminus 0 & \pminus 0 & \pminus 0 & \pminus 0 & \pminus 0 & \pminus 0 & \pminus 0 & \pminus 0 & \pminus 0 & \pminus 0 & \pminus 0 & \pminus 0 & \pminus 0 & \pminus 0 & \pminus 0 & \pminus 0 & \pminus 0 & \pminus 0 & \pminus 0 & \pminus 0 & \pminus 0 \\
 \pminus 0 & \minus  2 & \pminus 0 & \pminus 0 & \pminus 0 & \pminus 0 & \pminus 0 & \pminus 0 & \pminus 0 & \pminus 0 & \pminus 0 & \pminus 0 & \pminus 0 & \pminus 0 & \pminus 0 & \pminus 0 & \pminus 0 & \pminus 0 & \pminus 0 & \pminus 0 & \pminus 0 & \pminus 0 & \pminus 0 & \pminus 0 & \pminus 0 & \pminus 0 & \pminus 0 & \pminus 0 & \pminus 0 & \pminus 0 & \pminus 0 & \pminus 0 & \pminus 0 & \pminus 0 & \pminus 0 & \pminus 0 & \pminus 0 & \pminus 0 & \pminus 0 & \pminus 0 & \pminus 0 & \pminus 0 \\
 \pminus 0 & \pminus 0 & \pminus 0 & \pminus 0 & \pminus 0 & \pminus 0 & \pminus 0 & \pminus 0 & \pminus 0 & \pminus 0 & \pminus 0 & \pminus 0 & \pminus 0 & \pminus 0 & \pminus 0 & \pminus 0 & \pminus 0 & \pminus 0 & \pminus 0 & \pminus 0 & \pminus 0 & \pminus 0 & \pminus 0 & \pminus 0 & \pminus 0 & \pminus 0 & \pminus 0 & \pminus 0 & \pminus 0 & \pminus 0 & \pminus 0 & \pminus 0 & \pminus 0 & \pminus 0 & \pminus 0 & \pminus 0 & \pminus 0 & \pminus 0 & \pminus 0 & \pminus 0 & \pminus 0 & \pminus 0 \\
\end{array}
\right)
  }\,,
\end{align}

\begin{align}
\MM_7 = \scalemath{0.4}{
	\left(
\begin{array}{cccccccccccccccccccccccccccccccccccccccccc}
 \pminus 0 & \pminus 0 & \pminus 0 & \pminus 0 & \pminus 0 & \pminus 0 & \pminus 0 & \pminus 0 & \pminus 0 & \pminus 0 & \pminus 0 & \pminus 0 & \pminus 0 & \pminus 0 & \pminus 0 & \pminus 0 & \pminus 0 & \pminus 0 & \pminus 0 & \pminus 0 & \pminus 0 & \pminus 0 & \pminus 0 & \pminus 0 & \pminus 0 & \pminus 0 & \pminus 0 & \pminus 0 & \pminus 0 & \pminus 0 & \pminus 0 & \pminus 0 & \pminus 0 & \pminus 0 & \pminus 0 & \pminus 0 & \pminus 0 & \pminus 0 & \pminus 0 & \pminus 0 & \pminus 0 & \pminus 0 \\
 \pminus 0 & \pminus 0 & \pminus 0 & \pminus 0 & \pminus 0 & \pminus 0 & \pminus 0 & \pminus 0 & \pminus 0 & \pminus 0 & \pminus 0 & \pminus 0 & \pminus 0 & \pminus 0 & \pminus 0 & \pminus 0 & \pminus 0 & \pminus 0 & \pminus 0 & \pminus 0 & \pminus 0 & \pminus 0 & \pminus 0 & \pminus 0 & \pminus 0 & \pminus 0 & \pminus 0 & \pminus 0 & \pminus 0 & \pminus 0 & \pminus 0 & \pminus 0 & \pminus 0 & \pminus 0 & \pminus 0 & \pminus 0 & \pminus 0 & \pminus 0 & \pminus 0 & \pminus 0 & \pminus 0 & \pminus 0 \\
 \pminus 0 & \pminus 0 & \pminus 0 & \pminus 0 & \pminus 0 & \pminus 0 & \pminus 0 & \pminus 0 & \pminus 0 & \pminus 0 & \pminus 0 & \pminus 0 & \pminus 0 & \pminus 0 & \pminus 0 & \pminus 0 & \pminus 0 & \pminus 0 & \pminus 0 & \pminus 0 & \pminus 0 & \pminus 0 & \pminus 0 & \pminus 0 & \pminus 0 & \pminus 0 & \pminus 0 & \pminus 0 & \pminus 0 & \pminus 0 & \pminus 0 & \pminus 0 & \pminus 0 & \pminus 0 & \pminus 0 & \pminus 0 & \pminus 0 & \pminus 0 & \pminus 0 & \pminus 0 & \pminus 0 & \pminus 0 \\
\end{array}
\right)
}\,,
\end{align}

\begin{align}
\MM_8 = \scalemath{0.4}{
	\left(
\begin{array}{cccccccccccccccccccccccccccccccccccccccccc}
 \pminus 0 & \pminus 0 & \pminus 0 & \pminus 0 & \pminus 0 & \pminus 0 & \pminus 0 & \pminus 0 & \pminus 0 & \pminus 0 & \pminus 0 & \pminus 0 & \pminus 0 & \pminus 0 & \pminus 0 & \pminus 0 & \pminus 0 & \pminus 0 & \pminus 0 & \pminus 0 & \pminus 0 & \pminus 0 & \pminus 0 & \pminus 0 & \pminus 0 & \pminus 0 & \pminus 0 & \pminus 0 & \pminus 0 & \pminus 0 & \pminus 0 & \pminus 0 & \pminus 0 & \pminus 0 & \pminus 0 & \pminus 0 & \pminus 0 & \pminus 0 & \pminus 0 & \pminus 0 & \pminus 0 & \pminus 0 \\
 \pminus 0 & \pminus 0 & \pminus 0 & \pminus 0 & \pminus 0 & \pminus 0 & \pminus 0 & \pminus 0 & \pminus 0 & \pminus 0 & \pminus 0 & \pminus 0 & \pminus 0 & \pminus 0 & \pminus 0 & \pminus 0 & \pminus 0 & \pminus 0 & \pminus 0 & \pminus 0 & \pminus 0 & \pminus 0 & \pminus 0 & \pminus 0 & \pminus 0 & \pminus 0 & \pminus 0 & \pminus 0 & \pminus 0 & \pminus 0 & \pminus 0 & \pminus 0 & \pminus 0 & \pminus 0 & \pminus 0 & \pminus 0 & \pminus 0 & \pminus 0 & \pminus 0 & \pminus 0 & \pminus 0 & \pminus 0 \\
 \pminus 0 & \pminus 0 & \pminus 0 & \pminus 0 & \pminus 0 & \pminus 0 & \pminus 0 & \pminus 0 & \pminus 0 & \pminus 0 & \pminus 0 & \pminus 0 & \pminus 0 & \pminus 0 & \pminus 0 & \pminus 0 & \pminus 0 & \pminus 0 & \pminus 0 & \pminus 0 & \pminus 0 & \pminus 0 & \pminus 0 & \pminus 0 & \pminus 0 & \pminus 0 & \pminus 0 & \pminus 0 & \pminus 0 & \pminus 0 & \pminus 0 & \pminus 0 & \pminus 0 & \pminus 0 & \pminus 0 & \pminus 0 & \pminus 0 & \pminus 0 & \pminus 0 & \pminus 0 & \pminus 0 & \pminus 0 \\
\end{array}
\right)
}\,,
\end{align}

\begin{align}
\MM_9 = \scalemath{0.4}{
	\left(
\begin{array}{cccccccccccccccccccccccccccccccccccccccccc}
 \pminus 0 & \pminus 0 & \pminus 0 & \pminus 0 & \pminus 0 & \pminus 0 & \pminus 0 & \pminus 0 & \pminus 0 & \pminus 0 & \pminus 0 & \pminus 0 & \pminus 0 & \pminus 0 & \pminus 0 & \pminus 0 & \pminus 0 & \pminus 0 & \pminus 0 & \pminus 0 & \pminus 0 & \pminus 0 & \pminus 0 & \pminus 0 & \pminus 0 & \pminus 0 & \pminus 0 & \pminus 0 & \pminus 0 & \pminus 0 & \pminus 0 & \pminus 0 & \pminus 0 & \pminus 0 & \pminus 0 & \pminus 0 & \pminus 0 & \pminus 0 & \pminus 0 & \pminus 0 & \pminus 0 & \pminus 0 \\
 \pminus 0 & \pminus 0 & \pminus 0 & \pminus 0 & \pminus 0 & \pminus 0 & \pminus 0 & \pminus 0 & \pminus 0 & \pminus 0 & \pminus 0 & \pminus 0 & \pminus 0 & \pminus 0 & \pminus 0 & \pminus 0 & \pminus 0 & \pminus 0 & \pminus 0 & \pminus 0 & \pminus 0 & \pminus 0 & \pminus 0 & \pminus 0 & \pminus 0 & \pminus 0 & \pminus 0 & \pminus 0 & \pminus 0 & \pminus 0 & \pminus 0 & \pminus 0 & \pminus 0 & \pminus 0 & \pminus 0 & \pminus 0 & \pminus 0 & \pminus 0 & \pminus 0 & \pminus 0 & \pminus 0 & \pminus 0 \\
 \pminus 0 & \pminus 0 & \pminus 0 & \pminus 0 & \pminus 0 & \pminus 0 & \pminus 0 & \pminus 0 & \pminus 0 & \pminus 0 & \pminus 0 & \pminus 0 & \pminus 0 & \pminus 0 & \pminus 0 & \pminus 0 & \pminus 0 & \pminus 0 & \pminus 0 & \pminus 0 & \pminus 0 & \pminus 0 & \pminus 0 & \pminus 0 & \pminus 0 & \pminus 0 & \pminus 0 & \pminus 0 & \pminus 0 & \pminus 0 & \pminus 0 & \pminus 0 & \pminus 0 & \pminus 0 & \pminus 0 & \pminus 0 & \pminus 0 & \pminus 0 & \pminus 0 & \pminus 0 & \pminus 0 & \pminus 0 \\
\end{array}
\right)
}\,.
\end{align}
%
%
%


\bibliographystyle{JHEP}
\bibliography{references}

\end{document}